\documentclass[12pt]{article}
\usepackage[cmtip,arrow]{xy}\usepackage{pb-diagram,pb-xy}%

\input xy
\xyoption{all}
\input xy
\xyoption{all}
\usepackage{heck}
\usepackage{cite}
\usepackage{graphicx}
\usepackage{makeidx}
\usepackage{multicol}
\usepackage{geometry}
\usepackage{amsfonts}
\usepackage{mathrsfs}
\usepackage{amssymb}
\usepackage{amsmath}%

\usepackage{lscape}%package to print horizontally displaced tables

\providecommand{\U}[1]{\protect\rule{.1in}{.1in}}
\numberwithin{equation}{section}

\newcommand{\bea}{\begin{eqnarray}}
\newcommand{\eea}{\end{eqnarray}}
\newcommand{\be}{\begin{equation}}
\newcommand{\ee}{\end{equation}}

\newcommand{\bem}{\begin{pmatrix}}
\newcommand{\eem}{\end{pmatrix}}

\def\c{\gamma}

\def\U{\Upsilon}

\def\cd{{\cal D}}

\def\cn{{\cal N}}

\def\cq{{\cal Q}}

\def\cs{{\cal S}}

\def\cw{{\cal W}}

\def \N {{\mathcal N}}
\def \Z {{\mathbb Z}}
\def \C {{\mathbb C}}
\def \R {{\mathbb R}}

\xyoption{arc}

\begin{document}

\bibliographystyle{utphys}

\date{July, 2011}

%\preprint{arXiv:????.????}

\institution{SISSA}{\centerline{${}^{1}$SISSA, Scuola Internazionale Superiore di Studi Avanzati,
I-34100 Trieste, ITALY}}

\title{On Arnold's $14$ `exceptional' $\cn=2$ superconformal gauge theories}
%EndExpansion
%

\authors{Sergio Cecotti\worksat{\SISSA}\footnote{e-mail: {\tt cecotti@sissa.it}} and Michele Del Zotto\worksat{\SISSA}\footnote{e-mail: {\tt eledelz@gmail.com}}}%
%EndExpansion

\abstract{We study the four--dimensional superconformal $\cn=2$ gauge theories engineered by  the Type IIB superstring on Arnold's $14$ exceptional unimodal singularities (\textit{a.k.a.}\! Arnold's strange duality list), thus extending the methods of \texttt{arXiv:1006.3435} to singularities which are not the direct sum of minimal ones. In particular, we compute their BPS spectra in several `strongly coupled' chambers.

From the TBA side, we construct ten new \emph{periodic} $Y$--systems, providing additional evidence for the 
existence of a periodic $Y$--system for 
each isolated quasi--homogeneous singularity with $\hat c<2$ (more generally, for each $\cn=2$ superconformal theory with a finite BPS chamber whose chiral primaries have dimensions of the form $\mathbb{N}/\ell$).}

 %that to each isolated quasi--homogeneous singularity with $\hat c<2$ there corresponds a periodic $Y$--system.}

%TCIMACRO{\TeXButton{Maketitle}{\maketitle}}%
%BeginExpansion
\maketitle
%EndExpansion

\tableofcontents

\pagebreak

\section{Introduction}

Supersymmetric $\cn=2$ theories in four dimensions are an interesting laboratory to understand Quantum Field Theory at strong coupling \cite{Seiberg:1994rs,Seiberg:1994rs2,Gaiotto:2008cd,Gaiotto:2009hg,gaiotto}. In these theories many interesting physical quantities are protected by supersymmetry, and hence exactly calculable. In particular, the precise spectrum of BPS states may be determined (in principle) by a variety of methods \cite{Seiberg:1994rs,Seiberg:1994rs2,Gaiotto:2008cd,Gaiotto:2009hg,douglas1,douglas2,fiol1,lerche,fiol2,denef}. An especially simple and elegant technique was introduced in ref.\!\cite{cnv}, based on the analysis of the quantum monodromy $\mathbb{M}(q)$ ---
the basic wall--crossing invariant \cite{ks1,Gaiotto:2008cd,Gaiotto:2009hg,Dimofte:2009bv,Dimofte:2009tm,Cecotti:2009uf}
 --- in combination with the quantum cluster algebra formalism \cite{MR1887642,MR2004457,MR2383126,MR2132323,fominIV,cluster-intro,qd-cluster,clqd2,qd-pentagon}.\smallskip

In ref.\!\cite{cnv} a large class of $4d$  $\cn=2$ theories were discussed in detail. Those theories are labelled by a pair $(G,G^\prime)$ of simply--laced Lie algebras, and are UV superconformal. They belong to the more general class of $4d$ models which may be geometrically engineered by considering the Type IIB superstring on the geometry $\R^{3,1}\times \mathscr{H}$  \cite{shapere}, where $\mathscr{H}\subset \C^4$ is a local $3$--CY hypersurface specified by a polynomial equation $$\mathscr{H}\colon \ f(x_1,x_2,x_3,x_4)=0.$$
The resulting four--dimensional theory is $\cn=2$ \emph{superconformal} iff the defining polynomial of $\mathscr{H}$, $f(x_i)$, is \emph{quasi--homogeneous}, which implies that $\mathscr{H}$ is singular at the origin. The four--dimensional theory engineered on a \emph{smooth} hypersurface $f(x_i)=0$ is then physically interpreted as the superconformal $\cn=2$ theory associated to the
 the leading quasi--homogeneous part $f_0(x_i)$ of the polynomial $f(x_i)$, deformed by a set of \emph{relevant} operators corresponding to the lower degree part of the polynomial, \textit{i.e.}\! to $\Delta f\equiv f(x_i)-f_0(x_i)$.

In refs.\!\cite{Gukov:1999ya,shapere} it was shown that the singularity $f_0(x_i)$ is at finite distance in the complex moduli  if it satisfies the condition
\begin{equation}\label{gwv}
\sum_{i=1}^4 q_i >1,
\end{equation}
where the weights $q_i$ of the quasi--homogeneous polynomial $f_0(x_i)$ are defined trough the identity $\lambda\, f_0(x_i)=f_0(\lambda^{q_i}\,x_i)$, $\lambda\in \C^*$. In the $2d$ language \cite{min1,min2,rings,Cecotti:1993rm}, the condition \eqref{gwv} is equivalent to the statement that the Landau--Ginzburg model with superpotential $W\equiv f_0(x_i)$ has central charge $\hat{c}<2$.  As a consequence, if the homogeneous part of the defining polynomial, $f_0(x_i)$, satisfies eqn.\eqref{gwv}, the geometry $\mathbb{R}^{3,1}\times\mathscr{H}$ is a valid Type IIB background, and the geometrical engineering produces a consistent  $4d$ $\cn=2$ quantum field theory which, typically, has no weakly coupled Lagrangian description.

The $(G,G^\prime)$ models studied in ref.\!\cite{cnv} correspond to the special case in which $f_0(x_i)$ is the direct sum of two quasi--homogeneous polynomials 
\begin{equation}\label{decoupled}f_0(x_i)= W_G(x_1,x_2)+W_{G^\prime}(x_3,x_4),\end{equation} 
where $W_G(x,y)$ stands for the quasi--homogeneous polynomial describing the minimal singularity associated to the $ADE$ algebra $G$ \cite{ar,min1,min2}. Of course, the general polynomial $f_0(x_i)$ satisfying eqn.\eqref{gwv} has not the `decoupled' form of eqn.\eqref{decoupled}. Thus one is lead to ask for the extension of the methods and results of \cite{cnv} to singular hypersurfaces of more general form.

Such an extension is the main purpose of the present paper. There is a particularly important class of non--minimal singularities, namely Arnold's $14$ exceptional unimodal singularities \cite{ar,eb}. They have $\hat{c}<2$, and hence define superconformal $\cn=2$ theories in four dimensions. These $14$ models are, in a sense, the simplest $\cn=2$ superconformal gauge theories which are \emph{not} complete in the sense of ref.\!\cite{CV11}. The associated $14$ singularities naturally appear in many different areas of mathematics, and in particular in the representation theory of path algebras of quivers with relations \cite{le1,le2} (for a review \cite{le3}), which is a natural mathematical arena for understanding the BPS spectra of $\cn=2$ theories \cite{douglas1,douglas2,denef,CV11}. Hence this class of $\cn=2$ models appears to be `exceptional' from the mathematical side as well as from the physical one.

These $14$ `exceptional' gauge theories are the subject of our study. Actually, four of them are of the $(G,G^\prime)$ form, and were already analyzed in \cite{cnv}. The other $10$ are novel. In the process we introduce some combinatorial technique which may be of use in computing the BPS spectrum of many other interesting $\cn=2$ theories, as we illustrate in some examples.

When the hypersuface $\mathscr{H}$ has the special form $$0=f(x_i)\equiv g(x_1,x_2)+x_3x_4,$$ the four dimensional $\cn=2$ theory may also be engineered by considering the Abelian $(2,0)$ six dimensional theory on the curve $\{g(x_1,x_2)=0\}\subset \C^2$ \cite{cnv,shapere}. From the point of view of singularity and algebra representation theory, the equivalence of the two constructions from $10d$ and $6d$ is just the
Kn\"orrer--Solberg periodicity \cite{perrev} which directly implies the equality of BPS spectra.\smallskip

The results of \cite{cnv} and \cite{Gukov:1999ya,shapere}
 have a peculiar implication from the point of view of  
the Thermodynamical Bethe Ansatz \cite{zamolodchikovTBA}: They suggest the conjecture\footnote{\ From a physical viewpoint (\textit{i.e.}\! arguing trough string theory), this statement is equivalent to the conjecture that all $4d$ $\cn=2$ models engineered on such a singular hypersurface have at least one chamber with a finite BPS spectrum.} that to each isolated quasi--homogeneous hypersurface singularity, having $\hat{c}<2$, there is associated a TBA $Y$--system which is periodic (the two--Dynkin diagrams $Y$--systems \cite{keller-periodicity} corresponding to direct sums of minimal singularities as in eqn.\eqref{decoupled}, \cite{cnv}). Here we check this prediction for the $14$  Arnold exceptional singularities, including the precise value $\ell$ of the minimal period.
It will be highly desirable to have a direct proof of this correspondence, making explicit the underlying connection between singularity theory and cluster categories, in the spirit of ref.\!\cite{keller-periodicity}. 

More in general, one expects a $Y$--system of period $\ell$ to be associated to any $\cn=2$ superconformal model having a BPS chamber with a finite spectrum and whose chiral primary fields have dimensions of the form $\mathbb{N}/\ell$. 
\medskip

The paper is organized as follows. In section 2 we introduce Arnold's exceptional singularities and their properties. In section
3 we discuss the elementary properties of the corresponding $\cn=2$ theories, and in particular identify the quivers and superpotentials of the corresponding SQM. In section 4 we review the part of the results of \cite{cnv} we need. In section 5 we introduce the combinatorial techniques we use to implement the strategy of \cite{cnv}. Section 6 contains a side example, $SU(2)$ SQCD with $N_f=4$, which illustrates the power of the method in a gauge model not of the Arnold class. In section 7 we present the BPS spectra of the Arnold exceptional $\cn=2$ theories in diverse chambers. In section 8 we discuss the periodicity of the related $Y$--systems and check their consistence with the physical expectations. The three appendices contain some more technical detail for the benefit of the interested reader.

\section{Arnold's $14$ exceptional unimodal singularities}

The $14$ Arnold exceptional unimodal singularities (at the quasi--homogeneous value of the modulus) are written in table \ref{arnoldtable} as polynomials $W(x,y,z)$ in the three complex variables $x,y,z$. The local CY $3$--fold $\mathscr{H}$, on which we engineer the corresponding $\cn=2$ model, is then given by the hypersurface in $\C^4$
\begin{equation}
W(x,y,z)+u^2+\text{lower terms}=0.
\end{equation}

\begin{table}
\caption{Arnold's $14$ exceptional singularities}
\begin{center}
\begin{tabular}{|c|c|c|c|}\hline
name & polynomial $W(x,y,z)$ & weights $q_i$ & Coxeter--Dynkin diagram \\\hline
$\begin{matrix}\\ E_{12}\\ \phantom{a}\end{matrix}$ & $\begin{matrix}\\ x^3+y^7+z^2\\ \phantom{a}\end{matrix}$ & $\begin{matrix}\\ 1/3, 1/7, 1/2\\ \phantom{a}\end{matrix}$ &
$\begin{gathered}\xymatrix{\bullet \ar@{-}[r]\ar@{-}[d]\ar@{..}[dr]&\bullet \ar@{-}[r]\ar@{-}[d]\ar@{..}[dr]&\bullet \ar@{-}[r]\ar@{-}[d]\ar@{..}[dr]&\bullet \ar@{-}[r]\ar@{-}[d]\ar@{..}[dr]&\bullet \ar@{-}[r]\ar@{-}[d]\ar@{..}[dr]&\bullet \ar@{-}[d]\\
\bullet \ar@{-}[r]&\bullet \ar@{-}[r]&\bullet \ar@{-}[r]&\bullet \ar@{-}[r]&\bullet \ar@{-}[r]&\bullet }\end{gathered}$\\\hline
$\begin{matrix}\\ E_{13}\\ \phantom{a}\end{matrix}$ & $\begin{matrix}\\ x^3+xy^5+z^2\\ \phantom{a}\end{matrix}$ & $\begin{matrix}\\ 1/3, 2/15, 1/2\\ \phantom{a}\end{matrix}$ &
$\begin{gathered}\xymatrix{\bullet \ar@{-}[r]\ar@{-}[d]\ar@{..}[dr]&\bullet \ar@{-}[r]\ar@{-}[d]\ar@{..}[dr]&\bullet \ar@{-}[r]\ar@{-}[d]\ar@{..}[dr]&\bullet \ar@{-}[r]\ar@{-}[d]\ar@{..}[dr]&\bullet \ar@{-}[r]\ar@{-}[d]\ar@{..}[dr]&\bullet \ar@{-}[d]\ar@{..}[dr]&\\
\bullet \ar@{-}[r]&\bullet \ar@{-}[r]&\bullet \ar@{-}[r]&\bullet \ar@{-}[r]&\bullet \ar@{-}[r]&\bullet\ar@{-}[r]&\bullet }\end{gathered}$\\\hline
$\begin{matrix}\\ E_{14}\\ \phantom{a}\end{matrix}$ & $\begin{matrix}\\ x^3+y^8+z^2\\ \phantom{a}\end{matrix}$ & $\begin{matrix}\\ 1/3, 1/8, 1/2\\ \phantom{a}\end{matrix}$ &
$\begin{gathered}\xymatrix{\bullet \ar@{-}[r]\ar@{-}[d]\ar@{..}[dr]&\bullet \ar@{-}[r]\ar@{-}[d]\ar@{..}[dr]&\bullet \ar@{-}[r]\ar@{-}[d]\ar@{..}[dr]&\bullet \ar@{-}[r]\ar@{-}[d]\ar@{..}[dr]&\bullet \ar@{-}[r]\ar@{-}[d]\ar@{..}[dr]&\bullet \ar@{-}[d]\ar@{..}[dr]&\bullet\ar@{-}[l]\ar@{-}[d]\\
\bullet \ar@{-}[r]&\bullet \ar@{-}[r]&\bullet \ar@{-}[r]&\bullet \ar@{-}[r]&\bullet \ar@{-}[r]&\bullet\ar@{-}[r]&\bullet }\end{gathered}$\\\hline
$\begin{matrix}\\ Z_{11}\\ \phantom{a}\end{matrix}$ & $\begin{matrix}\\ x^3y+y^5+z^2\\ \phantom{a}\end{matrix}$ & $\begin{matrix}\\ 4/15, 1/5, 1/2\\ \phantom{a}\end{matrix}$ &
$\begin{gathered}\xymatrix{\bullet \ar@{-}[r]\ar@{-}[d]&\bullet \ar@{-}[r]\ar@{-}[d]\ar@{..}[dl]&\bullet \ar@{-}[d]\ar@{..}[dl] & \\
\bullet \ar@{-}[r]\ar@{-}[d]\ar@{..}[dr]&\bullet \ar@{-}[r]\ar@{-}[d]\ar@{..}[dr]&\bullet \ar@{-}[r]\ar@{-}[d]\ar@{..}[dr]&\bullet \ar@{-}[d]\\
\bullet \ar@{-}[r]&\bullet \ar@{-}[r]&\bullet \ar@{-}[r]&\bullet\\
}\end{gathered}$\\\hline
$\begin{matrix}\\ Z_{12}\\ \phantom{a}\end{matrix}$ & $\begin{matrix}\\ x^3y+xy^4+z^2\\ \phantom{a}\end{matrix}$ & $\begin{matrix}\\ 3/11, 2/11, 1/2\\ \phantom{a}\end{matrix}$ &
$\begin{gathered}\xymatrix{\bullet \ar@{-}[r]\ar@{-}[d]&\bullet \ar@{-}[r]\ar@{-}[d]\ar@{..}[dl]&\bullet \ar@{-}[d]\ar@{..}[dl] & &\\
\bullet \ar@{-}[r]\ar@{-}[d]\ar@{..}[dr]&\bullet \ar@{-}[r]\ar@{-}[d]\ar@{..}[dr]&\bullet \ar@{-}[r]\ar@{-}[d]\ar@{..}[dr]&\bullet \ar@{-}[d] \ar@{..}[dr]&\\
\bullet \ar@{-}[r]&\bullet \ar@{-}[r]&\bullet \ar@{-}[r]&\bullet \ar@{-}[r]&\bullet\\
}\end{gathered}$\\\hline
$\begin{matrix}\\ Z_{13}\\ \phantom{a}\end{matrix}$ & $\begin{matrix}\\ x^3y+y^6+z^2\\ \phantom{a}\end{matrix}$ & $\begin{matrix}\\ 5/18, 1/6, 1/2\\ \phantom{a}\end{matrix}$ &
$\begin{gathered}\xymatrix{\bullet \ar@{-}[r]\ar@{-}[d]&\bullet \ar@{-}[r]\ar@{-}[d]\ar@{..}[dl]&\bullet \ar@{-}[d]\ar@{..}[dl] & &\\
\bullet \ar@{-}[r]\ar@{-}[d]\ar@{..}[dr]&\bullet \ar@{-}[r]\ar@{-}[d]\ar@{..}[dr]&\bullet \ar@{-}[r]\ar@{-}[d]\ar@{..}[dr]&\bullet \ar@{-}[d] \ar@{..}[dr]\ar@{-}[r]&\bullet \ar@{-}[d]\\
\bullet \ar@{-}[r]&\bullet \ar@{-}[r]&\bullet \ar@{-}[r]&\bullet \ar@{-}[r]&\bullet\\
}\end{gathered}$\\\hline
\end{tabular}
\end{center}
\label{arnoldtable}
\end{table}

\begin{table*}
%\caption{Arnold's $14$ exceptional singularities}
\begin{center}
\begin{tabular}{|c|c|c|c|}\hline
name & polynomial $W(x,y,z)$ & weights $q_i$ & Coxeter--Dynkin diagram \\\hline
$\begin{matrix}\\ W_{12}\\ \phantom{a}\end{matrix}$ & $\begin{matrix}\\ x^4+y^5+z^2\\ \phantom{a}\end{matrix}$ & $\begin{matrix}\\ 1/4, 1/5, 1/2\\ \phantom{a}\end{matrix}$ &
$\begin{gathered}\xymatrix{\bullet \ar@{-}[r]\ar@{-}[d]&\bullet \ar@{-}[r]\ar@{-}[d]\ar@{..}[dl]&\bullet \ar@{-}[d]\ar@{..}[dl]\ar@{-}[r]& \bullet \ar@{-}[d]\ar@{..}[dl]\\
\bullet \ar@{-}[r]\ar@{-}[d]\ar@{..}[dr]&\bullet \ar@{-}[r]\ar@{-}[d]\ar@{..}[dr]&\bullet \ar@{-}[r]\ar@{-}[d]\ar@{..}[dr]&\bullet \ar@{-}[d]\\
\bullet \ar@{-}[r]&\bullet \ar@{-}[r]&\bullet \ar@{-}[r]&\bullet\\
}\end{gathered}$\\\hline
$\begin{matrix}\\ W_{13}\\ \phantom{a}\end{matrix}$ & $\begin{matrix}\\ x^4+xy^4+z^2\\ \phantom{a}\end{matrix}$ & $\begin{matrix}\\ 1/4, 3/16, 1/2\\ \phantom{a}\end{matrix}$ &
$\begin{gathered}\xymatrix{\bullet \ar@{-}[r]\ar@{-}[d]&\bullet \ar@{-}[r]\ar@{-}[d]\ar@{..}[dl]&\bullet \ar@{-}[d]\ar@{..}[dl]\ar@{-}[r]& \bullet \ar@{-}[d]\ar@{..}[dl] &\\
\bullet \ar@{-}[r]\ar@{-}[d]\ar@{..}[dr]&\bullet \ar@{-}[r]\ar@{-}[d]\ar@{..}[dr]&\bullet \ar@{-}[r]\ar@{-}[d]\ar@{..}[dr]&\bullet \ar@{-}[d] \ar@{..}[dr]&\\
\bullet \ar@{-}[r]&\bullet \ar@{-}[r]&\bullet \ar@{-}[r]&\bullet \ar@{-}[r]&\bullet\\
}\end{gathered}$\\\hline
$\begin{matrix}\\ Q_{10}\\ \phantom{a}\end{matrix}$ & $\begin{matrix}\\ x^2z+y^3+z^4\\ \phantom{a}\end{matrix}$ & $\begin{matrix}\\ 3/8, 1/3, 1/4\\ \phantom{a}\end{matrix}$ &
$\begin{gathered}\xymatrix{\bullet\ar@{-}[r]\ar@{-}@/_/[dd]&\bullet \ar@{..}[ddl]\ar@{-}@/_/[dd]&\\
\bullet \ar@{-}[r]\ar@{-}[d]&\bullet \ar@{-}[d]\ar@{..}[dl]& \\
\bullet \ar@{-}[r]\ar@{-}[d]\ar@{..}[dr]&\bullet \ar@{-}[r]\ar@{-}[d]\ar@{..}[dr]&\bullet \ar@{-}[d] \\
\bullet \ar@{-}[r]&\bullet \ar@{-}[r]&\bullet\\
}\end{gathered}$\\\hline
$\begin{matrix}\\ Q_{11}\\ \phantom{a}\end{matrix}$ & $\begin{matrix}\\ x^2z+y^3+yz^3\\ \phantom{a}\end{matrix}$ & $\begin{matrix}\\ 7/18, 1/3, 2/9\\ \phantom{a}\end{matrix}$ &
$\begin{gathered}\xymatrix{\bullet\ar@{-}[r]\ar@{-}@/_/[dd]&\bullet \ar@{..}[ddl]\ar@{-}@/_/[dd]&&\\
\bullet \ar@{-}[r]\ar@{-}[d]&\bullet \ar@{-}[d]\ar@{..}[dl]& &\\
\bullet \ar@{-}[r]\ar@{-}[d]\ar@{..}[dr]&\bullet \ar@{-}[r]\ar@{-}[d]\ar@{..}[dr]&\bullet \ar@{-}[d] \ar@{..}[dr]& \\
\bullet \ar@{-}[r]&\bullet \ar@{-}[r]&\bullet \ar@{-}[r]&\bullet\\
}\end{gathered}$\\\hline
$\begin{matrix}\\ Q_{12}\\ \phantom{a}\end{matrix}$ & $\begin{matrix}\\ x^2z+y^3+z^5\\ \phantom{a}\end{matrix}$ & $\begin{matrix}\\ 2/5, 1/3, 1/5\\ \phantom{a}\end{matrix}$ &
$\begin{gathered}\xymatrix{\bullet\ar@{-}[r]\ar@{-}@/_/[dd]&\bullet \ar@{..}[ddl]\ar@{-}@/_/[dd]&&\\
\bullet \ar@{-}[r]\ar@{-}[d]&\bullet \ar@{-}[d]\ar@{..}[dl]& &\\
\bullet \ar@{-}[r]\ar@{-}[d]\ar@{..}[dr]&\bullet \ar@{-}[r]\ar@{-}[d]\ar@{..}[dr]&\bullet \ar@{-}[r]\ar@{-}[d]\ar@{..}[dr]&\bullet\ar@{-}[d] \\
\bullet \ar@{-}[r]&\bullet \ar@{-}[r]&\bullet \ar@{-}[r]&\bullet\\
}\end{gathered}$\\\hline
\end{tabular}
\end{center}
\label{arnoldtable2}
\end{table*}%

\begin{table*}
%\caption{Arnold's $14$ exceptional singularities}
\begin{center}
\begin{tabular}{|c|c|c|c|}\hline
name & polynomial $W(x,y,z)$ & weights $q_i$ & Coxeter--Dynkin diagram \\\hline
$\begin{matrix}\\ S_{11}\\ \phantom{a}\end{matrix}$ & $\begin{matrix}\\ x^2z+yz^2+y^4\\ \phantom{a}\end{matrix}$ & $\begin{matrix}\\ 5/16, 1/4, 3/8\\ \phantom{a}\end{matrix}$ &
$\begin{gathered}\xymatrix{\bullet\ar@{-}[r]\ar@{-}@/_/[dd]&\bullet \ar@{..}[ddl]\ar@{-}@/_/[dd]&\\
\bullet \ar@{-}[r]\ar@{-}[d]&\bullet \ar@{-}[d]\ar@{..}[dl]\ar@{-}[r]&\bullet \ar@{-}[d]\ar@{..}[dl]\\
\bullet \ar@{-}[r]\ar@{-}[d]\ar@{..}[dr]&\bullet \ar@{-}[r]\ar@{-}[d]\ar@{..}[dr]&\bullet \ar@{-}[d] \\
\bullet \ar@{-}[r]&\bullet \ar@{-}[r]&\bullet\\
}\end{gathered}$\\\hline
$\begin{matrix}\\ S_{12}\\ \phantom{a}\end{matrix}$ & $\begin{matrix}\\ x y^3 + x^2 z + y z^2\\ \phantom{a}\end{matrix}$ & $\begin{matrix}\\ 4/13, 3/13, 5/13\\ \phantom{a}\end{matrix}$ &
$\begin{gathered}\xymatrix{\bullet\ar@{-}[r]\ar@{-}@/_/[dd]&\bullet \ar@{..}[ddl]\ar@{-}@/_/[dd]&\\
\bullet \ar@{-}[r]\ar@{-}[d]&\bullet \ar@{-}[d]\ar@{..}[dl]\ar@{-}[r]&\bullet \ar@{-}[d]\ar@{..}[dl]\\
\bullet \ar@{-}[r]\ar@{-}[d]\ar@{..}[dr]&\bullet \ar@{-}[r]\ar@{-}[d]\ar@{..}[dr]&\bullet \ar@{-}[d]\ar@{..}[dr] \\
\bullet \ar@{-}[r]&\bullet \ar@{-}[r]&\bullet \ar@{-}[r]&\bullet\\
}\end{gathered}$\\\hline
$\begin{matrix}\\ U_{12}\\ \phantom{a}\end{matrix}$ & $\begin{matrix}\\ x^3+y^3+z^4\\ \phantom{a}\end{matrix}$ & $\begin{matrix}\\ 1/3, 1/3, 1/4\\ \phantom{a}\end{matrix}$ &
$\begin{gathered}\xymatrix{\bullet\ar@{-}[r]\ar@{-}@/_/[dd]&\bullet \ar@{..}[ddl]\ar@{-}[r]\ar@{-}@/_/[dd]&\bullet \ar@{..}[ddl]\ar@{-}@/_/[dd]\\
\bullet \ar@{-}[r]\ar@{-}[d]&\bullet \ar@{-}[d]\ar@{..}[dl]\ar@{-}[r]&\bullet \ar@{-}[d]\ar@{..}[dl]\\
\bullet \ar@{-}[r]\ar@{-}[d]\ar@{..}[dr]&\bullet \ar@{-}[r]\ar@{-}[d]\ar@{..}[dr]&\bullet \ar@{-}[d] \\
\bullet \ar@{-}[r]&\bullet \ar@{-}[r]&\bullet\\
}\end{gathered}$\\\hline

\end{tabular}
\end{center}
\label{arnoldtable3}
\end{table*}%

From the explicit expressions for $W(x,y,z)$, we have the identifications
\begin{align}
E_{12}\equiv A_2\,\square\,A_6 && E_{14}\equiv A_2\,\square\,A_7\\
W_{12}\equiv A_3\,\square\,A_4 && U_{12}\equiv D_4\,\square\, A_3
\end{align}
of four Arnold's models with theories of the form $G\,\square\,G^\prime$, $G,G^\prime=ADE$, already studied in \cite{cnv}. In the present paper we focus on the remaining $10$ Arnold exceptional $\cn=2$ theories.

\subsection{Coxeter--Dynkin graphs, Coxeter transformations}

The last column of table \ref{arnoldtable} shows the Coxeter--Dynkin diagram of the singularity \cite{eb}. We recall its definition: The compact homology of the complex \emph{surface}\begin{equation*} \{W(x,y,z)+\cdots=0\}\subset \C^3\end{equation*} is generated by $\mu$ $2$--spheres \cite{milnor}, where $\mu$ is the Milnor number of the singularity (equal to the subfix in the singularity's name). Fixing a strongly distinguished basis of (vanishing) $2$--cycles $\delta_j$ \cite{ar,eb}, the negative of their intersection form, $-\delta_j\cdot \delta_k$, is an integral symmetric $\mu\times\mu$ matrix, with $2$'s along the main diagonal, that is naturally interpreted as a `Cartan matrix'. In fact, 
 for a minimal $ADE$ singularity\footnote{\ And a suitable choice of the basis $\delta_j$.}, $-\delta_j\cdot \delta_k$ is the Cartan matrix of the associated simply--laced Lie algebra. However, for a \emph{non}--minimal singularity, it is not true that $-\delta_j\cdot\delta_k \leq 0$ for $j\neq k$, and hence $-\delta_j\cdot\delta_k$ is not a standard Cartan matrix in the Kac sense \cite{kac}.
 
  Correspondingly, the Coxeter--Dynkin graph becomes a \emph{bi}--graph, \textit{i.e.}\! a graph with two kinds of edges, solid and dashed. Nodes $j$, $k$ are connected by $|\delta_j\cdot\delta_k|$ edges; the edges are solid if $\delta_j\cdot\delta_k>0$, and dashed if $\delta_j\cdot\delta_k<0$.
\smallskip

It should be stressed that the Coxeter--Dynkin diagram is not unique, since it depends on the particular choice of a (strongly distinguished) homology basis. Two such bases differ by the action of the braid group acting by Picard--Lefshetz transformations \cite{ar,eb}. The physical interpretation of this non--uniqueness is well known: In the $2d$ language the Picard--Lefshetz transformations correspond to BPS wall--crossings \cite{Cecotti:1993rm}, while from the $4d$ perspective they are understood as SQM Seiberg dualities \cite{cachazo}.
\medskip

One important invariant of the singularities is (the conjugacy class of) its Coxeter transformation, also known as the \emph{strong monodromy} $H$.
With respect to a strongly distinguished basis one has
\begin{equation}
H=-(S^{-1})^tS,
\end{equation}
where
\begin{equation}\label{whatS}
S_{jk}= \delta_{jk}-\begin{cases}\delta_j\cdot\delta_k & k>j\\ 0 &\text{otherwise,}\end{cases}
\end{equation}
and $S$ encodes the $2d$ BPS spectrum of the Landau--Ginzburg (LG) model with superpotential $W$ \cite{Cecotti:1993rm}.

\section{Arnold's $\cn=2$ superconformal theories}

\subsection{Quivers and superpotentials}\label{quivsup}

\subsubsection{Set up}

Being quantized, the conserved (electric, magnetic, and flavor) charges of a four dimensional $\cn=2$ theory take value in a lattice $\Gamma$. On general grounds such a lattice is endowed with a skew--symmetric integral form
\begin{equation}
\langle \gamma, \gamma^\prime\rangle_\text{Dirac}=-\langle \gamma^\prime, \gamma\rangle_\text{Dirac} \in \Z,\qquad \gamma,\gamma^\prime\in \Gamma,
\end{equation} 
given by the Dirac electric--magnetic pairing.

If the four dimensional $\cn=2$ theory is a \emph{quiver} theory in the sense of ref.\!\cite{CV11}, 
we may choose (non--uniquely, in general)
a distinguished set $\{\alpha_i\}$ of generators of the charge lattice, $\Gamma \simeq \oplus_i \,\Z\alpha_i$, having the physical properties specified in ref.\!\cite{CV11}, and, in particular, such that (in the given chamber) all stable BPS particles have charge vectors of the form $\pm \sum_i N_i \,\alpha_i \in \Gamma$ with $N_i$ non--negative integers. 

Given such a preferred basis $\{\alpha_i\}$,
we define the exchange matrix $B_{ij}$ to be
\begin{equation}
B_{ij}=\langle \alpha_i,\alpha_j\rangle_\text{Dirac},\qquad i,j=1,2,\dots,\mathrm{rank}\,\Gamma.
\end{equation}
To the skew--symmetric matrix $B_{ij}$ we associate a $2$--acyclic quiver $Q$ trough the following rule: $Q$ has one node
(labelled $i$) for each generator $\alpha_i$, whereas between nodes $i$ and $j$ we draw $B_{ij}$ arrows, a negative number standing for arrows in the opposite direction $i \leftarrow j$.

The quiver $Q$ so obtained has a direct physical meaning (see \cite{CV11}
	and references therein): A BPS state of the $4d$ $\cn=2$ theory having charge vector $\sum_i N_i\alpha_i\in \Gamma$ is identified with a supersymmetric state of the quiver SQM with quiver $Q$, gauge group at the $i$--th node $U(N_i)$, and a suitable $\prod_i U(N_i)$ gauge--invariant superpotential $\cw$. The SQM superpotentials $\cw$ which may appear are severely restricted by the structure of the quiver $Q$, and are often uniquely determined by it.

\subsubsection{$2d/4d$ correspondence revisited}
The $2d/4d$ correspondence of ref.\!\cite{cnv} states that the quiver $Q$ of the $4d$ theory engineered on a CY hypersurface
$W+u^2=0$ is equal to BPS quiver of the $2d$ LG model having superpotential $W+u^2$. Basically, the nodes of the $4d$ quiver $Q$ are in one--to--one correspondence with the \textsc{susy} vacua of the $2d$ model, and two nodes of $Q$, $j$ and $k$, are connected by a number of arrows equal to the \emph{signed} number of BPS states interpolating the corresponding $2d$ vacua, $|j\rangle$ and $|k\rangle$. To implement this rule, it is convenient to integrate away the decoupled free $2d$ superfield $u$, remaining with the LG $2d$ superpotential $W$. Then, as shown in \cite{Cecotti:1993rm,HIV}, the $2d$ \textsc{susy} vacua $|j\rangle$ are in one--to--one correspondence with the elements of a strongly distinguished basis $\{\delta_j\}$ of the vanishing homology of the hypersurface $\mathscr{H}\colon \{W=\mathrm{const.}\}$, and the signed number of BPS particles interpolating between $|j\rangle$ and $|k\rangle$ is given by the corresponding intersection number $\delta_j\cdot\delta_k$. Hence the $2d/4d$ correspondence predicts a quiver with $\delta_j\cdot\delta_k$ arrows between nodes $j$ and $k$, a negative number again meaning arrows in the opposite direction. In other words, the exchange matrix $B_{jk}$ of $Q$ is given by
\begin{equation}
B=S^t-S,
\end{equation}
where $S$ is as in eqn.\eqref{whatS}\footnote{\
Notice that the notions of a \emph{strongly distinguished homology basis} $\{\delta_j\}$ in the sense of \cite{ar,eb}, and that of a \emph{distinguished basis of the charge lattice} $\{\alpha_j\}$ in the sense of \cite{CV11} agree under the $2d/4d$ correspondence.}.
Equivalently, the $2d$ quantum monodromy in the sense of ref.\!\cite{Cecotti:1993rm} is minus the Coxeter transformation $H$ of the singular hypersurface (and thus $S$ is identified with the half--plane Stokes matrix of \cite{Cecotti:1993rm}).
\smallskip

The $2d/4d$ correspondence is rather subtle, since it depends on the correct identification of a \emph{strongly} distinguished basis, and it should be implemented with the necessary care.
For this reason, here we present a more intrinsic derivation of the $\cn=2$ Dirac quiver from the Coxeter--Dynkin diagram of the singularity; this method has the additional merit of predicting also the superpotential $\cw$ of the quiver (super)quantum mechanics whose \textsc{susy} vacua give  the $4d$ BPS states. One check that the proposed procedure is equivalent to the proper $2d/4d$ correspondence, is that it reproduces the correct $2d$ quantum monodromy $-H$, which is the mutation--invariant content of the $2d$ BPS quiver.
\smallskip

There is a standard dictionary \cite{Ring} between Dynkin \emph{bi}--graphs and (classes of) algebras which generalizes Gabriel's relation between representation--finite hereditary algebras and ordinary (simply--laced) Dynkin graphs \cite{AL1,AL2,AL3}. One picks an orientation of the solid arrows to get a quiver $Q$; then the dashed arrows are interpreted as a minimal set of relations generating an ideal $J$ in the path algebra $\C Q$ of that quiver. Finally, one considers the basic algebra $\C Q/J$. Of course, the orientation of $Q$ has to be chosen in such a way that the dashed lines make sense as relations in $\C Q$.

Let us illustrate this procedure in the example of the $E_{12}$ Coxeter--Dynkin \emph{bi}--graph
\begin{equation}
\begin{split}
&\begin{gathered}\xymatrix{\bullet \ar@{-}[r]\ar@{-}[d]\ar@{..}[dr]&\bullet \ar@{-}[r]\ar@{-}[d]\ar@{..}[dr]&\bullet \ar@{-}[r]\ar@{-}[d]\ar@{..}[dr]&\bullet \ar@{-}[r]\ar@{-}[d]\ar@{..}[dr]&\bullet \ar@{-}[r]\ar@{-}[d]\ar@{..}[dr]&\bullet \ar@{-}[d]\\
\bullet \ar@{-}[r]&\bullet \ar@{-}[r]&\bullet \ar@{-}[r]&\bullet \ar@{-}[r]&\bullet \ar@{-}[r]&\bullet }\end{gathered}\xrightarrow{\ \text{orientation}\ }\\
&\qquad\xrightarrow{\ \text{orientation}\ } 
\begin{gathered}\xymatrix{\bullet \ar[r]^{a_1}\ar[d]^{b_1}&\bullet \ar[r]^{a_2}\ar[d]^{b_2}&\bullet \ar[r]^{a_3}\ar[d]^{b_3}&\bullet \ar[r]^{a_4}\ar[d]^{b_4}&\bullet \ar[r]^{a_5}\ar[d]^{b_5}&\bullet \ar[d]^{b_6}\\
\bullet \ar[r]_{c_1}&\bullet \ar[r]_{c_2}&\bullet \ar[r]_{c_3}&\bullet \ar[r]_{c_4}&\bullet \ar[r]_{c_5}&\bullet }\end{gathered}
\end{split}\label{ssswww}
\end{equation}
where the quiver in the \textsc{rhs} is supplemented with the relations generating the ideal $J$ determined by the dashed edges in the \textsc{lhs}, namely
\begin{equation}
b_{j+1}\,a_j= c_j\, b_j,\qquad j=1,2,\dots, 5.
\end{equation} 
These relations just state that the squares in \eqref{ssswww} are commutative, and hence imply that the resulting algebra $\mathscr{A}_{E_{12}}\equiv \C Q/J$ is isomorphic to the product $\C \vec A_6 \otimes \C \vec A_2$, where $\C \vec A_n$ stands for the path algebra of the \emph{linear} $A_n$ Dynkin quiver
\begin{equation*}
\vec A_n\colon\qquad \overbrace{\xymatrix{\bullet\ar[r]& \bullet\ar[r] &\bullet\ar[r] & \cdots \ar[r] & \bullet\ar[r] &\bullet}}^{n\ \text{nodes}}.
\end{equation*}

The models having this tensor product form were solved in ref.\!\cite{cnv} by exploiting the isomorphism $\mathscr{A}_{E_{12}}\simeq \C \vec A_6 \otimes \C \vec A_2$ and its generalizations to $\C G\otimes \C G^\prime$ ($G,G^\prime$ being arbitrary $ADE$ Dynkin quivers).

Let\footnote{\ Here $P_i$ denotes the projective cover of the simple representation $S_i$ ($S_i$ is the representation with the one--dimensional space $\C$ at the $i$--th node, and zero elsewhere). Since all our algebras are basic, $\mathscr{A}\simeq\oplus_i P_i$ as (right) $\mathscr{A}$--modules. We also stress that our algebras satisfy $\mathrm{gl.dim.}\,\mathscr{A}\leq 2$. This property is absolutely crucial for the consistency of our manipulations.} $C_{ji}\equiv \dim \mathrm{Hom}(P_i,P_j)$ be the matrix counting the number of paths between the $i$--th and $j$--th node in the quiver $Q$ (identifying paths which differ by an element of $J$).
The Euler form of  $\mathscr{A}_{E_{12}}$ is the non--symmetric bilinear form on the dimension lattice $\Gamma_{E_{12}}$
defined by the matrix $C^{-1}$, that is
\begin{equation}
\langle X, Y\rangle_E\equiv \sum_{k=0}^2 (-1)^k\,\mathrm{Ext}^{k}(X,Y) = (\dim X)^t C^{-1} (\dim Y).
\end{equation}
The Cartan matrix, Dirac pairing, and Coxeter element of the algebra
$\mathscr{A}_{E_{12}}$ are, respectively,
\begin{equation}\begin{aligned}
& (C^{-1})^t+C^{-1} &&\text{(Cartan matrix)}\\
& (C^{-1})^t-C^{-1} &&\text{(Dirac pairing)}\\
& -C^tC^{-1} &&\text{(Coxeter element)}\\
\end{aligned}
\end{equation}
which agree with the predictions of the $2d/4d$ correspondence since $C^{-1}=S$, as it easy to check going trough the definitions.

However $\mathscr{A}_{E_{12}}\simeq \C \vec A_6 \otimes \C \vec A_2$ is not the final story. From the point of view of the quiver supersymmetric quantum mechanics, the relations of $J$ may arise only from the $F$--term flatness equations $\partial \cw=0$. Hence we have to introduce a SQM superpotential $\cw$ and additional Lagrange--multiplier superfields $\lambda_j$, one per fundamental relation of $J$, that is, one $\lambda_j$ per dashed edge in the \emph{bi}--graph. This is equivalent to replacing the dashed edges of the Coxeter--Dynkin diagram with arrows  going in the \emph{opposite} direction. Then, for the $E_{12}$ example, the superpotential is
\begin{equation}
\cw=\sum_{j=1}^5 \mathrm{Tr}\big(\lambda_j(b_{j+1}a_j-c_jb_j)\big).
\end{equation} 
In this way we get a completed quiver $\widetilde{Q}$, and the algebra
$\mathscr{A}_{E_{12}}$ gets completed to the Jacobian algebra
$\C\tilde Q/\partial \cw$ which is known as the \emph{$3$--Calabi--Yau completion of} $\mathscr{A}_{E_{12}}$, written $\Pi_3(\mathscr{A}_{E_{12}})$ \cite{kelP}. This completed algebra is the one relevant for the SQM theory describing the $4d$ BPS states.

\subsubsection{The square and the Coxeter--Dynkin forms of the quiver}\label{canonical}
By repeated mutations (Seiberg dualities) we eliminate all diagonal arrows from the completed quiver $\widetilde{Q}$, and we end up with the \emph{square} form of the quiver
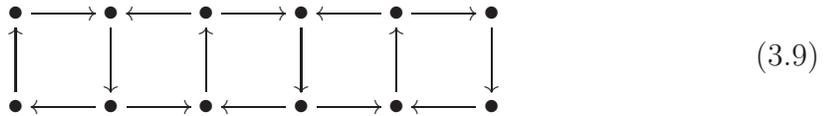
\begin{equation}\begin{gathered}\xymatrix{\bullet \ar[r]&\bullet \ar[d]&\bullet \ar[r]\ar[l]&\bullet \ar[d]&\bullet \ar[r]\ar[l]&\bullet \ar[d]\\
\bullet \ar[u]&\bullet \ar[r]\ar[l]&\bullet \ar[u]&\bullet \ar[r]\ar[l]&\bullet \ar[u]&\bullet\ar[l] }\end{gathered}
\end{equation}
where all squares are cyclically oriented\footnote{\ The claim is easily checked with the help of Keller's quiver mutation applet \cite{kelapp}.}. Then the superpotential $\cw$ is simply given  by the sum of the traces of the products of Higgs fields along each oriented square.

This procedure may be repeated word--for--word for all the Coxeter--Dynkin diagrams of the $14$ exceptional singularities.
Then

\textbf{(The square form of the quiver with superpotential)} \textit{The quiver of the corresponding $\cn=2$ theory is obtained from the Coxeter--Dynkin diagram in the form of table \ref{arnoldtable} by eliminating the dashed arrows and orienting all the squares. The superpotential $\cw$ is the sum of the traces of the cycles corresponding to the oriented squares. }
\smallskip

\begin{figure}
\begin{gather*}
\begin{gathered}
\xymatrix{& & & & \bullet & & & & & &\\
a_1 \ar[r] & a_2 \ar[r] &\cdots \ar[r]&a_{p-1}\ar[r] &\bullet \ar@<0.4ex>[d]\ar@<-0.4ex>[d] \ar[u]& b_{q-1}\ar[l] &\cdots\ar[l] &b_1\ar[l]\\
& & & & \bullet \ar[ul]\ar[ur]\ar[r] & c_{r-1}\ar[r]\ar[ul] &c_{r-2}\ar[r] & \cdots \ar[r] & c_2 \ar[r] & c_1}
\end{gathered}
\end{gather*}
\begin{gather*}
\text{values of }p,q,r\\ 
\begin{array}{|c|c||c|c||c|c||c|c||c|c|}\hline
E_{12} & 2,3,7& Z_{11} & 2,4,5& Q_{10} &3,3,4 & W_{12} & 2,5,5& S_{11} & 3,4,5\\\hline
E_{13} & 2,3,8& Z_{12} & 2,4,6& Q_{11} & 3,3,5& W_{13} & 2,5,6& S_{12} & 3,4,5\\\hline
E_{14} & 2,3,9& Z_{13} & 2,4,7& Q_{12}  & 3,3,6& & & U_{12} & 4,4,4\\\hline
\end{array}
\end{gather*}
\caption{\label{canalgepqr}The quiver corresponding to the $3$--Calabi--Yau completion of the Coxeter--Dynkin algebra of extended canonical type $\widehat{D}(p,q,r)$; the table gives the correspondence (singularity type) $\longleftrightarrow (p,q,r)$.}
\end{figure}

Of course, the quiver is not unique, and indeed each mutation class contains infinitely many different cluster--equivalent quivers. The one described above is particularly convenient for `strong coupling' calculations. There is also a `Coxeter--Dynkin' form of the quiver whose Jacobian algebra corresponds to the $3$--CY completion of a \emph{Coxeter--Dynkin algebra of extended canonical type}\footnote{\ The the Coxeter--Dynkin algebra of extended canonical type, $\widehat{D}(p,q,r)$ is identified with the path algebra of the quiver in figure \ref{canalgepqr} with the Kronecher subquiver replaced by two dashed lines (\textit{i.e.}\! by two relations) bounded by the ideal $J$ generated by the two relations. Calling $\alpha_i,\beta_i$, $i=1,2,3$, the single arrows forming the oriented triangles in figure \ref{canalgepqr}, the two relations are
\begin{equation*}
\alpha_2\beta_2+\alpha_3\beta_3=0 \qquad\text{and}\qquad \alpha_1\beta_1=\alpha_3\beta_3,
\end{equation*} from which we deduce the superpotential of the $3$--CY completed \emph{canonical} quiver SQM
\begin{equation*}
\cw=\lambda_1(\alpha_2\beta_2+\alpha_3\beta_3)+\lambda_2(\alpha_1\beta_1-\alpha_3\beta_3).
\end{equation*}} $\widehat{D}(p,q,r)$ which is a tilting of (and hence derived equivalent to) the one--point extension of the canonical algebra $C(p,q,r)$ at a projective indecomposable. The quiver of $\Pi_3(\widehat{D}(p,q,r))$ is presented in figure \ref{canalgepqr}. For a discussion of the relevant {extended} canonical algebras and Coxeter--Dynkin algebras, and their relations to Arnold's exceptional singularities, see refs.\!\cite{le1,le2}. 
\smallskip

The \emph{bi}--graph obtained by replacing in figure \ref{canalgepqr} the double arrows by dashed lines and all other arrows by solid edges was show by Ebeling \cite{eb2} to correspond to the Coxeter--Dynkin diagram of the singularity with respect to a \emph{strongly distinguished} homology basis (related to the previous one by a braid transformation). This is another check of the $2d/4d$ correspondence in the stronger version used here.
 
\subsection{Minimal non--complete models}\label{MinNC}

The models discussed in the present paper are \emph{not} complete theories in the sense of \cite{CV11}.  To contrast them, we start by recalling the definition of complete $\cn=2$ theories. Let $\cd$ be the domain  in parameter space which corresponds to consistent quantum field theories. The central charge function $Z\big(\sum_iN_i\alpha_i\big)= \sum_i Z_i\, N_i$ defines a holomorphic map $\varpi$
\begin{equation}
\cd\xrightarrow{\ \varpi\ } \C^r,\qquad \lambda_a\mapsto Z_i\in \C^r.
\end{equation}
An $\cn=2$ model is called \emph{complete} \cite{CV11} if the image of $\varpi$ has dimension $r$, that is, codimension zero.
 
 For a \emph{non}--complete theory, the computation of the BPS spectrum
by any method related to the KS wall--crossing formula --- such as cluster--combinatorics \cite{cnv}, or the stability conditions on quiver representations \cite{CV11} ---  is questionable on the grounds that the particular mathematically--defined BPS chamber $\mathscr{C}_\text{BPS}\subset \C^r$ in which we are computing --- typically picked up for its technical simplicity --- may have \emph{no} overlap with the image $\varpi(\cd)$, and hence be outside the physical region of the parameters. In this case, the spectrum we compute does not correspond to any physically realizable regime. Of course, the computation is still mathematically correct, and all the chamber independent quantities, like the conjugacy class of the quantum monodromy and the related UV invariants $\mathrm{Tr}[\,\mathbb{M}(q)^k]$ \cite{cnv}, have their physically correct values, and we can always recover the physical spectrum (in principle) by applying the KS wall--crossing formula.
However, as physicists, we are interested in knowing whether the spectrum we compute has a direct physical meaning, or if some further mathematical work is required to extract the physically relevant informations. 

The purpose of the present subsection is to present some general remark on the question of the \emph{physical realizability} of the special symmetric BPS chambers we use in our computations. The reader may prefer to skip the following qualitative discussion, and jump ahead to the more formal arguments.
\medskip

 The present theories, although non--complete, are \emph{minimally} so, in the sense that the codimension of the image $\varpi(\cd)\subset \C^r$ is just $1$. In other words, there is only one quantum--obstructed variation $(\delta Z_i)_\mathrm{obs}$ of the central charge function, normal to the physical submanifold $\varpi(\cd)\subset \C^r$. In general, modifications $Z_i\rightarrow Z_i+\delta Z_i$ correspond to infinitesimal deformations of the periods of the  holomorphic $3$--form $\Omega$ associated to deformations $\delta t_j$ of the complex structure of the hypersurface $\mathscr{H}$ of the form
 \begin{equation}
 W(x,y,z)+u^2+\sum\nolimits_j \delta t_j\,\phi_j=0,
 \end{equation}
 where $\{\phi_j\}$ is a basis of chiral primaries for the $2d$ LG model with superpotential $W(x,y,z)$. The offending deformation $(\delta Z_i)_\mathrm{obs}$ is the one associated to the unique chiral primary of dimension $>1$, namely the Hessian $\mathfrak{H}=\det \partial_\alpha\partial_\beta W$. The problematic deformation is precisely the one defining the $1$--parameter family of inequivalent singularities\footnote{\ By definition, a \emph{unimodal} singularity has  a $1$--parameter family of inequivalent singular deformations.}, which is the only primary perturbation which changes the behavior at infinity in field--space (and hence may spoil the quantum consistency). 
 
 To address the physical realizability question, we have to make sure that, in the chamber we compute, the Hessian deformation is not switched on. There are two arguments: a mathematical one which is only partly conclusive, and a more stringent physical one.
 
 Mathematically, if we can show that our spectral computation holds true in (an open neighborhood of) a complex submanifold $\mathscr{S}\subset \C^r$ of dimension $\geq 1$, we would expect that
  \emph{generically} the intersection $\mathscr{S}\cap\varpi(\cd)\neq \emptyset$, and hence the computed spectrum \emph{is} actually realized in some physical regime. Our computations are typically valid in submanifolds $\mathscr{S}$ of large dimension, and so, as long as their position is not too special, the physical subset $\mathscr{S}\cap\varpi(\cd)$ also has positive dimension. However, our $\mathscr{S}$'s will correspond to particularly simple situations (otherwise the computations would not be \emph{that} easy), and hence are typically \emph{non}--generic.
 \smallskip
 
Looking to the Coxeter--Dynkin diagrams in table \ref{arnoldtable}, we see that they are all subgraphs of two kinds of (bi)graphs associated to direct sums of minimal singularities of the two forms
 \begin{equation}
 \begin{array}{c|c}
 A_n\boxtimes A_m & x^{n+1}+y^{m+1}+z^2\\\hline
 A_n\boxtimes D_4 & x^3+y^3+z^{n+1}\end{array}
 \end{equation}
 to which the arguments of \cite{cnv} directly apply.
 Physically, the $10$ Arnold superconformal models which are not already of the form $G\,\square\,G^\prime$ may be obtained as follows: one starts with a suitable `big' $G\,\square\,G^\prime$ theory, and perturbs it by a certain \emph{relevant} operator (that is, relevant at the UV fixed point described by the $G\,\square\,G^\prime$ theory), in such a way that the corresponding $\cn=2$ theory will flow in the IR to the Arnold superconformal theory we are interested in.

 \begin{table}
\begin{center}\begin{small}
\begin{tabular}{|c|c|c|c|}\hline
$E_{13}$ & $A_7\,\square\,A_2\colon\: x^3+y^8+z^2\: \left(\frac{13}{12}\right)$ & $xy^5\: \left(\frac{23}{24}\right)$ & $\frac{1}{11}$\\\hline 
$Z_{11}$ & $A_4\,\square\,A_3\colon\: x^4+y^5+z^2\: \left(\frac{11}{10}\right)$ &
$x^3y\: \left(\frac{19}{20}\right)$ & $\frac{1}{9}$\\\hline 
$Z_{13}$ & $A_5\,\square\,A_3\colon\: 
x^4+y^6+z^2\: \left(\frac{7}{6}\right)$ & $x^3y\: \left(\frac{11}{12}\right)$ & $\frac{1}{5}$ \\\hline
$W_{13}$ & $A_5\,\square\, A_3\colon\:  x^4+y^6+z^2\: \left(\frac{7}{6}\right)$ & $xy^4\: \left(\frac{11}{12}\right)$ & $\frac{1}{5}$\\\hline
$Q_{10}$ & $A_3\,\square\,D_4\colon\: x^3+y^3+z^4\:\left(\frac{7}{6}\right)$ &
$x^2z\:\left(\frac{11}{12}\right)$ & $\frac{1}{5}$ \\\hline
$Q_{12}$ & $A_4\,\square\,D_4\colon\: x^3+y^3+z^5\:\left(\frac{19}{15}\right)$ & $x^2z\: \left(\frac{13}{15}\right)$
& $\frac{4}{11}$ \\\hline
$S_{11}$ & $A_3\,\square\,D_4\colon\: x^2z+z^3+y^4\: \left(\frac{7}{6}\right)$ &
$ yz^2\ \left(\frac{11}{12}\right)$ & $\frac{1}{5}$\\\hline
\end{tabular}\end{small}
\end{center}
\caption{Arnold's \emph{superconformal} gauge theories as IR fixed points of superconformal square tensor models \cite{cnv} perturbed by the less relevant operator. }
\label{arnoldaRG}
\end{table}%

In table \ref{arnoldaRG} we list some convenient choices of UV $G\,\square\,G^\prime$ theories and relevant perturbations $\phi_\star$ for seven of the $10$ non--product Arnold theories. The first number in parenthesis is the central charge $\hat c$ of the $2d$ UV Landau--Ginzburg; one has $\hat c<2$, and hence the corresponding $4d$ $\cn=2$ quantum theories exist by the criterion of refs.\!\cite{Gukov:1999ya,cnv,shapere}. The second number in parenthesis is the UV dimension (in the $2d$ sense\,!) of the perturbing chiral primary $\phi_\star$; notice that it is always $2d$ relevant (at the UV fixed point). The last column of table \ref{arnoldaRG} is the mass dimension of the $4d$ coupling $t_\star$ corresponding to the deformation $\phi_\star$ (at the UV fixed point) given by
\cite{shapere}
\begin{equation}
[t_\star]= \frac{2(1-q(\phi_\star))}{(2-\hat{c})}. 
\end{equation}
The two theories $Z_{12}$ and $Q_{11}$ are better described as the final IR fixed points of RG `cascades'
 \begin{gather}
 A_5\,\square\,A_3\xrightarrow{\ x^3y\ (11/12)\ } Z_{13} \xrightarrow{\ xy^4\ (17/18)\ } Z_{12}\\
A_4\,\square\, D_4\xrightarrow{\ x^2z\ (13/15)\ } Q_{12}\xrightarrow{\ yz^3\ (14/15)\ } Q_{11}
 \end{gather} 
where the perturbing monomials $\phi_\star$ and their dimensions are written over the corresponding arrow. $S_{12}$ is more tricky; however we may still consider it as the IR fixed point of the model defined by the hypersurface $(y^2 z+z^3+ x^5)+xz^2+x^3y$ whose UV limit is
$A_4\,\square\, D_4$.

The above RG discussion applies directly to the Arnold $\cn=2$ theories at their \emph{superconformal point}, that is with all relevant deformations switched off. We are, of course, interested in the massive deformations of the theory which produce interesting chamber--dependent BPS spectra.
For the massive case, we may argue as follows: we start with the $A_n\,\square\, G$ deformed hypersurface
\begin{equation}\label{defpolll}
 \lambda\,y^{n+1}+W_G(x,z)+\phi_\star +\sum\nolimits_i^*t_i\phi_i +v^2=0,
\end{equation}
 where the sum $\sum^*$ is over chiral primaries of dimension $q$ less than $q(\phi_\star)$. By the criterion of \cite{Gukov:1999ya,shapere} , the hypersurface \eqref{defpolll} corresponds to a physical regime of the (non--complete) $A_n\,\square\, G$ theory for all $\lambda,\, t_i$ provided $\lambda\neq 0$. As $\lambda\rightarrow 0$, some states become infinitely massive and decouple.  The decoupling limit produces a physically realizable regime of the mass--deformed $\cn=2$ Arnold theory we are interested in. 

The physical idea is then to control the realizability of a given BPS chamber for an Arnold theory by starting from the $A_n\,\square\, G$ theory \eqref{defpolll}, at large $\lambda$, in a BPS chamber which is known to be physical, and then continuously deform $\lambda$ to zero, while ensuring that no wall of marginal stability is crossed in the process. By construction, we end up into a physical chamber of the (massive) Arnold theory, whose BPS spectrum differs from the one of the original $A_n\,\square\, G$ theory only because some particle got an infinite mass in the $\lambda\rightarrow 0$ limit and decoupled.

As an initial reference chamber of the $A_n\,\square\,G$ theory we take one of those considered in \cite{cnv}. In general, for a $G^\prime\,\square\,G$ model there is a chamber with a finite spectrum consisting of hypermultiplets with charge vectors
\begin{equation}\label{spectCNV}
 \alpha \otimes \beta_a \in \Gamma \simeq \Gamma_G \otimes \Gamma_{G^\prime},\qquad \alpha\ \text{positive root of }G,\ \beta_a\ \text{simple roots of }G^\prime.
\end{equation}
There is an obvious duality $G\leftrightarrow G^\prime$ which produces a second finite chamber with the role of $G$ and $G^\prime$ interchanged. It is believed \cite{cnv} that these two BPS chambers do correspond to physical situations, and hence they may be used as the starting points at large $\lambda$ for the family of theories \eqref{defpolll}. 

Let us sketch the argument of \cite{cnv} for $\cn=2$ models of the form $A_n\,\square\, G$, where $G$ is any $ADE$ Dynkin diagram. Such models are engineered by Type IIB on a hypersurface $\mathscr{H}\colon W_G(x,y,z)+P_{n+1}(v)=0$, where $W_G(x,y,z)$ stands for the usual $G$ minimal singularity and $P_{n+1}(v)$ is a degree $n+1$ polynomial that we take of the Chebyshev form. We can see this geometry as a compactification of IIB down to $6$ dimensions on a deformed $G$--singularity whose deforming parameters depend (adiabatically) on the complex coordinate $v$. As in ref.\!\cite{shapere}, the compact $3$--cycles on the hypersurface $\mathscr{H}$ are seen as vanishing $2$--cycles of the $G$--type singularity fibered over a curve in the $v$--plane connecting two zeros of $P_{n+1}(v)$.
The $G$--singularity produces tensionless strings in one--to--one correspondence with the positive roots of $G$. Let $\delta_\alpha(v)$ be the vanishing cycle over $v$ associated to the positive root $\alpha$. We define and effective SW differential
\begin{equation}
 \lambda_\alpha(v)=\int\limits_{\delta_\alpha(v)} \Omega\sim \big(P_{n+1}(v)\big)^{\Delta_\alpha}\, dv,
\end{equation}
vanishing at the zeros of $P_{n+1}(v)$.  
 Then for each $\alpha\in \Delta_+(G)$ we may repeat the analysis of \cite{shapere}, showing that the spectrum \eqref{spectCNV} corresponds to a physically realizable chamber.
\smallskip

In practice, it may be difficult to check the existence, in the complex $\lambda$--plane, of a path from zero to infinity which avoids all wall--crossings while keeping control of the possible mixing between the conserved quantum currents. Therefore, we shall mostly use the above idea in a weak sense, namely, we shall consider a mathematically correct BPS spectrum which is naturally interpreted as the result of the decoupling of some heavy states from the known physical spectrum of the appropriate $A_n\,\square\, G$ model, as a physically sound BPS spectrum which, having a simple physical interpretation, also provides circumstantial evidence for the physical realizability of the corresponding chamber.
 
 \subsection{Flavor symmetries}

The number $n_f$ of flavor charges, or more precisely the dimension of the Cartan subalgebra of the flavor symmetry group $G_f$, is an important invariant of the theory, which is independent of the parameters (however, at particular points in the physical domain $\cd$ we may have a non--Abelian enhancement of the flavor symmetry, $U(1)^{n_f}\rightarrow G_{n_f}$, which preserves its rank).

A general consequence of $2d/4d$ correspondence \cite{cnv,CV11} is that $n_f$, which is (by definition) the number of zero eigenvalues of the Dirac pairing matrix $B_{ij}$, is equal to the number of $2d$ chiral primary operators whose UV dimension $q_i$ is equal to $\hat c/2$. In particular, $n_f=\mathrm{rank}\,\Gamma \!\mod 2$. The eigenvalues of $H$ are equal to $-\exp[2\pi i (q_i-\hat c/2)]$ \cite{Cecotti:1993rm}, and so $n_f$ is equal to the multiplicity of $-1$ as an eigenvalue of $H$. Then $n_f$ may be directly read from the factorization of the characteristic polynomial of $H$ into cyclotomic polynomials, see table 2 of ref.\!\cite{le1}: $n_f$ is just the number of $\Phi_2$ factors in the product. Thus
\begin{equation}
n_f=\begin{cases}2 & Z_{12}, U_{12}\\
1 & \text{odd rank}\\
0 & \text{otherwise}.\end{cases}
\end{equation}

\subsection{Order of the quantum monodromy}\label{monodromy}

In ref.\!\cite{cnv} it was shown that the quantum monodromy $\mathbb{M}(q)$ of a $\cn=2$ model engineered by Type IIB on a non--compact CY hypersurface $\mathscr{H}\subset \C^4$, given by the zero locus of a (relevant deformation of a) quasi--homogeneous polynomial $f_0(x_i)$, has a finite order $\ell$, that is,
\begin{equation}
\mathbb{M}(q)^\ell=1,\qquad \text{(identically in }q\in \C^*)
\end{equation}
in the sense of equality of adjoint actions on the quantum torus algebra $\mathbb{T}_\Gamma$ (see \S.\,\ref{qclumut} for precise definitions).

The minimal value of the integer $\ell$ is easy to predict. Let $d,w_1,w_2,w_3,w_4$ be integers such that
\begin{equation}
\lambda^d\, f_0(x_i)= f_0(\lambda^{w_i}x_i)\quad \forall\, \lambda\in \C,
\end{equation}
normalized so that $\gcd\{d,w_1,w_2,w_3,w_4\}=1$. The redefinition $x_i\rightarrow \lambda^{w_i}\,x_i$ transforms the CY holomorphic $3$--form $\Omega$ into $\lambda^{\sum_iw_i-d}\Omega$. Hence the monodromy corresponds to replacing
$\lambda$ with $\exp[2\pi i t/(\sum_i w_i-d)]$ and  continuously taking $t$ from $0$ to $1$. In terms of the original variables, this is
\begin{equation}
x_i\rightarrow \exp\!\left(2\pi i \frac{w_i}{\sum_iw_i-d}\right)x_i
\end{equation}
and the monodromy order $\ell$ is 
\begin{equation}
\ell= \frac{\sum_i w_i-d}{\gcd\{\sum_iw_i-d, w_1,w_2,w_3,w_4\}}.
\end{equation}
In the case of a singularity of the form $f_0(x,y)+ z^2+v^2$ it is more convenient to consider the reduced order, corresponding to the engineering of the model from the $6d$ $(2,0)$ theory. It corresponds to setting $w_3=d,w_4=0$ in the above formula.

 \begin{table}\begin{center}
  \begin{tabular}{|c|c||c|c||c|c||c|c||c|c|}\hline
  $E_{13}$ & 7 & $Z_{11}$ & 7 & $Z_{12}$ & 5 & $Z_{13}$ & 8 & $W_{13}$ &  7\\\hline
  $Q_{10}$ & 11 & $Q_{11}$ & 8 & $Q_{12}$ & 13 & $S_{11}$ & 7 & $S_{12}$ & 11\\\hline
  \end{tabular} 
\end{center}\vglue -12pt
\caption{(Reduced) orders of the quantum monodromy for the 10 Arnold's exceptional theories which are not of the tensor from $G\,\square\, G^\prime$.}
\label{monorders}
 \end{table}

\subsection{Arnold's exceptional $\cn=2$ models as gauge theories}

In the title we referred to the Arnold's exceptional models as \emph{gauge} theories. Up to now, the gauge aspect of these models has not manifested itself. Although in the present paper we are mainly interested in `strong coupled' regimes in which the BPS spectrum contains just finitely many hypermultiplets, these theories do have `weakly coupled' phases where BPS vector--multiplets are present. At least in the simplest situations, the couplings of these vector--multiplets may be physically interpreted as a super--Yang--Mills sector weakly gauging  a subgroup $G$ of the global symmetry group of some `matter' system (which is non--Lagrangian, in general). Hence the Arnold exceptional $\cn=2$ theories behave as gauge theories in some corner of their parameter space, although a full understanding of the phases with stable BPS vector--multiplets requires a more in--depth study which we leave for future work.\smallskip

Counting dimensions, we see that a minimal non--complete $\cn=2$ model which has, in some limit, the structure of a $G$ SYM  weakly coupled to some other sector, the gauge group $G$ must have one of the following forms
\begin{equation}
SU(2)^k,\quad SU(2)^k\times SU(3),\quad SU(2)^k\times SO(5),\quad SU(2)^k\times G_2,
\end{equation}
for some $k\in \mathbb{N}$. For the exceptional Arnold models, it is easy to prove the existence of physical limits with $G=SU(2)$, while larger gauge groups are not at all excluded. 

To produce a physical regime with a weakly coupled $SU(2)$ SYM sector, it is enough to deform the Arnold singularity 
with suitable lower--order monomials (corresponding to a particular choice of the central charge function $Z_i$ inside the physical region $\varpi(\cd)$) 
 which causes the flow, in the IR, to one of the elliptic--$E$ complete superconformal gauge theories \cite{CV11}, and specifically  
\begin{align*}&\text{to the }E_8^{(1,1)}\equiv A_2\,\square\,A_5\ \text{model}&&\text{for }E_{12}, E_{13}, E_{14},\\
&\text{to the }E_7^{(1,1)}\equiv A_3\,\square\, A_3\ \text{model}&&\text{for }Z_{11}, Z_{12}, Z_{13}, W_{12}, W_{13},\\
&\text{to the }E_6^{(1,1)}\equiv A_2\,\square\, D_4\ \text{model} 
&&\text{for }Q_{10}, Q_{11}, Q_{12}, S_{11}, S_{12}, U_{12}.
\end{align*} 
The IR effective theory is known to have physical chambers with a stable $SU(2)$ gauge vector coupled to three $D$--type Argyres--Douglas systems \cite{CV11}. Since the IR theory is complete, we can tune the coefficients of the defining polynomial of $\mathscr{H}$ to get an arbitrarily weak gauge coupling.\smallskip

One way to prove the existence of a \emph{mathematically--defined}\footnote{\ By mathematically--defined we mean a chamber defined by some choice of the complex numbers $Z_i$ which may or may not be in the image $\varpi(\cd)$ of the physical domain in parameter space, $\cd$.} BPS chamber with a stable BPS vector--multiplet is to look for a (non--necessarily full) subquiver $\cs$ of $Q$ which is mutation equivalent to an acyclic affine quiver. This generalizes the strategy of looking for Kronecker, \textit{i.e.}\! $\widehat{A}(1,1)$, subquivers used in \cite{CV11}.

If $\cs$ is a \emph{full} subquiver of $Q$, the existence of a mathematical BPS chamber with a stable BPS vector--multiplet is guaranteed:
Indeed, the quantization of the $\mathbb{P}^1$ family of brick representations\footnote{\ The existence of this family of brick representations follows directly from Kac's theorem \cite{kacthm}. For details see \textit{e.g.}\! \cite{Ring,sko,cox,craw}.} of $\cs$ with dimension vector $\sum_i N_i \alpha_i$ equal to the minimal imaginary root $\delta$, extended by zero to a representation of the total quiver $Q$, produces --- for suitable choices of the complex numbers $Z_i$ --- a stable BPS vector--multiplet.

 If $\cs$ is not a full subquiver, the statement remains true, provided the above $\mathbb{P}^1$ family of representations of $\cs$, when seen as representations of the total quiver $Q$,  has the following two properties: 1) it satisfies the relations $\partial\cw=0$ induced from the arrows in $Q\setminus \cs$, and 2) it does not admit further continuous deformations corresponding to switching on non--trivial maps along the arrows of the full subquiver over the nodes $\cs_0$ which are not in $\cs$.  
Indeed, if this \emph{no--extra--deformation} condition is not verified, we have to quantize a moduli space of dimension larger than one, possibly producing higher spin representations of $\cn=2$ supersymmetry, instead than just vector--multiplets.

The quivers of the exceptional Arnold models always have affine subquivers (as it is already evident from the Coxeter--Dynkin form of the quiver, see figure \ref{canalgepqr}) and we may even find \emph{pairs} of non--overlapping such affine subquivers, leading to the possibilities of chambers with more than one BPS vector--multiplet.

As an (intriguing) example, take the model $E_{13}$ and consider the following pair of $\mathbb{P}^1$ families of representations with mutually disjoint support
\begin{gather}\label{e131}
\begin{gathered}
\xymatrix{0 \ar[r] & 0 \ar[d] & 0 \ar[l]\ar[r] & 0 \ar[d] & \C\ar[l]\ar[r]^0 &\C\ar[d]^{[1,0]} &\\
0 \ar[u]& 0\ar[l]\ar[r] & 0\ar[u] &\C\ar[l]\ar[r]_i & \C^2\ar[u]^{[0,1]^t} & \C^2\ar[l]^1\ar[r]_{[1,1]^t} & \C}
\end{gathered}\\
\begin{gathered}
\xymatrix{\C \ar[r]^i & \C^2 \ar[d]_{[0,1]^t} & \C^2 \ar[l]_1\ar[r]^{[1,1]^t} & \C \ar[d] & 0\ar[l]\ar[r] &0\ar[d] &\\
0 \ar[u]& \C\ar[l]\ar[r]_0 & \C\ar[u]_{[1,0]} &0\ar[l]\ar[r] & 0\ar[u] & 0\ar[l]\ar[r] & 0}
\end{gathered}\label{e132}
\end{gather}
where the map $\C\xrightarrow {i}\C^2$ defines a line in $\C^2$ and hence a point in $\mathbb{P}^1$. Both representations are pulled back from a representation of a $\widehat{D}_5$ \emph{non}--full subquiver having dimension vector the minimal imaginary root.
Note that the representations satisfy the constraints from the $F$--term flatness conditions $\partial\cw=0$, with $\cw$ as in \S.\,\ref{quivsup}. It remains to check that there are no continuous deformations of these $\mathbb{P}^1$ families obtained by giving non--zero values to the omitted arrow (the arrow with an explicit $0$ in eqns.\eqref{e131}\eqref{e132}). Indeed, these arrows are constrained to remain zero by the $F$--term relations $\partial\cw=0$. Hence, the $\mathbb{P}^1$ family is not further enlarged, and the corresponding BPS vector--multiplet is stable for a suitable choice of the $Z_i$'s. We write $\delta_1$, $\delta_2$ for the charge vectors of the resulting vector--multiplets. Counting arrows, we see that
\begin{equation}
\langle \delta_1,\delta_2\rangle_\text{Dirac} = 1
\end{equation}
Hence the two vector--multiplets are \emph{not} mutually local.
If the mathematical chamber in which both vectors are stable is physically realizable --- which is certainly \emph{not} guaranteed, and perhaps unlikely --- the physics will not be that of a conventional gauge theory.

\section{The cluster strategy of CNV}

\subsection{Quantum cluster mutations}\label{qclumut}

Let $Q, \cw$ be the SQM quiver and superpotential\footnote{\ Technically, $(Q,\cw)$ should be a non degenerate quiver with superpotential in the sense of refs.\cite{zel1,zel2,zel3}. This is automatically true for our $\cw$'s produced by the strong $2d/4d$ correspondence. } of a $4d$ $\cn=2$ theory and $\Gamma=\oplus_i \Z\alpha_i$ its charge lattice.  We write $Y_i\equiv Y_{\alpha_i}$ for the generators of the associated quantum torus algebra $\mathbb{T}_Q$ defined by the relations
\begin{equation}
 Y_i\,Y_j = q^{\langle\alpha_i, \alpha_j\rangle}\, Y_j \, Y_i \equiv q^{B_{ij}}\, Y_j\, Y_i.
\end{equation}

\subsubsection{Quantum mutations}

By a \textit{quantum mutation} of the quantum torus algebra $\mathbb{T}_Q$ we mean the composition of an (ordered) sequence of {elementary} mutations at various nodes of $Q$.
The {elementary quantum mutation}, $\mathcal{Q}_k$, at the $k$--th node of the $2$--acyclic quiver $Q$ is the composition of {two} transformations \cite{qd-cluster,clqd2,qd-pentagon,cnv,kel}:\smallskip

\textbf{(1)} a {basic mutation} of the quiver at the $k$--th node, $Q\rightarrow \mu_k(Q)$. The incidence matrix $\mu_k(B_{ij})$ of the mutated quiver $\mu_k(Q)$ is
\begin{equation}\label{mutMatB}
 B_{ij}\rightarrow \mu_k(B_{ij})=\begin{cases}
                - B_{ij} & \text{if } i=k\ \text{or }j=k;\\
B_{ij}+ \mathrm{sign}(B_{ik})\,\max\{B_{ik}B_{kj},0\} &\text{otherwise}
               \end{cases}
\end{equation}
together with a suitable mutation of the superpotential, $\cw\rightarrow \mu_k(\cw)$ \cite{zel1,zel2,zel3}, whose explicit form we do not need.
To compare elements of $\mathbb{T}_Q$ and $\mathbb{T}_{\mu_k(Q)}$, the mutation of the quiver $Q\rightarrow \mu_k(Q)$ should be supplemented by a change of basis in the charge lattice $\Gamma$, which corresponds to choosing a different set of generators of the algebra $\mathbb{T}_Q$ according to the rule\footnote{\ This is the \emph{right mutation}. There is also a \emph{left mutation} differing by a twist \cite{kel}.}
\begin{align}\label{signcon}
	&Y_i\rightarrow Y_i^\prime =q^{-\langle \alpha_i,\alpha_k\rangle\, [\langle \alpha_i,\alpha_k\rangle]_+/2}\: Y_i\, Y_k^{[\langle \alpha_i,\alpha_k\rangle]_+}& & i\neq k\\
	&Y_k\rightarrow Y_k^\prime = Y_k^{-1}\qquad\  \text{where } [a]_+\equiv \max\{a,0\},
\end{align}
or, equivalently,
\begin{equation}
 Y^\prime_i= \begin{cases}
              Y_k^{-1} & \text{if } i=k\\
Y_i & \text{if there are no arrows } i\rightarrow k \ \text{in } Q\\
q^{-m^2/2}\, Y_i Y_k^m & \text{if there are }m\geq 1\ \text{arrows }i\rightarrow k \ \text{in } Q.
             \end{cases}
\end{equation}
Notice that
\begin{equation}
 q^{-m^2/2}\, Y_i Y_k^m = N[Y_i Y_k^m] \equiv Y_{\alpha_i+m\alpha_j},
\end{equation}
where $N[\cdots]$ is the usual normal order \cite{cnv}.
More generally, 
\begin{equation}
\phantom{\Big|} \mu_k(Y_\alpha)= Y_{\sigma_k(\alpha)}
\end{equation}
where $\sigma_k$ is the matrix
\begin{equation}
(\sigma_k)_{ij}=\delta_{ij}-2\,\delta_{i,k}\,\delta_{k,j}-\delta_{i,k}\,\max\big\{0, - B_{k,j}\big\}.
\end{equation}

\smallskip

We stress that $\mu_k^2$ is the identity at the quiver level, $\mu^2_k(Q)\equiv Q$, but a non--trivial transformation on the set of generators of $\mathbb{T}_Q$
\begin{equation}\label{twiceeee}
 \mu_k^2\colon Y_i \mapsto q^{-\langle \alpha_i,\alpha_k\rangle^2/2}\, Y_i\,Y_k^{\langle \alpha_i, \alpha_k\rangle}\equiv Y_{\alpha_i+\langle \alpha_i, \alpha_k\rangle\,\alpha_k}.
\end{equation}
$\mu^2_k$ is called the Seidel--Thomas \emph{twists} $t_k$ \cite{kel}.

 $\mu_k$ is not in general an automorphism of the algebra $\mathbb{T}_Q$; a composition of $m_k$'s is an algebra automorphism iff it is the identity on the underlying quiver $Q$ since only in this case it leaves invariant the commutation relations. The following special cases hold:
\begin{align}\label{source}
&\text{1. the node $k$ is a \emph{source} in $Q$ $\Rightarrow$}\quad \mu_k(Y_j)=\begin{cases}
             Y_j & j\neq k\\
Y_j^{-1} & j= k;
            \end{cases}\\
&\text{2. the node $k$ is a \emph{sink} in $Q$\quad\  $\Rightarrow$}\quad
\mu_k(Y_j)\equiv \mu_k(Y_{\alpha_j})= Y_{s_k(\alpha_j)}\label{sink}\end{align}
where
\begin{equation}
 s_k(\alpha_j) = \alpha_j - (\alpha_j, \alpha_k)\, \alpha_k,
\end{equation}
is the elementary reflection in $\mathrm{Weyl}(Q)$ associated to the simple root $\alpha_k$.

\smallskip

\textbf{(2)} the adjoint action on $\mathbb{T}_Q$ of the quantum dilogarithm of $Y_k\equiv Y_{\alpha_k}$
\begin{equation}\label{adjact}
	Y_\gamma \mapsto \Psi(Y_k; q)^{-1}\,Y_\gamma\, \Psi(Y_k;q).
\end{equation}

Thus, explicitly, the elementary quantum cluster mutation at the $k$--th node is
\begin{gather}\label{qkexp}
\cq_k = \mathrm{Ad}(\Psi(Y_k)^{-1})\circ \mu_k.
\end{gather}
%\smallskip

The elementary quantum mutations are {involutions}
of $\mathbb{T}_Q$, \textit{i.e.}\!
one has the identity \cite{qd-cluster,clqd2,qd-pentagon,cnv,kel}
\begin{equation}
 \cq_k^2=\text{identity on }\mathbb{T}_Q.
\end{equation}

\subsubsection{Sink (sources) sequences \cite{BGP,sko,cox}}\label{sinksourcesse}

A sequence of nodes $\Lambda=\{i_1,i_2,\cdots, i_k\}$ of a quiver $Q$ is called a \emph{sink} sequence (resp.\! a \emph{source} sequence)
it the $i_s$ node is a sink (resp.\! a source) in the mutated quiver $\mu_{i_{s-1}}\mu_{i_{s-2}}\cdots \mu_{i_1}(Q)$ for all $1\leq s\leq k$. Let $\boldsymbol{m}_\Lambda =\mu_{i_k}\mu_{i_{k-1}}\cdots \mu_{i_1}$ be the mutation defined by the sequence $\Lambda$. From eqns.\eqref{sink}\eqref{source} one has
\begin{equation}
 \boldsymbol{m}_\Lambda(Y_\alpha)= Y_{t_\Lambda(\alpha)},
\end{equation}
where, for a sink sequence, $t_\Lambda$ is the element of Weyl$(Q)$
\begin{equation}
 t_\Lambda= s_{i_k}s_{i_{k-1}}\cdots s_{i_1}.
\end{equation}
For a source sequence one has
\begin{equation}
 t_\Lambda =\mathrm{diag}\big((-1)^{m_s}\big),\qquad m_s\equiv \#\, \text{of times }i_s\ \text{is repeated in }\Lambda.
\end{equation}
A sink (resp.\! source) sequence $\Lambda$ is called \emph{full} if contains each node of $Q$ exactly once. If $\Lambda$ is a full sink sequence, $\Lambda^{-1}$ is a full source sequence. If $\Lambda$ is a full sink sequence $t_\Lambda=c$, the Coxeter element, while for a full source sequence $t_\Lambda=I$ (the inversion).

\subsection{Quantum monodromy and the CNV strategy}\label{CNV}

 We summarize the method of ref.\!\cite{cnv}. Assume that our quiver (endowed with a non--degenerate superpotential $\cw$) $Q$ admits a sequence of elementary quiver mutations $\mu_{k(i)}$
($i=1,\dots, s$) such that
\begin{equation}
 \mu_{k(s)}\mu_{k(s-1)}\cdots \mu_{k(1)}(Q,\cw)=(Q,\cw),
\end{equation}
that is, the mutation $\boldsymbol{m}\equiv{\prod\limits^\curvearrowright}\mu_{k(i)}$ is the identity at the quiver level. 
In particular, $\boldsymbol{m}$ preserves the commutation relations in $\mathbb{T}_Q$, and hence is an automorphism of the quantum torus algebra of the form
\begin{equation}\begin{split}
 &\boldsymbol{m}\colon Y_\gamma = Y_{m(\gamma)}
\end{split}\end{equation}

We write 
\begin{equation}
 \boldsymbol{m}(Y_\gamma)= Y_{m(\gamma)} = \boldsymbol{V}\, Y_\gamma\, \boldsymbol{V}^{-1}, \qquad \forall\, \gamma\in \Gamma.
\end{equation}
for some $\boldsymbol{V}$.

 The corresponding cluster quantum mutation is 
\begin{equation}\label{parff}
\cq_{k(s)}\circ \cq_{k(s-1)}\circ \cdots \circ\cq_{k(1)}=\mathrm{Ad}(\mathbb{Y}(q)^{-1})\end{equation}
where
\begin{equation}\begin{split}
 \mathbb{Y}(q)&= \boldsymbol{V}^{-1}\cdot\Psi(\mu_{k(s)}\mu_{k(s-1)}\cdots \mu_{k(2)}Y_{k(1)})\,\cdot
\Psi(\mu_{k(s)}\mu_{k(s-1)}\cdots \mu_{k(3)}Y_{k(2)})\times\\
&\qquad\qquad \times \cdots\cdot \Psi(\mu_{k(s)}Y_{k(s-1)})\cdot\Psi(Y_{k(s)}).
\end{split}\end{equation}

Assume that $\boldsymbol{V}^m=1$ for some (finite) integer. Then $\mathbb{Y}(q)$ has precisely the general form of the $1/m$--fractional monodromy defined in ref.\!\cite{cnv}. Then the full quantum monodromy is
\begin{equation}
 \mathbb{M}(q)=\mathbb{Y}(q)^m,
\end{equation}
and since the monodromy is already written as an order product of quantum dilogarithms
\begin{equation*}
 \prod^\curvearrowright_{a=1,\dots 2\ell} \Psi(Y_{\gamma_a}),
\end{equation*}
by comparing with the 
Kontsevich--Soibelman WCF \cite{ks1}, we deduce that, in this particular chamber, the BPS spectrum consists of hypermultiplets of charge $\gamma_a\in \Gamma$ (counting both the states and their PCT conjugates).
If the identification is correct, this corresponds to a $\Z_m$--symmetric BPS chamber \cite{cnv}.
We have many ways to check the identification:
\begin{itemize}
 \item check that the original $\cn=2$ does have a $\Z_m$ symmetry acting in the correct way on the central charge in the regime in which that finite BPS spectrum is expected;
\item check that the predicted spectrum satisfies the PCT constraint: that is, if $\alpha$ is a charge vector associated to the BPS angle $\theta$, $-\alpha$ is also a charge vector with phase $\theta+\pi$;
\item check that the simple roots are among the charge vectors (they are always stable BPS states);
\item check that the phase order is consistent with one of the orders induced from a linear central charge $Z(\cdot)$ (this gives a set of triangle--like inequalities which should be satisfied; it is quite a strong constraint; if true, it gives a very compelling evidence).
\end{itemize}

Moreover, 
when, as in the models of interest here, we know from physical grounds that $\mathbb{M}$ has a given finite order $\ell$,
we have an equation
\begin{equation}
 \mathbb{M}(q)^\ell= \mathbb{Y}(q)^{m\,\ell} = 1\qquad \text{(in the sense of adjoint action).}
\end{equation}
If our putative fractional monodromy $\mathbb{Y}(q)$ also satisfies $\mathbb{Y}(q)^{m\ell}=1$, it should be the correct solution since this is already an overdetermined infinite set of equations.

In fact, one can show directly from the first principles that the results of the above CNV strategy should agree with the direct computation of the BPS spectrum (see \textit{e.g.} \cite{kel}).

Then, in order to find solutions to the spectral problem, we are lead to the following

\textbf{CNV method. }\label{CNVcrit} \textit{To get the BPS spectrum (in a particular finite chamber), look to the quiver mutations $\boldsymbol{m}$ which are the identity at the quiver level and are of finite order $m$ on $\mathbb{T}_\Gamma$. Then the BPS spectrum consists of \emph{half}--hypermultiplets having charge vectors $\alpha$
\begin{equation}
\{Y_\alpha\}_{\alpha\in A} = \{ \boldsymbol{m}^t \mu_{k(s)}\cdots \mu_{k(j+1)}Y_{\alpha_{k(j)}}\,|\, j= 1,\dots,s,\ t=0,\dots m-1\}
\end{equation}
(where
$
\boldsymbol{m}= \mu_{k(s)}\cdots\mu_{k(1)})$.}

The method may be generalized to infinite chambers as well, but we shall not need the more sophisticated version.

 \section{Complete families of Dynkin subquivers}\label{sinkseq}
The main technical tool used in ref.\!\cite{cnv} to construct the quantum monodromy of the $(G,G^\prime)$ models was the combinatorics of  mutations for square tensor products of alternating quivers, $G\,\square\,G^\prime$, \cite{keller-periodicity}. We start by generalizing that structure; we introduce the class of quivers admitting sink--factorized sequences with respect to a complete family of subquivers and, in particular, Weyl--factorized sequences. The quivers of the $14$ exceptional Arnold models are (mutation equivalent to) quivers admitting several such special sequences. To illustrate the power and versatilily of the technique, in the next section we shall use it to determine the spectrum in the finite chamber of $SU(2)$ SQCD with four fundamental flavors.
 
 \subsection{Sink--factorized sequences of mutations}\label{sinkseq1}
 
 We adopt the following notation: if $S\subset Q_0$ is a subset of the node set $Q_0$ of the quiver $Q$, by $Q|_S$ we mean the \emph{full subquiver} of $Q$ over the nodes in $S$.
 
We assume that the node set $Q_0$ of our quiver $Q$ is the disjoint union of a family of sets $\{q_a\}_{a\in A}$ of nodes
\begin{equation}
Q_0=\coprod_{a\in A} q_a.
\end{equation}
We shall refer to $\{Q|_{q_a}\}_{a\in A}$ as a \emph{complete family of subquivers}.
We write $q_{a(i)}$ for the unique node subset in the complete family $\{q_a\}_{a\in A}$ containing the $i$--th node.
 \smallskip

Let $Y_{\alpha,a}$, $\alpha=1,2,\dots, n_a$ denote the generators of the quantum torus algebra $\mathbb{T}_Q$ associated with the nodes in $q_a$. We write $\mathbb{T}_a\subset \mathbb{T}_Q$ for the \emph{quantum torus subalgebra} generated by the $Y_{i,a}$ with fixed subquiver label $a$.

\smallskip

We consider finite sequences of nodes, 
$\Lambda=\{i_1,i_2,\cdots, i_m\}$, $i_\ell\in Q_0$, which satisfy the following conditions:
\begin{enumerate}
\item we allow repetitions in the node list, $\{i_1,i_2,\cdots, i_m\}$, but each node should be present at least once;
\item $i_\ell\neq i_{\ell+1};$
\item at the quiver level $\mu_{i_m}\cdots \mu_{i_1}(Q)=Q$, while the corresponding map $\mathbb{T}_Q \rightarrow \mu_{i_m}\cdots \mu_{i_1}(\mathbb{T}_Q)$ has a finite order $s$.
\end{enumerate}

We write $\Lambda_a$ for the subsequence of $\Lambda$ obtained from $\Lambda$ by omitting all nodes $i_\ell\not\in q_a$.
As a matter of notation, given an (ordered) sequence of nodes $\Lambda$, we write $\boldsymbol{m}_\Lambda$ for the composite mutation corresponding to the sequence, \textit{i.e.} $\boldsymbol{m}_\Lambda\equiv \mu_{i_m}\cdots \mu_{i_1}$. We use the same symbol $\boldsymbol{m}_\Lambda$ both for the quiver mutation and the corresponding (right) mutation in $\mathbb{T}_Q$.
\medskip

\textsc{Definition.} We say that such a sequence $\Lambda$ is \emph{factorized with respect to the family $\{q_a\}_{a\in A}$ of subquivers} iff, for all $\ell=1,2,\dots, m$,
the $\ell$--th node in the sequence, $i_\ell$ is
 a source in the full subquiver 
 \begin{equation}
 \mu_{i_{\ell-1}}\mu_{i_{\ell-2}}\cdots \mu_{i_1}(Q)\Big|_{(Q_0\setminus q_{a(i_\ell)})\cup \{i_\ell\}}.
 \end{equation}
 We say that the sequence is \emph{sink--factorized with respect to the family} if it is factorized and moreover the node $i_\ell$ is a sink in the full subquiver
 \begin{equation}
  \mu_{i_{\ell-1}}\mu_{i_{\ell-2}}\cdots \mu_{i_1}(Q)\Big|_{q_{a(i_\ell)}}.
 \end{equation} 
\smallskip
  
  \textsc{Example.} Let $G,G^\prime$ be alternating Dynkin quivers. We have
  \begin{equation}
  G\,\square\,G^\prime = \coprod_{i=1}^{r(G^\prime)} G_{(i)},\qquad
 \big( G\,\square\,G^\prime\big)^\mathrm{op} = \coprod_{i=1}^{r(G)} G_{(i)}^\prime, 
  \end{equation} 
   where
   \begin{equation}
   G_{(i)}= \begin{cases}G & i\ \text{odd}\\
   G^\mathrm{op} & i\ \text{even},\end{cases}
   \end{equation}
   and the usual sequence in which the nodes $(i,a)$ with $i+a$ even are before those with $i+a$ odd is sink--factorized. \medskip

   The main properties of the factorized sequences are \medskip
   
   \textsc{Lemma.} 
 \textit{Let $\Lambda=\{i_1,i_2,\dots,i_m\}$ be a \emph{factorized} sequence, and let
   \begin{equation}
   \boldsymbol{m}_\Lambda\equiv \mu_{i_m}\cdots \mu_{i_2}\mu_{i_1},
   \end{equation} 
   be the corresponding quiver mutation. Then
   \begin{equation}
   \boldsymbol{m}_\Lambda(Y_{\alpha,a})\in \mathbb{T}_a. 
   \end{equation} 
Moreover, if $\Lambda$ is \emph{sink--factorized}
  \begin{equation}\label{sinkfa}
   \boldsymbol{m}_\Lambda(Y_{\alpha,a})=\boldsymbol{m}_{\Lambda_a}(Y_{\alpha,a})\equiv Y_{t_a(\alpha),a}, 
   \end{equation}
   where $t_a$ is the element of the Kac's Weyl group of the subquiver $Q|_{q_a}$
   \begin{equation}
   t_a= \prod_{(\alpha,a)\in\Lambda_a}^\curvearrowright s_{\alpha} 
   \end{equation}
   where $s_\alpha$ is the simple reflection associated to the $\alpha$--th simple root and the product is taken in the order dictated by the sub--sequence $\Lambda_a$.}  \smallskip
   
 \textsc{Proof.} The first part is clear from the definitions: The mutation $\mu_{i_\ell}$ acts on the quiver $Q^{(\ell-1)}\equiv\mu_{i_{\ell-1}}\cdots \mu_{i_1}(Q)$; let $Y^{(\ell-1)}_{\alpha,a}$ be the elements of $\mathbb{T}_Q$ associated to the nodes of the mutated quiver $Q^{(\ell-1)}$. One has
 \begin{equation}
Y_{\alpha,a}^{(\ell)}\equiv \mu_{i_\ell}(Y^{(\ell-1)}_{\alpha,a})=\begin{cases} \big(Y^{(\ell-1)}_{i_\ell}\big)^{-1} & (\alpha,a)=i_\ell\\
 Y^{(\ell-1)}_{\alpha, a(i_\ell)} \big(Y_{i_\ell}^{(\ell-1)}\big)^{\langle i_\ell,(\alpha,a)\rangle} & i_\ell\in q_a,\ i_\ell\neq (\alpha,a)\\
 Y_{\alpha,a} & \text{otherwise.}\end{cases}
 \end{equation}
 Hence, by induction on $\ell$,
 \begin{equation}
 Y_{\alpha,a}^{(m)}\in \mathbb{T}_a,
 \end{equation}
 which shows the first claim. Moreover,
 \begin{equation}\label{ooop}
 \mu_{i_\ell}\Big|_{\mathbb{T}_b} =\mathrm{Id},\quad \text{for }b\neq a(i_\ell).
 \end{equation}
 Next, we show that, \emph{if $\Lambda$ is a sink--factorized sequence,} 
 \begin{align}
& Q^{(\ell)}\big|_{q_a} = \boldsymbol{m}_{\Lambda_a(\ell)}\big(Q\big|_{q_a}\big),\\
 &\text{where }
 \Lambda_a(\ell)= \Lambda_a\cap \{i_1,i_2,\dots, i_\ell\},
 \end{align}
 that is, the $q_a$ subquiver of the mutated subquiver $Q^{(\ell)}$ is the same as the quiver obtained starting with the subquiver $Q|_{q_a}$ and mutating it with the sub--sequence of mutations $\Lambda_a(\ell)$, {completely ignoring the rest of the quiver $Q$.} This follows from the fact that, acting on $Q^{(\ell-1)}$, the mutation $\mu_{i_\ell}$ may introduce new arrows from a node in $q_{a(i_\ell)}$ to a node in $q_b$ with $b\neq a(i_\ell)$, but not new arrows connecting two nodes in the same subset $q_b$. The only effect of $\mu_{i_\ell}$ on the arrows in the subquivers is to invert those ending at $i_\ell$.  In particular, $\mu_{i_\ell}$ does not mutate the arrows of the full subquivers $Q^{(\ell-1)}|_{q_b}$ with $b\neq a(i_\ell)$. Therefore, the mutated subquiver $Q^{(\ell)}|_{q_a}$ is obtained from the subquiver $Q|_{q_a}$ by acting with the mutations in $\Lambda$ associated to nodes in $q_a$, \textit{i.e.}\! by acting with $\boldsymbol{m}_{\Lambda_a(\ell)}$. In view of eqn.\eqref{ooop}, this shows the first equality in eqn.\eqref{sinkfa}. The second one follows from \S.\,\ref{sinksourcesse}.   \hfill $\square$

  \subsection{Weyl--factorized sequences}\label{sinkseq2}
    A special case is when all the subquivers in the complete family are $ADE$ Dynkin quivers $\{G_a\}_{a\in A}$.
In this case, we have an isomorphism
\begin{equation}\label{latisp}
 \Gamma\simeq \bigoplus_{a\in A} \Gamma_{G_a},
\end{equation}
where $\Gamma_G$ stands for the root lattice of the Lie algebra $G$.
\smallskip

 Let  $\Lambda$ be a sink--factorized sequence. Then the elements $t_a=\prod_{\Lambda_a} s_\alpha\in \mathrm{Weyl}(G_a)$.
  In this case we say that $\Lambda$ is a sink--factorized sequence of type 
  \begin{equation}
  (G_1,t_1;G_2,t_2;\cdots G_k, t_k), \qquad (k=\# A).
  \end{equation}  
       %\smallskip

  \textsc{Example.} The square tensor product quiver $G\,\square\,G^\prime$ with the usual sequence of mutation is a Weyl--factorized sequence of type
  \begin{equation}
  \big(\overbrace{G,c;G,c;\cdots; G,c}^{r(G^\prime)\ \text{terms}}\big),
  \end{equation}
  where $c\in \mathrm{Weyl}(G)$ is the Coxeter transformation. 
\medskip

If $\Lambda$ is a sink--factorized sequence of type $(G_1,t_1;\cdots, G_k, t_k)$, we have    \begin{equation}
  \boldsymbol{m}_\Lambda(Y_{\alpha,a})=Y_{t_a(\alpha),a}. 
  \end{equation}
 Let $s=\mathrm{lcm}\big\{\text{order }t_a\ \text{in Weyl}(G_a)\big\}$. We have $\boldsymbol{m}_\Lambda^s=$identity on $\mathbb{T}_Q$. In view of the CNV strategy, one is lead to identify
 \begin{equation}\label{fractionamon}
\mathbb{Y}_\Lambda(q)\equiv{\prod_{i_\ell\in \Lambda}^\curvearrowright}\cq_{i_\ell},
\end{equation}
 with the $1/s$--fractional monodromy.  In order this identification to be consistent, $\mathbb{Y}_\Lambda(q)^s$ must have a structure compatible with PCT and with the BPS phase ordering. The first condition require that charge vectors $\alpha$ and $-\alpha$ should appear with the same multiplicities while the second puts a series of conditions. These conditions are automatically satisfied if $t_a$ is of the form $c_a^{r_a}$, where $c_a$ is the Coxeter element $c_a\in\mathrm{Weyl}(G)$ and the exponent $r_a$ is such that $sr_a=h_a$, $h_a$ being the Coxeter element of $G_a$. In this case, under the isomorphism \eqref{latisp},
the spectrum consists of one hypermultiplet per charge vector of the direct--sum form
\begin{equation}\label{goodspectrum}
 0\oplus 0 \oplus \cdots \oplus \alpha^{(a)} \oplus 0 \oplus \cdots \oplus 0,\qquad \alpha^{(a)}\in \Delta_+(G_a),
\end{equation}
having only one non--zero component (equal to a positive root of the corresponding Lie algebra $G_a$).
In this case, the consistency of the mass spectrum follows from comparison with the (obviously consistent) mass spectrum of the $G_a$--type Argyres--Douglas model in the maximal chamber.  In fact, the spectrum coincides with the disjoint union of the maximal spectra of the $G_a$ Argyres--Douglas theories.\smallskip

For a more general Weyl--factorized sequence,  the BPS charge vectors are
 \begin{equation}\label{genspectral}
 \beta_{k, \ell}^{(a)}\equiv t_a^k \,s_{j_{n_a},a}\cdots s_{j_{\ell+1},a}\,\alpha_{j_\ell,a},\qquad \ell=1,2,\dots, n_a,\  k=0,1,\dots, s-1,
 \end{equation}
 where $\Lambda_a=\{j_1,j_2,\cdots, j_{n_a}\}$. 
 
 The spectra we compute in this paper happen all to be of the especially nice direct--sum form
in eqn.\eqref{goodspectrum}.
\smallskip

We may extend the method of the present subsection to more general situations by allowing in the complete family of full subquivers $\{q_a\}_{a\in A}$ \emph{affine} Dynkin subquivers, besides the $ADE$ ones.

 \subsection{A baby example ($E_7$ AD in diverse chambers)}\label{baby}
 We illustrate the method and its physical meaning in a very baby model:
we consider the rank $7$ quiver
 \begin{equation}\label{D5quiver}
 \begin{gathered}
 \xymatrix{1,1\ar[r]^a & 1,2 \ar[d]^b & 1,3\ar[l]_e & \\
 2,1\ar[u]^d & 2,2 \ar[l]^c\ar[r]_f & 2,3\ar[u]_g & 0\ar[l]}
 \end{gathered}
 \end{equation}
 with superpotential
 \begin{equation}\label{quipot}
 \cw=\mathrm{Tr}(dcba)+\mathrm{Tr}(fgeb).
 \end{equation}

\subsubsection{Decoupling limits and all that}
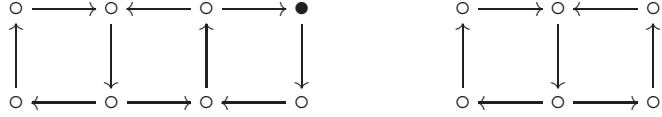
\begin{figure}
\begin{center}
\begin{equation*}
\begin{gathered}
\xymatrix{\circ\ar[r] & \circ \ar[d] & \circ \ar[l]\ar[r] & \bullet\ar[d]\\
 \circ\ar[u] & \circ \ar[l] \ar[r] & \circ\ar[u] &\circ\ar[l]}
\end{gathered}
\qquad\qquad
\begin{gathered}
\xymatrix{\circ\ar[r] & \circ \ar[d] & \circ \ar[l]\\
 \circ\ar[u] & \circ \ar[l] \ar[r] & \circ\ar[u]}
\end{gathered}
\end{equation*}
\caption{The $A_4\,\square\,A_2$ and the $A_3\,\square\,A_2$ quivers.}
\label{aaaquivers}
\end{center}
\end{figure}

Physically, we may realize the $\cn=2$ superconformal theory described by \eqref{D5quiver}\eqref{quipot} as the IR fixed point of the theory associated to the $A_4\,\square\,A_2$ quiver (the first quiver in figure \ref{aaaquivers}) perturbed by a suitable relevant operator which corresponds to giving a large central charge $|Z_\bullet|\gg \Lambda$ to the black node in figure \ref{aaaquivers}, with the effect of decoupling (in the IR) all the degrees of freedom carrying a non--zero $\bullet$ charge. In the same way, the $A_3\,\square\,A_2$ theory (second quiver in figure \ref{aaaquivers}) may be seen as a suitable IR limit of the theory \eqref{D5quiver} where we take $|Z_0|\rightarrow\infty$. Since $A_4\,\square\,A_2$ and $A_3\,\square\,A_2$ are, respectively, the type--$E_8$ and type--$E_6$ Argyres--Douglas models \cite{cnv}, the theory \eqref{D5quiver} should be a rank $7$ Argyres--Douglas theory, and hence the type--$E_7$ one.

Of course, there is an elementary direct proof of this last identification: mutating \eqref{D5quiver} one gets the $E_7$ quiver in its standard Dynkin form. However, here we are interested in the Dynkin subquiver viewpoint which will turn useful for the more complicated Arnold models. The present baby example is conceptually simpler, since the theory is actually complete \cite{CV11}, and all formal manipulations at the quiver level do have a direct physical meaning, and we are allowed to be naive. 

From the previous discussion, we see than the theory
\eqref{D5quiver} is a decoupling limit of the $4d$ $\cn=2$ theory geometrically engineered by the $E_8$--singular local Calabi--Yau hypersurface
\begin{equation}\label{E8sing}
x^5+y^3+ uv=0,
\end{equation} 
deformed by the relevant perturbation
\begin{equation}
x^5+y^3 +\epsilon\, x^3 y+ uv=0
\end{equation}
equal to the Hessian $\mathfrak{h}$ of the \textsc{lhs} of \eqref{E8sing} (\textit{i.e.}\! the \emph{less relevant} relevant deformation, from both the $2d$ and $4d$ viewpoints). In the IR the theory flows to the fixed point corresponding to the singular hypersurface $y^3+y x^3+uv=0$ (after a rescaling of the $x$ coordinate). 

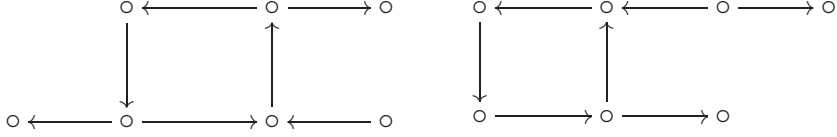
\begin{figure}
\begin{center}
\begin{equation*}
\begin{xy} 0;<1pt,0pt>:<0pt,-1pt>:: 
(141,0) *+{\circ} ="0",
(0,43) *+{\circ} ="1",
(141,43) *+{\circ} ="2",
(43,0) *+{\circ} ="3",
(98,43) *+{\circ} ="4",
(43,43) *+{\circ} ="5",
(98,0) *+{\circ} ="6",
"6", {\ar"0"},
"5", {\ar"1"},
"2", {\ar"4"},
"3", {\ar"5"},
"6", {\ar"3"},
"5", {\ar"4"},
"4", {\ar"6"},
\end{xy}\qquad
\begin{xy} 0;<1pt,0pt>:<0pt,-1pt>:: 
(132,0) *+{\circ} ="0",
(92,41) *+{\circ} ="1",
(48,0) *+{\circ} ="2",
(0,0) *+{\circ} ="3",
(0,41) *+{\circ} ="4",
(48,41) *+{\circ} ="5",
(92,0) *+{\circ} ="6",
"6", {\ar"0"},
"5", {\ar"1"},
"2", {\ar"3"},
"5", {\ar"2"},
"6", {\ar"2"},
"3", {\ar"4"},
"4", {\ar"5"},
\end{xy}
\end{equation*}
\caption{Other quivers in the $E_7$ class with a complete family of Dynkin subquivers}
\label{dynsub}
\end{center}
\end{figure}

The quiver \eqref{D5quiver} has an obvious decomposition into the complete family of Dynkin subquivers
\begin{equation}\label{a2xx}
A_2\coprod A_2\coprod A_2\coprod A_1,
\end{equation}
where the three copies of $A_2$ correspond to the subquivers over the nodes $\{(1,a),(2,a)\}$, $a=1,2,3$. Dually, we have the complete family
\begin{equation}
A_4\coprod A_3
\end{equation}
By mutating at $0$ we get a quiver which admits the complete families
\begin{equation}
A_3\coprod A_3\coprod A_1\quad \text{and}\quad A_3\coprod A_2\coprod A_2 .
\end{equation}
Other quivers in the $E_7$ mutation class may admit a complete family of Dynkin subquivers, see \textit{e.g.}\! the two quivers in figure \ref{dynsub} (as well as, of course, the $E_7$ Dynkin quiver itself, and the seven $A_1$'s: these two cases being already covered in \cite{cnv}).

All these decompositions into Dynkin subquivers correspond to BPS chambers of the $E_7$ AD theory whose BPS spectrum may be easily derived using the combinatorial methods of the present section. Of course, this is not so interesting for Argyres--Douglas theories, but it becomes relevant when applied to more general theories, see the next two sections.
\smallskip

Suppose we start with the $E_8$ Argyres--Douglas in a BPS chamber where the spectrum is given by a set of hypermultiplets with charge vectors in\footnote{\ $\Gamma_G$ denotes the root lattice of the simple simply--laced Lie group $G$.} $\Gamma\simeq \Gamma_{A_2}\otimes \Gamma_{A_4}$ of the form
\begin{equation}\label{spectE8cham}
\alpha \otimes \beta_a,\qquad \begin{aligned}&\alpha\in \Delta_+(A_2), &&\text{(\underline{all} positive roots of }A_2)\\ &\beta_a,\ a=1,2,3,4, &&\text{(\underline{simple} roots of }A_4)\end{aligned}
\end{equation}
Such a chamber of the $E_8$ AD theory is physically realizable since the model is complete. Now switch on the perturbation corresponding to the Hessian $\mathfrak{h}$; by what we saw above, the deformation is expected to give a large mass to the states having a charge vector
$\sum_{i,a} d_{i,a}\, \alpha_i\otimes\beta_a$ with $d_{1,4}\neq 0$.  Tuning the phase of the parameter $\epsilon$ in such a way\footnote{\ This is possible since the model is complete and has a finite spectrum in all chambers, the chamber themselves being finite in number.} that the deformation does not trigger spurious wall--crossings, the net effect is just that in the IR two of the states \eqref{spectE8cham} decouple, namely those of charge $\alpha_1\otimes \beta_4$ and
 $(\alpha_1+\alpha_2)\otimes \beta_4$, the others remaining unaffected.
 
\subsubsection{Stability of quiver representations}
 
 Of course, the same spectrum could be obtained directly from the analysis of the stable representations of the quiver \eqref{D5quiver}: if $\arg Z_0 < \arg Z_{i,a}$, it is easy to see\footnote{\ Let $X$ be a \emph{stable} representation of the (bound) quiver
\eqref{D5quiver} with $(\dim X)_0=n\neq 0$. Then there is an exact sequence of the form
\begin{equation*}
 0\rightarrow Y \rightarrow X \rightarrow S(0)^{\oplus n}\rightarrow 0,
\end{equation*}
where $S(0)$ is the simple representation with vector space $\C$ at the $0$--th node, and zero elsewhere. If $Y\neq 0$, one has
$\arg Z(Y)=\arg (Z(X)-n Z_0) > \arg Z(X)$. Since $X$ is stable, we get a contradiction. Hence $Y=0$.} that the stable representations are the ones with charge vector $\alpha_0$ plus the stable representations of the subquiver $A_3\,\square\,A_2$, which (in the corresponding BPS chamber) have charge vectors $\alpha\otimes \beta_a$, $a=1,2,3$. In this case the theory is complete \cite{CV11} and the two methods give equivalent information. However, in general, the decoupling analysis is more powerful: computing the BPS spectrum mathematically, we get a spectrum which is wall--crossing equivalent to the physical one, but we do not know whether the particular chamber in which we computed it may or may not be physically realized (see discussion in \cite{CV11}). The decoupling analysis, instead, gives us a {physical definition} of the BPS chamber we are computing in, and we are guaranteed that our finding have a direct physical meaning as actual particle spectra. Of course, it is also more delicate, since as we move in parameter space we have to control both the wall--crossings and the potential mixing of conserved charges.  
 
 \subsubsection{Cluster combinatorics}
 Let us compare the results of the decoupling (or quiver representation theoretical) analysis with those of the cluster--combinatorial CNV method. We consider only the decomposition \eqref{a2xx} which corresponds to the decoupling limit of the chamber \eqref{spectE8cham}, leaving the other cases as an amusement for the reader.
\smallskip

 Let $\Lambda$ be the node sequence
 \begin{multline}
\Lambda=\{ (1,1),(1,3),(2,2),0,(1,2),(2,1),(2,3),(1,1),(1,3),(2,2),\\
 (2,1),(2,3),(1,2),0,(2,2),(1,1),(1,3),(2,1),(2,3),(1,2)\}.
 \end{multline}
 It is easy to check (using, say, Keller's applet \cite{kelapp}) that this is a sink--factorized sequence with respect to the complete family of Dynkin subquivers \eqref{a2xx}. $\boldsymbol{m}_\Lambda$ acts as the identity on $\mathbb{T}_Q$, and hence the above sequence corresponds to the full quantum monodromy $\mathbb{M}(q)$.
 The general spectral formula \eqref{genspectral} gives the
 charge vectors
 \begin{equation}\label{e7str}
 \gamma_0,\ -\gamma_0, \quad \text{and}\quad \left\{\begin{aligned}&(s_2s_1)^{k-1}\alpha_2\otimes \beta_a\\
 &(s_2s_1)^{k-1}s_2\alpha_1\otimes \beta_a && k=1,2,3,\ a=1,2,3\end{aligned}\right.
 \end{equation}
 where $\alpha_i$ (resp.\! $s_i$) are the simple roots (resp.\! simple reflections) of $A_2$. Note that $s_2s_1$ is a Coxeter element of $A_2$. Then, for a fixed $a$, we get the spectrum of $A_2$ AD in the maximal chamber (which, in particular, shows that both the PCT and phase--ordering requirements are automatically satisfied). Hence the spectrum consists of a set of hypermultiplets with charge $\gamma_0$ and
 $\alpha\otimes\beta_a$, $a=1,2,3$ ($\alpha$ any positive root of $A_2$) which is precisely the spectrum predicted by the other two methods. \smallskip

This illustrates as the cluster combinatoric captures the BPS spectrum without going trough a detailed analysis.

\section{Example: the finite chamber of $SU(2)$ $N_f=4$}\label{seNf4}

In order not to give the false impression that the present methods are limited to the models which may be geometrical enginereed by a quasi--homogeneous singularity, in this section we discuss quite a different model, namely $SU(2)$ SQCD with four fundamental flavors. 

According to ref.\!\cite{Gaiotto:2009hg} $SU(2)$ SQCD with $N_f=4$ has a BPS chamber with a finite spectrum consisting of $12$ hypermultiplets. Let us see how this result follows from the existence of a complete family of Dynkin subquivers.

 We write the quiver of $SU(2)$ SQCD with four flavors in the form
 \begin{equation}
 \begin{gathered}
 \xymatrix{& & 1\ar[dl]\ar[dll] & &\\
 2\ar[drr] & 3\ar[dr] && 4\ar[ul] & 5\ar[ull]\\
 & & 6\ar[ur]\ar[urr]}
 \end{gathered}
 \end{equation}
 which admits the complete family of Dynkin subquivers
 \begin{equation}
 A_3\coprod A_3,
 \end{equation}
 where the two $A_3$ are the full subquivers over the nodes $\{1,2,3\}$ and, respectively, $\{4,5,6\}$; the sink--factorized sequence of nodes is
 \begin{equation}
 \Lambda=\{2,3,4,5,6,1\}
 \end{equation}
 having type
\begin{equation}
 (A_3,c\,;A_3,c)
\end{equation}
($c$ is Coxeter element of $A_3$)
as it is easy to check using Keller's applet \cite{kelapp}. Under the identification
 $\Gamma= \Gamma_{A_3}\oplus \Gamma_{A_3}$ we have
 \begin{equation}
 \boldsymbol{m}_\Lambda(Y_{\alpha\oplus \beta})= Y_{c(\alpha)\oplus c(\beta)}.
 \end{equation} 
 Since $h(A_3)=4$, one has $\boldsymbol{m}_\Lambda^4=1$ and $\prod_\Lambda \cq_k$ is the $1/4$--monodromy. The corresponding monodromy $\mathbb{M}(q)$ satisfies all the physical constraints by comparison with the $A_3$ AD model.
 
In conclusion, $SU(2)$ SQCD with $N_f=4$ has a $\Z_4$--symmetric finite BPS chamber with 12 hypermultiplets whose charge vectors, under the isomorphism $\Gamma= \Gamma_{A_3}\oplus \Gamma_{A_3}$, are
\begin{equation}\label{sser}
\{\alpha\oplus 0\ \text{and } 0\oplus\alpha\:|\: \alpha\in \Delta_+(A_3)\}.
\end{equation}  
 
\medskip
 
 It seems likely that all finite chambers of any $\cn=2$ model may be interpreted in terms of complete families of Dynkin subquivers.
 \medskip
 
 $SU(2)$ SQCD with $N_f=4$ is a superconformal theory (setting the mass parameters to zero) whose chiral primary operators have integer dimension. Hence the quantum monodromy should have period $1$, that is should be the identity on $\mathbb{T}_Q$,
 \begin{equation}
 \mathrm{Ad}(\mathbb{M}(q)) =\boldsymbol{1}.
 \end{equation}  
 Using the known spectrum \eqref{sser}, we shall check this statement in section \ref{Ys}.

\section{BPS spectra}\label{Spectras}

Now we determine the BPS spectra of the Arnold's exceptional theories in some `strongly coupled' finite chamber using the CNV strategy. 
\medskip

\subsection{$E_{13}$ spectrum from the CNV method}
We start with the $E_{13}$ model. Its quiver
\begin{equation}\label{equie13}
Q_{E_{13}}\colon \quad\begin{gathered}\begin{xy} 0;<0.7pt,0pt>:<0pt,-0.7pt>:: 
(0,55) *+{1,1} ="0",
(59,55) *+{1,2} ="1",
(121,55) *+{1,3} ="2",
(191,55) *+{1,4} ="3",
(262,55) *+{1,5} ="4",
(326,55) *+{1,6} ="5",
(391,55) *+{0} ="6",
(0,1) *+{2,1} ="7",
(59,1) *+{2,2} ="8",
(121,1) *+{2,3} ="9",
(191,1) *+{2,4} ="10",
(262,1) *+{2,5} ="11",
(326,1) *+{2,6} ="12",
"0", {\ar"1"},
"7", {\ar"0"},
"2", {\ar"1"},
"1", {\ar"8"},
"2", {\ar"3"},
"9", {\ar"2"},
"4", {\ar"3"},
"3", {\ar"10"},
"4", {\ar"5"},
"11", {\ar"4"},
"6", {\ar"5"},
"5", {\ar"12"},
"8", {\ar"7"},
"8", {\ar"9"},
"10", {\ar"9"},
"10", {\ar"11"},
"12", {\ar"11"},
\end{xy}
\end{gathered}
\end{equation}
admits two obvious complete families of Dynkin quivers of type $$I\colon\ (A_7,A_6),\qquad II\colon\ (A_2,A_2,A_2,A_2,A_2,A_2,A_1).$$
Mutating at $0$ we get a quiver with a complete family of type $(A_6,A_6,A_1)$. We call the corresponding three chambers, $I$, $II$, and $III$, respectively.

With respect to these decompositions, the following sequences are sink--factorized
\begin{align*}
 \Lambda_I=&\big\{(1,2),(1,4),(1,6),(2,1),(2,3),(2,5),(1,1),(1,3),(1,5),0,\\
  &(2,2),(2,4),(2,6),(1,2),(1,4),(1,6),(2,1),(2,3),(2,5),(1,1),\\
  &(1,3),(1,5),0,(2,2),(2,4),(2,6),(1,2),(1,4),(1,6),(2,1),\\
  &(2,3),(2,5),(1,1),(1,3),(1,5),0,(2,2),(2,4),(2,6),(1,2),\\
  &(1,4),(1,6),(2,1),(2,3),(2,5),(1,1),(1,3),(1,5),0\big\}\\
\Lambda_{II}=&\big\{ (1,1),(2,2),(1,3),(2,4),(1,5),(2,6),0,(2,1),(1,2),(2,3),\\
&(1,4),(2,5),(1,6),(1,1),(2,2),(1,3),(2,4),(1,5),(2,6)\big\}\\
\Lambda_{III}=&\big\{(2,1),(2,3),(2,5),(1,2),(1,4),(1,6),(2,2),(2,4),(2,6),(1,1),\\
&(1,3),(1,5),(2,1),(2,3),(2,5),(1,2),(1,4),(1,6),(2,2),(2,4),\\
&(2,6),(1,1),(1,3),(1,5),(2,1),(2,3),(2,5),(1,2),(1,4),(1,6),\\
&(2,2),(2,4),(2,6),(1,1),(1,3),(1,5),(2,1),(2,3),(2,5),(1,2),\\
&(1,4),(1,6),0\big\}.
\end{align*}
Their effect on the quiver $Q_{E_{13}}$ is to give back the same quiver up to an involutive permutation of the nodes   
\begin{align*}
 &P_I=<(\c_{1,1},\c_{0}),(\c_{1,2},\c_{1,6}), (\c_{1,3},\c_{1,5}),(\c_{2,1},\c_{2,6}),(\c_{2,2},\c_{2,5}),(\c_{2,3},\c_{2,4})> \\
& P_{II}=<(\c_{1,a},\c_{2,a})> a=1,...,6 \\
&P_{III}= <(\c_{a,1},\c_{a,6}),(\c_{a,3},\c_{a,4}),(\c_{a,2},\c_{a,5}),a=1,2>
\end{align*}
For instance, $\mu_{\Lambda_I}(Q_{E_{13}})$ is the quiver
\begin{equation}
\mu_{\Lambda_I}(Q_{E_{13}})\colon\qquad\begin{gathered}\begin{xy} 0;<0.7pt,0pt>:<0pt,-0.7pt>:: 
(0,50) *+{1,1} ="0",
(50,50) *+{1,2} ="1",
(100,50) *+{1,3} ="2",
(150,50) *+{1,4} ="3",
(200,50) *+{1,5} ="4",
(250,50) *+{1,6} ="5",
(300,50) *+{0} ="6",
(50,0) *+{2,1} ="7",
(100,0) *+{2,2} ="8",
(150,0) *+{2,3} ="9",
(200,0) *+{2,4} ="10",
(250,0) *+{2,5} ="11",
(300,0) *+{2,6} ="12",
"0", {\ar"1"},
"2", {\ar"1"},
"1", {\ar"7"},
"2", {\ar"3"},
"8", {\ar"2"},
"4", {\ar"3"},
"3", {\ar"9"},
"4", {\ar"5"},
"10", {\ar"4"},
"6", {\ar"5"},
"5", {\ar"11"},
"12", {\ar"6"},
"7", {\ar"8"},
"9", {\ar"8"},
"9", {\ar"10"},
"11", {\ar"10"},
"11", {\ar"12"},
\end{xy}
\end{gathered}
\end{equation}
Then the two--fold reiteration of the above sequences give full Weyl--factorized sequences of types
\begin{align}
&I\colon &&(A_7,c^{-8};A_6,c^{-7})\\
&II\colon &&(A_2,c^{-3};A_2,c^{-3};A_2,c^{-3};A_2,c^{-3};A_2,c^{-3};A_2,c^{-3};A_1,c^{-2})\\
&
III\colon&&(A_6,c^{-7};A_6,c^{-7};A_1,c^{-2}).
\end{align}
Note that each Coxeter element $c$ is raised to a power equal to minus the Coxeter number of the corresponding Lie algebra. Hence the type of the sequence is always of the form $(G_1,1;G_2,1,\cdots;G_r, 1)$ which corresponds to the full quantum monodromy $\mathbb{M}(q)$ (cfr.\! \S.\,\ref{sinkseq2}). Then, under the isomorphism
$\Gamma= \oplus_a \Gamma_{G_a}$, the spectrum in all three chambers is given by one hypermultiplet per each charge vector of the form in eqn.
\eqref{goodspectrum}, that is, the spectrum has the direct sum form.

\subsection{Three independent checks of the $E_{13}$ spectrum}

We wish to show that the above results are consistent with what we know --- both from a physical and a mathematical point of view --- about the spectrum of the $E_{13}$ model.

Mathematically, the stable BPS particles correspond to the stable representations of the quiver \cite{CV11}.
The quiver \eqref{equie13} is a one--point extension \cite{Ring} of the $A_6\,\square\,A_2$ quiver. Let\footnote{\ $I$ stands for the set of nodes of the quiver.} $\Gamma=\oplus_{i\in I} \Z\,\alpha_i$ be the charge lattice, identified with the dimension lattice of the
representations $X$ of the quiver $Q_{E_{13}}$. Assume that the central charge function
\begin{equation}
Z(\cdot)\colon \Gamma\rightarrow \C,\qquad X\mapsto \sum_{i\in I} Z_i\, (\dim X)_i,
\end{equation}
satisfies the conditions
\begin{equation}\label{ZW13}
\mathrm{Im}\,Z_i >0,\quad \text{and}\quad \arg Z_0< \arg Z_j,\ j\neq 0.
\end{equation}
Consider a representation $X$ with $(\dim X)_0=k\neq 0$. Clearly, we have\footnote{\ $S(0)$ denotes the simple representation with $S(0)_0=\C$ and $S(0)_j=0$ for $j\neq 0$.}
\begin{equation}
0\rightarrow M \rightarrow X\rightarrow S(0)^{\oplus k}\rightarrow 0,
\end{equation}
for a certain sub--representation $M$ having $(\dim M)_0=0$. From eqn.\eqref{ZW13}, we see that, if $M\neq 0$,
\begin{equation}
\arg M > \arg X,
\end{equation}
and $X$ is not stable and hence does not correspond to a BPS particle in a chamber of the form \eqref{ZW13}. Therefore, the only stable representation $X$ with $(\dim X)_0\neq 0$ is $S(0)$ with charge vector $\alpha_0$. In a BPS chambers satisfying eqn.\eqref{ZW13}, all other stable representations have support in the $A_6\,\square\,A_2$ subquiver, and hence may be identified with BPS states of the corresponding theory, already studied in \cite{cnv}. In particular, there are two chambers whose spectrum consists of a single hypermultiplet of charge vector $\alpha_0$ plus the hypermultiplets of one of the two canonical chambers of $A_6\,\square\,A_2$ computed in \cite{cnv}. This reproduces the spectra of chambers $II$ and $III$ above.
\medskip

Physically, we may understand the $E_{13}$ model as a decoupling limit of the $A_7\,\square\, A_2$ model in which we give infinite mass to the node
$\alpha_7\otimes \beta_2$. Although this is a bit heuristic, we expect that of all the BPS states in the two canonical $A_7\,\square\,A_2$ chambers of \cite{cnv}, consisting of hypermultiplets with charge vectors
$$\alpha \otimes \beta_i,\quad\text{resp.}\quad \alpha_j\otimes \beta,\qquad\quad \left|\begin{aligned}&\alpha\in  \Delta_+(A_7),\: \beta\in \Delta_+(A_2),\\ &\alpha_j,\beta_i\ \text{simple},\end{aligned}\right.$$
the only states which decouple are precisely those with charge vector $\sum_{i,j} N_{i,j}\,\alpha_i\otimes \beta_j$ with
$N_{7,2}\neq 0$. The remaining states precisely form the spectra in our chambers $I$ and $II$, respectively, while $III$ is easily understood as a natural subset of $I$. In view of the discussion in section 
\ref{MinNC}, we see the fact that the BPS spectra have a simple physical interpretation as circumstantial evidence for the physical realizability of the corresponding BPS chambers. \smallskip

Mathematically, the strongest indication of the correctness of the above BPS spectra is that the resulting quantum monodromy,
$\mathbb{M}(q)$, does have order $7$, as predicted by the Type IIB construction (see next section). This, however, says nothing about the physical reality of the corresponding chambers.

\subsection{The other nine models}

The other nine models follow the same pattern as $E_{13}$. Again we have several decompositions into complete families of Dynkin quivers which admit full Weyl--factorized sequences of the standard type, namely,
\begin{equation}\label{standardtype}
 (G_1, c^{-h_1}; G_2, c^{-h_2}; G_3, c^{-h_3};\cdots ; G_s, c^{-h_s}),
\end{equation}
where $h_a$ is the Coxeter number of $G_a$. Hence the spectrum is always given by equation \eqref{goodspectrum}, and in particular is consistent with both PCT and the phase--ordering inequalities.
Again, this result is confirmed by the stability analysis of the quiver representations, as well as by the physical idea of decoupling states from a parent $A_m\,\square\,G$ theory. The fact that the corresponding monodromies do have the right order $\ell$ (cfr.\! table
\ref{monorders}) guarantees the correctness of the result.

A list of Weyl--factorized sequence types is presented in table \ref{arnoldatype}. For the details of these Weyl--factorized sequences, see appendix \ref{sedetail}.

\begin{table}
\begin{center}
\begin{tabular}{|c|c||c|c|}\hline
$E_{13}$&$(A_7,A_6)$&$Q_{10}$&$(D_4,D_4,A_2)$\\
&$(A_6,A_6,A_1)$&&$(A_3,A_3,A_2,A_2)$\\
&$(A_2,A_2,A_2,A_2,A_2,A_2,A_1)$&$\star$&$(D_4,D_4,A_1,A_1)$\\
\cline{1-2}
$Z_{11}$&$(A_4,A_4,A_3)$&$\star$&$(A_2,A_2,A_2,A_2,A_1,A_1)$\\\cline{3-4}
&$(A_3,A_3,A_3,A_2)$&$Q_{11}$&$(D_4,D_4,A_2,A_1)$\\
\cline{1-2}
$Z_{12}$&$(A_4,A_4,A_3,A_1)$&&$(A_3,A_3,A_2,A_2,A_1)$\\
&$(A_3,A_3,A_3,A_2,A_1)$&&$(A_4,A_3,A_2,A_2)$\\\cline{3-4}
&$(A_5,A_4,A_3)$&$Q_{12}$&$(D_4,D_4,A_2,A_2)$\\
\cline{1-2}
$Z_{13}$&$(A_5,A_5,A_3)$&&$(A_4,A_4,A_2,A_2)$\\\cline{3-4}
&$(A_3,A_3,A_3,A_2,A_2)$&$S_{11}$&$(D_4,D_4,A_3)$\\
\cline{1-2}
$W_{13}$&$(A_3,A_3,A_3,A_3,A_1)$&&$(A_3,A_3,A_3,A_2)$\\\cline{3-4}
&$(A_4,A_4,A_4,A_1)$&$S_{12}$&$(D_4,D_4,A_3,A_1)$\\
&$(A_5,A_4,A_4)$&&$(A_3,A_3,A_3,A_2,A_1)$\\
&&&$(A_4,A_3,A_3,A_2)$\\
\hline
\end{tabular}
\end{center}
\caption{Types of some Weyl--factorized sequences for the Arnold exceptional $\cn=2$ theories. (The ones denoted with a $\star$ correspond to the quiver \eqref{exother}). Each of them correspond to a BPS chamber of the corresponding $\cn=2$ theory with a finite spectrum having the direct--sum form \eqref{goodspectrum}.}\label{arnoldatype}
\end{table}
\medskip

Table \ref{arnoldatype} presents a list of chambers with finite BPS spectra, which have natural physical interpretations, and hence are expected to be physically realized. Of course, as for the $E_7$ AD model in \S.\ref{baby}, there exist other chambers in which the spectrum has the `Weyl--factorized' form \eqref{goodspectrum}. Indeed, the (infinite)  mutation classes of the exceptional Arnold quivers contain many quivers which admit complete families of Dynkin subquivers: Our combinatoric methods apply to all these chambers in a straightforward way.

However, in general it is difficult to establish whether a given chamber is physical or not, even at the heuristic level. This is one reason why here we have not attempted a full classification of all `Weyl--factorized' chambers, but limited ourselves to the set of those chambers simply related to the analysis of ref.\!\cite{cnv}. As an example of a mutated quiver with an obvious Weyl--factorized structure, consider the following quiver in the mutation class of $Q_{10}$:
\begin{equation}\label{exother}
\begin{gathered}\begin{xy} 0;<1pt,0pt>:<0pt,-1pt>:: 
(59,0) *+{4,1} ="0",
(102,0) *+{4,2} ="1",
(29,100) *+{1,1} ="2",
(72,100) *+{1,2} ="3",
(59,73) *+{2,1} ="4",
(102,73) *+{2,2} ="5",
(86,43) *+{3,1} ="6",
(131,43) *+{3,2} ="7",
(0,129) *+{2,3} ="8",
(45,129) *+{1,3} ="9",
"1", {\ar"0"},
"0", {\ar"4"},
"5", {\ar"1"},
"3", {\ar"2"},
"2", {\ar"4"},
"2", {\ar"8"},
"5", {\ar"3"},
"9", {\ar"3"},
"4", {\ar"5"},
"6", {\ar"4"},
"5", {\ar"7"},
"7", {\ar"6"},
\end{xy}\end{gathered}
 \end{equation}
 which, in algebraic terms, corresponds to a one--point co--extension of a one--point extension of the algebra $\C\vec D_4\otimes \C\vec A_2$ (at an injective and projective indecomposable, respectively).
On the nose, this quiver has two complete families of Dynkin subquivers of types $$(A_2,A_2,A_2,A_2,A_1,A_1)\quad\text{and}\quad (D_4,D_4,A_1,A_1)$$ 
with respect to which
we have Weyl--factorized sequences of the standard type \eqref{standardtype},
leading, in both cases, to direct--sum BPS spectra of the form \eqref{goodspectrum} (see appendix \ref{sedetail} for the details).

 \section{The $Y$--systems and their periodicity}\label{Ys}

A general consequence of ref.\!
\cite{cnv} is that the quantum (fractional) monodromy $\mathbb{M}(q)$ (resp.\! $\mathbb{Y}(q)$) of a $4d$ $\cn=2$ model geometrically engineered by Type IIB on an isolated  quasi--homogeneous singularity with $\hat{c}<2$ has finite order. Moreover, in each BPS chamber of such a theory with a \emph{finite} BPS spectrum the (fractional) monodromy is written as a finite product of elementary quantum cluster mutations.

In the classical limit $q\rightarrow 1$, the action of $\mathbb{M}(q)$ (resp.\! $\mathbb{Y}(q)$) on the quantum torus algebra $\mathbb{T}_Q$ reduces to the corresponding KS \emph{rational} symplecto--morphism of the complex torus $T\sim (\C^*)^{\text{rank}\,\Gamma}$ \cite{ks1,Gaiotto:2008cd}, which is directly related to to the hyperK\"ahler geometry of the $3d$ dimensional version of the theory \cite{Gaiotto:2008cd}. As explained in \cite{Gaiotto:2008cd,cnv}, the resulting symplectic rational maps form a $Y$--system in the sense of the Thermodynamical Bethe Ansatz\cite{zamolodchikovTBA}.

\medskip

The usual TBA periodic $Y$--systems \cite{zamolodchikovTBA,keller-periodicity,Ravanini:1992fi,Kuniba:1993cn} correspond to `decoupled' singularities of the form $W_G+W_{G^\prime}$. Ref.\!\cite{cnv} predicts the existence of many others such periodic $Y$--systems associated to non--decoupled singularities. Here we explain how we have checked this prediction for the  Arnold's exceptional singularities.

\medskip

We start by reviewing the construction of the $Y$--system from the quantum monodromy. Recall from section \ref{CNV} that the (fractional) quantum monodromy, as computed from the BPS data in a \emph{finite} chamber, may be seen as the result of a sequence of quantum mutations of the torus algebra $\mathbb{T}_Q$ which happens to have the particular form of eqn.\eqref{parff}.
The action of $\mathbb{M}(q)$ on the quantum torus algebra is specified by the action on the set of generators
$ \{Y_i \}_{i \in Q_0}$
where, as usual, we write $Q_0$ for the set of nodes of $Q$,
\begin{equation}
 Y_i\rightarrow Y^\prime_i \equiv \textrm{Ad}(\mathbb{M}^{-1})Y_i \equiv N[R_i(Y_j)],
\end{equation}
here $N[R_i(Y_j)]$ stands for the normal--order version of the rational function $R_i(Y_j)$ of the operators $Y_j$ \cite{cnv}.
$R_i$ has the log--symplectic property
\begin{equation}
 \langle \alpha_i, \alpha_j\rangle_\text{Dirac}\: d\log Y_i\wedge d\log Y_j =  
\langle \alpha_i, \alpha_j\rangle_\text{Dirac}\: d\log R_i\wedge d\log R_j.  
\end{equation}
The rational map $Y_i\rightarrow R_i$ is simply the classical limit of the monodromy action, from which we may recover the full quantum action by taking the normal--order prescription for the operators (this is true \cite{cnv} for all simply--laced, \textit{i.e.}\! $|B_{ij}| \leq 1$, quivers).
This shows that the quantum monodromy, acting on $\mathbb{T}_Q$ in the adjoint fashion, has finite order $\ell$ if and only if the rational map $\C^r\rightarrow \C^r$
\begin{equation}R\colon\quad Y_i\mapsto R_i(Y_j)\label{mapex}\end{equation} has order $\ell$.

On the other hand, $\mathrm{Ad}(\mathbb{M}^{-1})$ is an ordered product of basic quantum  mutations of the form $\prod\limits^\curvearrowright \cq_k$. The rational map $Y_j\rightarrow R_j$ coincides with the composition of the rational functions $R_j^{(k)}$ giving the classical limit of each basic mutation $\cq_k$ in the product. The map $Y_j\rightarrow R^{(k)}_j$ is just the elementary mutation at the $k$--th node of the $Y$--seed in the sense of Fomin--Zelevinsky\footnote{\ For $Y$ variables in the universal semi--field.} \cite{fominIV,cluster-intro} but for $Q$ replaced by the opposite quiver $Q^\mathrm{op}$ (see \textit{e.g.}\! \cite{kel}). The Keller applet \cite{kelapp} automatically generates the $Y$--seed mutations for any quiver $Q$, and hence, although the actual form of the rational map \eqref{mapex} is typically quite cumbersome, it is easily generated by a computer procedure.
\smallskip

By definition, the $Y$--system associated to a finite chamber of a $\cn=2$ model is simply the recursion relation generated by the iteration of the rational map \eqref{mapex}, namely
\begin{equation}\label{ysssi}
 Y_j(s+1) = R_j\big(Y_k(s)\big),\qquad s\in \Z.
\end{equation}
Specializing to the $\cn=2$ $(G,G^\prime)$ theories studied in \cite{cnv}, eqn.\eqref{ysssi} reproduces the well--known TBA $Y$--systems for the integrable $2d$ $(G,G^\prime)$ models \cite{zamolodchikovTBA,keller-periodicity,Ravanini:1992fi,Kuniba:1993cn}.

Although we have generated at the computer the $Y$--systems for all the exceptional Arnold models, to avoid useless vaste of paper, here we limit ourselves to present the explicit form of just a couple of examples: see the appendices. All the others may be straightforwardly generated, using the explicit Weyl--factorized sequences listed in appendix \ref{sedetail}, by the same computer procedure.\smallskip

We stress that, although the explicit form of the $Y$--system depends on the particular finite BPS chamber we use to write the map \eqref{mapex}, two $Y$--systems corresponding to different chambers of the \emph{same} $\cn=2$ theory are equivalent, in the sense that they are related by a rational change of variables $Y_j\rightarrow Y^\prime_j(Y_k)$. Indeed, the monodromy $\mathbb{M}(q)$ is independent of the chamber up to conjugacy, and so is its classical limit map $Y_j\rightarrow R_j$. Hence the rational maps $R_j$ obtained in different chambers are conjugate in the Cremona group.\smallskip 

In conclusion, the (adjoint action of the) quantum monodromy $\mathbb{M}(q)$ has a finite order $\ell$ if and only if the corresponding $Y$--system is periodic with (minimal) period $\ell$, that is
\begin{equation}
 \mathrm{Ad}\big[\mathbb{M}(q)^\ell\big]=\mathrm{Id}\quad\Longleftrightarrow\quad Y_j(s+\ell)=Y_j(s),\ \:\forall\, j\in Q_0,\:s\in \Z.
\end{equation}
 
For the Arnold exceptional models, we know from string theory that $\mathbb{M}(q)$ has the finite orders $\ell$ listed in table \ref{monorders}.
This proves that the corresponding $Y$--systems are periodic of period $\ell$. It should be possible to give an interpretation of these new periodic $Y$--systems in terms of exactly solvable $2d$ theories in analogy with the $(G,G^\prime)$ ones \cite{zamolodchikovTBA,Ravanini:1992fi,Kuniba:1993cn}.

At the mathematical level, we get an unexpected relation between singularity theory and cyclic subgroups of the Cremona groups $\mathrm{Cr}(n)$ of birational automorphisms $\mathbb{P}^n\rightarrow \mathbb{P}^n$, both interesting subjects in Algebraic Geometry (the second one being notoriously hard for $n\geq 3$ \cite{dolgachev}).
\medskip

\subsection{Checking the periodicity}\label{Schper}

Type IIB engineering of the model together with our computation of the BPS spectrum \emph{proves} that the corresponding $Y$--systems
are periodic with the periods listed in table \ref{monorders}. However, as a check, we wish to give an independent proof of the periodicity. 

In  principle, to prove periodicity, one has just to iterate $\ell$ times the rational map $R$ of eqn.\eqref{mapex}, and check that the resulting rational map is the identity. Unfortunately, at the intermediate stages of the recursion, one typically gets rational functions so cumbersome that no computer can handle them analytically \cite{FomLecture}. 
Luckily, there is an alternative strategy advocated by Fomin in \cite{FomLecture}. The $\ell$--fold interation of $R$, $R^\ell$, is a rational map whose fixed--point subvariety $\mathscr{F}$ has some codimension $n$ in $\C^r$. Periodicity is just the statement that $n=0$.

 If we specialize the $Y_i$'s to randomly chosen numbers uniformely distributed  in some disk of radius $\rho$, compute numerically the transformation $R^\ell(Y_i)$, and get back the original point $Y_i$, we conclude that our randomly chosen point $Y_i$ lays on the fixed--point subvariety $\mathscr{F}$ within the computational numerical accuracy $\epsilon$.
The probability that a randomly chosen point appears to be on the fixed locus $\mathscr{F}$ is then of order $(\epsilon/\rho)^{2n}$. 

Therefore, the probability that applying $R^\ell$ to a sequence of $k$ random points we get back the same sequence of points, is
of order $(\epsilon/\rho)^{2nk}$. Since $\epsilon/\rho\sim 10^{-11}$, for $n\neq 0$ the probability goes quite rapidly to zero as we increase $k$.
If we do get back the original sequence of points for, say, $k=5$, we may conclude that $n=0$ with a confidence level which differs from $100\%$ by a mere $10^{-108}\,\%$.
\medskip

Using this strategy, we have checked all the periodicities listed in table \ref{monorders}. The interested reader, may find the details of the check for the $Q_{12}$ model in appendix \ref{chekper}.

We have also checked the order of the $1/4$--fractional monodromy for $SU(2)$ SQCD with four flavors, getting $4$, namely order $1$ for the full monodromy $\mathbb{M}(q)$, in agreement with the physical prediction based on the fact that all chiral primaries have integral dimension.

\section*{Acknowledgements}

We thank Bernhard Keller and Cumrun Vafa for many enlightening discussions and valuable comments.

 \appendix

\section{The $E_7$ $Y$--system from the chamber \eqref{e7str} }
We illustrate the kind of $Y$--system one gets from Weyl--factorized sequence using the baby example of \S.\,\ref{baby}. There we presented a Weyl--factorized sequence of nodes for the family of subquivers \eqref{a2xx}. Written in terms of the BPS data in the corresponding chamber, the classical monodromy is equal to 
the $Y$--seed mutation (for the opposite quiver) associated to this Weyl--factorized sequence. This $Y$--seed mutation, as generated by the Keller mutation applet \cite{kelapp} is
(we set $Y_{i,a,s}\equiv Y_{i,a}(s)$)
$$Y_{1,1,s+1}=\frac{1 + Y_{2,3,s}}{Y_{2,3,s} Y_{0,s}}\qquad Y_{2,1,s+1}=\frac{Y_{1,2,s} Y_{1,3,s} (1 + Y_{2,3,s} + Y_{2,3,s} Y_{0,s})}{1 + Y_{1,2,s}}$$
$$Y_{1,2,s+1}=\frac{Y_{2,1,s} (1 + Y_{1,2,s}) Y_{2,2,s} Y_{2,3,s} Y_{0,s}}{(1 + Y_{2,3,s} + Y_{2,1,s}(1+ (1 + Y_{2,2,s}(1 + Y_{1,2,s})) Y_{2,3,s}) (1 + Y_{2,3,s}(1 + Y_{0,s})))}$$
$$Y_{2,2,s+1}=\frac{1 + Y_{2,3,s} + Y_{2,2,s} Y_{2,3,s} + Y_{1,2,s} Y_{2,2,s} Y_{2,3,s}}{Y_{1,2,s} + Y_{1,2,s} Y_{2,3,s}}$$
$$Y_{1,3,s+1}=\frac{1 + Y_{2,3,s}}{Y_{2,1,s} + Y_{2,1,s} (1 + Y_{2,2,s} + Y_{1,2,s} Y_{2,2,s}) Y_{2,3,s}}$$
$$Y_{2,3,s+1}=\frac{Y_{1,1,s} Y_{1,2,s} (1 + Y_{2,1,s} + Y_{2,3,s} + Y_{2,1,s} (1 + Y_{2,2,s} + Y_{1,2,s} Y_{2,2,s}) Y_{2,3,s})}{1 + Y_{1,2,s} + (1 + Y_{1,2,s}) (1 + (1 + Y_{1,2,s} + Y_{1,1,s} Y_{1,2,s}) Y_{2,2,s}) Y_{2,3,s}}$$
$$Y_{0,s+1}=\frac{1 + Y_{2,3,s} + Y_{2,2,s} Y_{2,3,s} + Y_{1,2,s} Y_{2,2,s} Y_{2,3,s}}{Y_{1,1,s} Y_{1,2,s} Y_{2,2,s} Y_{2,3,s}}.$$

This $Y$--system should be equivalent to the usual $E_7$ $Y$--system, differing only by a change of variables $Y_i\rightarrow \widetilde{Y}_i$. In particular, it must have the same minimal period $\ell$ as the usual one, namely $5$.
We have checked this using the strategy of \S.\,\ref{Schper}.

 \section{Periodicity of the $Y$--systems: the $Q_{12}$ example}\label{chekper}
In this appendix we illustrate in one example, $Q_{12}$, the details of our check of the periodicity of the $Y$--systems which we have performed for \emph{all} the models discussed in this paper getting full agreement with table \ref{monorders}. 

The main tool we have used is the Keller java applet \cite{kelapp}. For a sufficiently simple quiver, the test of the periodicity of the monodromy can be done algebraically via the Keller applet. However, quite typically, the rational functions appearing at intermediate stages in the iteration of the $Y$--system (even for Dynkin quivers\,!) are so cumbersome that no computer can handle them. Therefore we adopt the numerical strategy explained in \S.\,\ref{Schper} using \texttt{Mathematica}. 

\medskip

Here is our \texttt{Mathematica} program to test the periodicity for $Q_{12}$. The numbering of the nodes $Q_{Q_{12}}$ for the numerical iteration is the following one
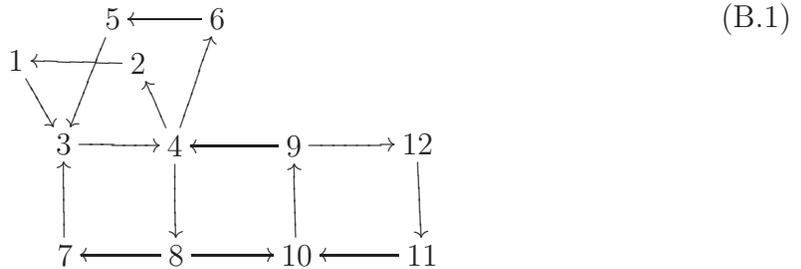
\begin{equation}
\begin{xy} 0;<0.6pt,0pt>:<0pt,-0.6pt>:: 
(0,26) *+{1} ="0",
(77,28) *+{2} ="1",
(30,79) *+{3} ="2",
(100,80) *+{4} ="3",
(61,0) *+{5} ="4",
(127,0) *+{6} ="5",
(31,149) *+{7} ="6",
(101,149) *+{8} ="7",
(175,80) *+{9} ="8",
(177,149) *+{10} ="9",
(256,149) *+{11} ="10",
(253,79) *+{12} ="11",
"1", {\ar"0"},
"0", {\ar"2"},
"3", {\ar"1"},
"2", {\ar"3"},
"4", {\ar"2"},
"6", {\ar"2"},
"3", {\ar"5"},
"3", {\ar"7"},
"8", {\ar"3"},
"5", {\ar"4"},
"7", {\ar"6"},
"7", {\ar"9"},
"9", {\ar"8"},
"8", {\ar"11"},
"10", {\ar"9"},
"11", {\ar"10"},
\end{xy}
\end{equation}

We start by defining the variables we will need. \texttt{Y} will be the variable we will use to reproduce the $Y$--system, while \texttt{A} is just a dummy variable we use to store the initial set of values for our iteration. In the array \texttt{Y} only the second index matters: it is the number of nodes of the quiver + 1 (notice that we are starting from 0), while the first index is there just to help us generating the interation.

\footnotesize{}

\begin{verbatim}
Clear[i, j,k,l, A, Y]
Array[Y, {4, 13}, 0];
Array[A, 12];
\end{verbatim}

\normalsize{}
After that, we proceed by generating the random intitial set of conditions uniformly distributed over a disk of radius \texttt{rho} centered at the origin and we are saving it in the dummy variable $\texttt{A}$. Notice that we are using the random number generator of \texttt{Mathematica}, so if you want to run our program remember that you should put in a separate cell at the beginning of your sheet the command that seeds the random number generator, say \texttt{RandomSeed[5]}.

\footnotesize{}
\begin{verbatim}

rho = 10.0;

For[j = 1, j < 2,
   For[k = 1, k < 13, k++, Y[0, k] = rho(2Random[] - 1);]
   If[Sum[Y[0, l]^2, {l, 1, 12}] <= rho^2, j = 3]
]

For[k = 1, k < 13, k++, A[k] = Y[0, k];]
\end{verbatim}

\normalsize{}
\texttt{r} is our prediction for the order of the monodromy, see table \ref{monorders}

\footnotesize{}\begin{verbatim}
r = 13;
\end{verbatim}

\normalsize{}
\normalsize{}
Then there is the iteration corresponding to the shortest sink--sequence we have found for $Q_{12}$: we are chopping it into the maximal pieces we are able to generate with the Keller applet. Since the sink--sequence we have found has order 2, we repeat it \texttt{2r} times. In general, if a sink--sequence have order $s$ one will have to repeat the iteration \texttt{sr} times. After the \texttt{For} instruction, there are the mutations of the $Y$--variables. Here you can see how we have used the first index of the array \texttt{Y}: we have splitted the sink--sequence into three parts. The variables \texttt{Y[1,k]} contains the mutations corresponding to the first part:  7 10 12 4 5 1 9 3 8; the variables \texttt{Y[2,k]} corresponds to the second one: 11 12 10 4 9 6 2 7 8; while the variables \texttt{Y[3,k]} to the third one: 3 11 7 4 10 12 1 5.

\footnotesize{}
\begin{verbatim}
For[i = 1, i < 2r + 1, i++,
  
Y[1,1]=(1 + (1 + (1 + Y[0,1])*Y[0,3])*Y[0,4] + (((1 + Y[0,1])*Y[0,3])
*Y[0,4])*Y[0,5] + (((1 + Y[0,1])*Y[0,3])*Y[0,4] + (((1 + Y[0,1])
*Y[0,3])*Y[0,4])*Y[0,5])*Y[0,7])/(Y[0,1] + Y[0,1]*Y[0,4]);

Y[1,2]=(Y[0,1]*Y[0,2] + Y[0,1]*Y[0,2]*Y[0,4] + (Y[0,1]*Y[0,2] + 
Y[0,1]*Y[0,2]*Y[0,4] + Y[0,1]*Y[0,2]*Y[0,4]*Y[0,9] + Y[0,1]*Y[0,2]
*Y[0,4]*Y[0,9]*Y[0,10])*Y[0,12])/(1 + Y[0,1] + (1 + Y[0,1])*
Y[0,12]);

Y[1,3]=(1 + Y[0,4])/(((1 + Y[0,1])*Y[0,3])*Y[0,4] + (((1 + Y[0,1])*
Y[0,3])*Y[0,4])*Y[0,5] + (((1 + Y[0,1])*Y[0,3])*Y[0,4] + (((1 + 
Y[0,1])*Y[0,3])*Y[0,4])*Y[0,5])*Y[0,7]);

Y[1,4]=(((((1 + Y[0,1])*Y[0,3])*Y[0,4] + (((1 + Y[0,1])*Y[0,3])*
Y[0,4])*Y[0,5] + (((1 + Y[0,1])*Y[0,3])*Y[0,4] + (((1 + Y[0,1])*
Y[0,3])*Y[0,4])*Y[0,5])*Y[0,7])*Y[0,9] + ((((1 + Y[0,1])*Y[0,3])*
Y[0,4] + (((1 + Y[0,1])*Y[0,3])*Y[0,4])*Y[0,5] + (((1 + Y[0,1])*
Y[0,3])*Y[0,4] + (((1 + Y[0,1])*Y[0,3])*Y[0,4])*Y[0,5])*Y[0,7] + ((((1
 + Y[0,1])*Y[0,3])*Y[0,4] + ((1 + Y[0,1])*Y[0,3])*Y[0,4]^2 + (((1 +
Y[0,1])*Y[0,3])*Y[0,4] + ((1 + Y[0,1])*Y[0,3])*Y[0,4]^2)*Y[0,5])*
Y[0,7])*Y[0,8])*Y[0,9])*Y[0,10])*Y[0,12])/(1 + (2 + (1 + Y[0,1])*
Y[0,3])*Y[0,4] + (1 + (1 + Y[0,1])*Y[0,3])*Y[0,4]^2 + (((1 + Y[0,1])
*Y[0,3])*Y[0,4] + ((1 + Y[0,1])*Y[0,3])*Y[0,4]^2)*Y[0,5] + (((1 + 
Y[0,1])*Y[0,3])*Y[0,4] + ((1 + Y[0,1])*Y[0,3])*Y[0,4]^2 + (((1 + 
Y[0,1])*Y[0,3])*Y[0,4] + ((1 + Y[0,1])*Y[0,3])*Y[0,4]^2)*Y[0,5])*
Y[0,7] + (1 + (2 + (1 + Y[0,1])*Y[0,3])*Y[0,4] + (1 + (1 + Y[0,1])*
Y[0,3])*Y[0,4]^2 + (((1 + Y[0,1])*Y[0,3])*Y[0,4] + ((1 + Y[0,1])*
Y[0,3])*Y[0,4]^2)*Y[0,5] + (((1 + Y[0,1])*Y[0,3])*Y[0,4] + ((1 + 
Y[0,1])*Y[0,3])*Y[0,4]^2 + (((1 + Y[0,1])*Y[0,3])*Y[0,4] + ((1 + Y
[0,1])*Y[0,3])*Y[0,4]^2)*Y[0,5])*Y[0,7] + (Y[0,4] + (1 + (1 + Y[0,1])
*Y[0,3])*Y[0,4]^2 + (((1 + Y[0,1])*Y[0,3])*Y[0,4]^2)*Y[0,5] + (((1 + 
Y[0,1])*Y[0,3])*Y[0,4]^2 + (((1 + Y[0,1])*Y[0,3])*Y[0,4]^2)*Y[0,5])
*Y[0,7])*Y[0,9] + ((Y[0,4] + (1 + (1 + Y[0,1])*Y[0,3])*Y[0,4]^2 + 
(((1 + Y[0,1])*Y[0,3])*Y[0,4]^2)*Y[0,5] + (((1 + Y[0,1])*Y[0,3])*
Y[0,4]^2 + (((1 + Y[0,1])*Y[0,3])*Y[0,4]^2)*Y[0,5])*Y[0,7])*Y[0,9])
*Y[0,10])*Y[0,12]);

Y[1,5]=(1 + (1 + (1 + Y[0,1])*Y[0,3])*Y[0,4] + (((1 + Y[0,1])*Y[0,3])
*Y[0,4])*Y[0,5] + (((1 + Y[0,1])*Y[0,3])*Y[0,4] + (((1 + Y[0,1])*
Y[0,3])*Y[0,4])*Y[0,5])*Y[0,7])/((1 + Y[0,4])*Y[0,5]);

Y[1,6]=(((1 + Y[0,4])*Y[0,5])*Y[0,6] + (((1 + Y[0,4])*Y[0,5])*Y[0,6]
+ Y[0,4]*Y[0,5]*Y[0,6]*Y[0,9] + Y[0,4]*Y[0,5]*Y[0,6]*Y[0,9]*
Y[0,10])*Y[0,12])/(1 + Y[0,5] + (1 + Y[0,5])*Y[0,12]);

Y[1,7]=(((1 + (1 + (1 + Y[0,1])*Y[0,3])*Y[0,4] + (((1 + Y[0,1])*
Y[0,3])*Y[0,4])*Y[0,5] + (((1 + Y[0,1])*Y[0,3])*Y[0,4] + (((1 + 
Y[0,1])*Y[0,3])*Y[0,4])*Y[0,5])*Y[0,7])*Y[0,8])*Y[0,10])/(1 + Y[0,7]
 + (1 + Y[0,7] + ((1 + Y[0,4])*Y[0,7])*Y[0,8])*Y[0,10]);

Y[1,8]=(1 + Y[0,7] + (1 + Y[0,7])*Y[0,10])/((((1 + Y[0,4])*Y[0,7])*
Y[0,8])*Y[0,10]);

Y[1,9]=(1 + Y[0,4] + (1 + Y[0,4])*Y[0,12])/((Y[0,4]*Y[0,9] + Y[0,4]
*Y[0,9]*Y[0,10])*Y[0,12]);

Y[1,10]=(((1 + Y[0,4])*Y[0,7])*Y[0,8] + (((1 + Y[0,4])*Y[0,7])*
Y[0,8] + Y[0,4]*Y[0,7]*Y[0,8]*Y[0,9] + Y[0,4]*Y[0,7]*Y[0,8]*Y[0,9]
*Y[0,10])*Y[0,12])/(1 + Y[0,7] + (1 + Y[0,7] + ((1 + Y[0,4])*Y[0,7])
*Y[0,8])*Y[0,10] + (1 + Y[0,7] + (1 + Y[0,7] + ((1 + Y[0,4])*Y[0,7])
*Y[0,8])*Y[0,10])*Y[0,12]);

Y[1,11]=(Y[0,10]*Y[0,11] + Y[0,10]*Y[0,11]*Y[0,12])/(1 + Y[0,10]);

Y[1,12]=(Y[0,4]*Y[0,9] + Y[0,4]*Y[0,9]*Y[0,10])/(1 + Y[0,4] + (1 + 
Y[0,4] + Y[0,4]*Y[0,9] + Y[0,4]*Y[0,9]*Y[0,10])*Y[0,12]);


Y[2,1]=(Y[1,1]*Y[1,2] + Y[1,1]*Y[1,2]*Y[1,4] + (Y[1,1]*Y[1,2] + Y
[1,1]*Y[1,2]*Y[1,4] + Y[1,1]*Y[1,2]*Y[1,4]*Y[1,9] + (Y[1,1]*Y[1,2] + 
Y[1,1]*Y[1,2]*Y[1,4] + Y[1,1]*Y[1,2]*Y[1,4]*Y[1,9] + Y[1,1]*Y[1,2]
*Y[1,4]*Y[1,9]*Y[1,10])*Y[1,11])*Y[1,12])/(1 + Y[1,2] + (Y[1,2] + Y
[1,2]*Y[1,11])*Y[1,12]);

Y[2,2]=1/(Y[1,2] + (Y[1,2] + Y[1,2]*Y[1,11])*Y[1,12]);

Y[2,3]=(Y[1,3]*Y[1,4] + Y[1,3]*Y[1,4]*Y[1,7])/(1 + Y[1,4]);

Y[2,4]=(((1 + Y[1,7])*Y[1,9] + ((1 + Y[1,7])*Y[1,9] + ((1 + Y[1,7] + 
((1 + Y[1,4])*Y[1,7])*Y[1,8])*Y[1,9])*Y[1,10])*Y[1,11])*Y[1,12])/(1
 + Y[1,4] + (1 + Y[1,4])*Y[1,7] + (1 + Y[1,4] + (1 + Y[1,4])*Y[1,7] +
  (Y[1,4] + Y[1,4]*Y[1,7])*Y[1,9] + (1 + Y[1,4] + (1 + Y[1,4])*Y[1,7]
   + (Y[1,4] + Y[1,4]*Y[1,7])*Y[1,9] + ((Y[1,4] + Y[1,4]*Y[1,7])*
   Y[1,9])*Y[1,10])*Y[1,11])*Y[1,12]);

Y[2,5]=(((1 + Y[1,4])*Y[1,5])*Y[1,6] + (((1 + Y[1,4])*Y[1,5])*Y[1,6]
 + Y[1,4]*Y[1,5]*Y[1,6]*Y[1,9] + (((1 + Y[1,4])*Y[1,5])*Y[1,6] + 
 Y[1,4]*Y[1,5]*Y[1,6]*Y[1,9] + Y[1,4]*Y[1,5]*Y[1,6]*Y[1,9]*Y[1,10])
 *Y[1,11])*Y[1,12])/(1 + Y[1,6] + (Y[1,6] + Y[1,6]*Y[1,11])*Y
 [1,12]);

Y[2,6]=1/(Y[1,6] + (Y[1,6] + Y[1,6]*Y[1,11])*Y[1,12]);

Y[2,7]=((((1 + Y[1,4])*Y[1,8])*Y[1,10])*Y[1,11])/(1 + Y[1,7] + (1 +

 Y[1,7] + (1 + Y[1,7] + ((1 + Y[1,4])*Y[1,7])*Y[1,8])*Y[1,10])*
 Y[1,11]);

Y[2,8]=(1 + Y[1,7] + (1 + Y[1,7] + (1 + Y[1,7])*Y[1,10])*Y[1,11])/
(((((1 + Y[1,4])*Y[1,7])*Y[1,8])*Y[1,10])*Y[1,11]);
Y[2,9]=(1 + Y[1,4] + (1 + Y[1,4] + (1 + Y[1,4])*Y[1,11])*Y[1,12])/
((Y[1,4]*Y[1,9] + (Y[1,4]*Y[1,9] + Y[1,4]*Y[1,9]*Y[1,10])*Y[1,11])
*Y[1,12]);

Y[2,10]=(((1 + Y[1,4])*Y[1,7])*Y[1,8] + (((1 + Y[1,4])*Y[1,7])*Y
[1,8])*Y[1,11] + (((1 + Y[1,4])*Y[1,7])*Y[1,8] + Y[1,4]*Y[1,7]*Y[1,8]
*Y[1,9] + (((2 + 2*Y[1,4])*Y[1,7])*Y[1,8] + 2*Y[1,4]*Y[1,7]*Y[1,8]
*Y[1,9] + Y[1,4]*Y[1,7]*Y[1,8]*Y[1,9]*Y[1,10])*Y[1,11] + (((1 + Y
[1,4])*Y[1,7])*Y[1,8] + Y[1,4]*Y[1,7]*Y[1,8]*Y[1,9] + Y[1,4]*Y[1,7]
*Y[1,8]*Y[1,9]*Y[1,10])*Y[1,11]^2)*Y[1,12])/(1 + Y[1,7] + (1 + Y
[1,7] + (1 + Y[1,7] + ((1 + Y[1,4])*Y[1,7])*Y[1,8])*Y[1,10])*Y[1,11] 
+ (1 + Y[1,7] + (2 + 2*Y[1,7] + (1 + Y[1,7] + ((1 + Y[1,4])*Y[1,7])*Y
[1,8])*Y[1,10])*Y[1,11] + (1 + Y[1,7] + (1 + Y[1,7] + ((1 + Y[1,4])*Y
[1,7])*Y[1,8])*Y[1,10])*Y[1,11]^2)*Y[1,12]);

Y[2,11]=(Y[1,10] + (Y[1,10] + Y[1,10]*Y[1,11])*Y[1,12])/(1 + (1 + 
Y[1,10])*Y[1,11]);

Y[2,12]=(((1 + Y[1,2])*Y[1,4] + ((1 + Y[1,2])*Y[1,4])*Y[1,6])*Y[1,9] 
+ (((1 + Y[1,2])*Y[1,4] + ((1 + Y[1,2])*Y[1,4])*Y[1,6])*Y[1,9] + (((1 
+ Y[1,2])*Y[1,4] + ((1 + Y[1,2])*Y[1,4])*Y[1,6])*Y[1,9])*Y[1,10])*Y
[1,11] + ((Y[1,2]*Y[1,4] + ((1 + 2*Y[1,2])*Y[1,4])*Y[1,6])*Y[1,9] + 
((2*Y[1,2]*Y[1,4] + ((2 + 4*Y[1,2])*Y[1,4])*Y[1,6])*Y[1,9] + ((Y[1,2]
*Y[1,4] + ((1 + 2*Y[1,2])*Y[1,4])*Y[1,6])*Y[1,9])*Y[1,10])*Y[1,11] 
+ ((Y[1,2]*Y[1,4] + ((1 + 2*Y[1,2])*Y[1,4])*Y[1,6])*Y[1,9] + ((Y[1,2]
*Y[1,4] + ((1 + 2*Y[1,2])*Y[1,4])*Y[1,6])*Y[1,9])*Y[1,10])*Y[1,11]
^2)*Y[1,12] + (Y[1,2]*Y[1,4]*Y[1,6]*Y[1,9] + (3*Y[1,2]*Y[1,4]*Y
[1,6]*Y[1,9] + Y[1,2]*Y[1,4]*Y[1,6]*Y[1,9]*Y[1,10])*Y[1,11] + (3*Y
[1,2]*Y[1,4]*Y[1,6]*Y[1,9] + 2*Y[1,2]*Y[1,4]*Y[1,6]*Y[1,9]*Y[1,10])
*Y[1,11]^2 + (Y[1,2]*Y[1,4]*Y[1,6]*Y[1,9] + Y[1,2]*Y[1,4]*Y[1,6]*Y
[1,9]*Y[1,10])*Y[1,11]^3)*Y[1,12]^2)/(1 + Y[1,4] + (1 + Y[1,4])*Y
[1,11] + (1 + Y[1,4] + Y[1,4]*Y[1,9] + (2 + 2*Y[1,4] + 2*Y[1,4]*Y
[1,9] + Y[1,4]*Y[1,9]*Y[1,10])*Y[1,11] + (1 + Y[1,4] + Y[1,4]*Y[1,9] 
+ Y[1,4]*Y[1,9]*Y[1,10])*Y[1,11]^2)*Y[1,12]);


Y[3,1]=1/Y[2,1];

Y[3,2]=(Y[2,1]*Y[2,2] + (Y[2,1]*Y[2,2] + Y[2,1]*Y[2,2]*Y[2,11])*Y
[2,12])/(1 + Y[2,1]);

Y[3,3]=(Y[2,4] + ((1 + Y[2,3])*Y[2,4])*Y[2,7])/(1 + Y[2,3] + Y[2,3]
*Y[2,4]);

Y[3,4]=(1 + Y[2,3])/(Y[2,3]*Y[2,4]);

Y[3,5]=1/Y[2,5];

Y[3,6]=(Y[2,5]*Y[2,6] + (Y[2,5]*Y[2,6] + Y[2,5]*Y[2,6]*Y[2,11])*Y
[2,12])/(1 + Y[2,5]);

Y[3,7]=1/((1 + Y[2,3])*Y[2,7]);

Y[3,8]=(((((1 + Y[2,3] + Y[2,3]*Y[2,4])*Y[2,7])*Y[2,8])*Y[2,10])*Y
[2,11])/(1 + (1 + Y[2,3])*Y[2,7] + (1 + (1 + Y[2,3])*Y[2,7] + (1 + (1 
+ Y[2,3])*Y[2,7])*Y[2,10])*Y[2,11]);

Y[3,9]=(((((1 + Y[2,1])*Y[2,3])*Y[2,4] + (((1 + Y[2,1])*Y[2,3])*Y
[2,4])*Y[2,5])*Y[2,9] + ((((1 + Y[2,1])*Y[2,3])*Y[2,4] + (((1 + Y[2,1])
*Y[2,3])*Y[2,4])*Y[2,5])*Y[2,9] + ((((1 + Y[2,1])*Y[2,3])*Y[2,4] + 
(((1 + Y[2,1])*Y[2,3])*Y[2,4])*Y[2,5])*Y[2,9])*Y[2,10])*Y[2,11])*Y
[2,12])/(1 + Y[2,3] + Y[2,3]*Y[2,4] + (1 + Y[2,3] + Y[2,3]*Y[2,4] + 
(1 + Y[2,3] + Y[2,3]*Y[2,4])*Y[2,11])*Y[2,12]);

Y[3,10]=(1 + Y[2,11])/(Y[2,10]*Y[2,11]);

Y[3,11]=(Y[2,10] + (Y[2,10] + Y[2,10]*Y[2,11])*Y[2,12])/(1 + (1 + 
Y[2,10])*Y[2,11]);

Y[3,12]=1/((1 + Y[2,11])*Y[2,12]);

\end{verbatim}

\normalsize{}
After one has implemented all the sink--sequence, one has to change the basis of the charge lattice according to the expression of $P_{\Lambda}$:

\footnotesize{}
\begin{verbatim}
  
 Y[0, 1] = Y[3, 2];
  Y[0, 2] = Y[3, 1];
  Y[0, 3] = Y[3, 12];
  Y[0, 4] = Y[3, 9];
  Y[0, 5] = Y[3, 6];
  Y[0, 6] = Y[3, 5];
  Y[0, 7] = Y[3, 11];
  Y[0, 8] = Y[3, 10];
  Y[0, 9] = Y[3, 4];
  Y[0, 10] = Y[3, 8];
  Y[0, 11] = Y[3, 7];
  Y[0, 12] = Y[3, 3];
  ]  
\end{verbatim}

\normalsize{}
 Having changed the basis of the charge lattice, the iteration goes back to the beginning. When  the iteration finishes we check our results with the following line:

\footnotesize{}
\begin{verbatim}
For[k = 1, k < 14, k++, Print[{k, Chop[A[k] - Y[0, k]]}]]
\end{verbatim}

\normalsize{}
Notice that we are using the \texttt{Chop} command in order to avoid problems with the machine precision. And that's it. With this very very simple program one is able, with the help of the Keller applet for generating the corresponding mutations of the  $Y$--variables, to check all the monodromy orders of table \ref{monorders}.

%%%%%%%%%%%%

\section{Details on the Weyl--factorized sequences}\label{sedetail} 
In this appendix we present the details of the computations summarized in table \ref{arnoldatype}. For each model  we specify the quiver used and the associated Weyl--factorized sequences, with their types and involutive permutations.  For the quivers arising from one--point extensions of known algebras --- $Z_{12}$, $Q_{11}$, $S_{12}$, $W_{13}$ --- we report only the sequences associated with the quivers in figures \ref{Zquivers}, \ref{Qquivers}, \ref{SWquivers}: obtaining the sequences corresponding to these quivers mutated at $0$ is straightforward.

\begin{figure}
\begin{center}
 \begin{equation*}
Q_{Z_{11}}\colon\quad\begin{gathered}
\xymatrix{3,1\ar[r] & 3,2\ar[d] & 3,3 \ar[l] & \\
2,1 \ar[u]\ar[d] & 2,2 \ar[l]\ar[r] & 2,3 \ar[u]\ar[d] & 2,4\ar[l]\\
1,1 \ar[r] & 1,2\ar[u] & 1,3 \ar[l]\ar[r] & 1,4\ar[u]}
\end{gathered}
\end{equation*}

\begin{equation*}
Q_{Z_{12}}\colon\qquad\begin{gathered}
\xymatrix{3,1\ar[r] & 3,2\ar[d] & 3,3 \ar[l] & &\\
2,1 \ar[u]\ar[d] & 2,2 \ar[l]\ar[r] & 2,3 \ar[u]\ar[d] & 2,4\ar[l]&\\
1,1 \ar[r] & 1,2\ar[u] & 1,3 \ar[l]\ar[r] & 1,4\ar[u] & 0\ar[l]}
\end{gathered}
\end{equation*}

\begin{equation*}
Q_{Z_{13}}\colon\qquad\begin{gathered}\begin{xy} 0;<0.9pt,0pt>:<0pt,-0.9pt>:: 
(0,0) *+{3,1} ="0",
(0,50) *+{2,1} ="1",
(0,100) *+{1,1} ="2",
(50,0) *+{3,2} ="3",
(50,50) *+{2,2} ="4",
(50,100) *+{1,2} ="5",
(100,0) *+{3,3} ="6",
(100,50) *+{2,3} ="7",
(100,100) *+{1,3} ="8",
(150,50) *+{2,4} ="9",
(150,100) *+{1,4} ="10",
(200,50) *+{2,5} ="11",
(200,100) *+{1,5} ="12",
"1", {\ar"0"},
"0", {\ar"3"},
"1", {\ar"2"},
"4", {\ar"1"},
"2", {\ar"5"},
"3", {\ar"4"},
"6", {\ar"3"},
"5", {\ar"4"},
"4", {\ar"7"},
"8", {\ar"5"},
"7", {\ar"6"},
"7", {\ar"8"},
"9", {\ar"7"},
"8", {\ar"10"},
"10", {\ar"9"},
"9", {\ar"11"},
"12", {\ar"10"},
"11", {\ar"12"},
\end{xy}\end{gathered}
\end{equation*}

\caption{Quivers for the Z familiy.}\label{Zquivers}
\end{center}
\end{figure}
 
\begin{table}%[!ht]
\begin{center}
\begin{tabular}{|c|c|p{12.5cm}|}\hline
$Z_{11}$&&(3,2),(2,1),(2,3),(1,2),(1,4),(3,1),(2,2),(2,4),(1,1),(1,3),(3,3),(2,1),(2,3), (1,2),(1,4),(3,2),(2,2),(2,4),(1,1),(1,3),(3,1),(3,3),(2,1),(2,3),(1,2),(1,4)\\ \cline{2-3}
&Type:&$(A_4,s_4 s_2 (c_{A_4})^{2}:A_4,s_3 s_1(c_{A_4})^{2}:A_3,(c_{A_3})^{2})$\\ \cline{2-3}
&$P_{\Lambda}$&$<(\gamma_{3,1},\gamma_{3,3}),(\gamma_{a,4},\gamma_{a,1}),(\gamma_{a,2},
\gamma_{a,3}), a =1,2>$\\
\cline{2-3}
&&(3,1),(3,3),(1,1),(1,3),(2,2),(2,4),(3,2),(1,2),(1,4),(2,1),(2,3),(3,1),(3,3), (1,1),(1,3),(2,2),(2,4),(3,2),(1,2),(2,1),(2,3)\\\cline{2-3}
&Type:&$(A_3,(c_{A_3})^2: A_3,(c_{A_3})^2:A_3,(c_{A_3})^2: A_2, u)$\\\cline{2-3}
&$P_{\Lambda}$&$<(\gamma_{1,4},\gamma_{2,4}),(\gamma_{1,a},\gamma_{3,a}), a = 1,2,3> $\\\hline
\hline
$Z_{12}$&&(3,2),(2,1),(2,3),(1,2),(1,4),(3,1),(2,2),(1,1),(1,3),0,(2,1), (1,2),(1,4),(1,1),(1,3),(1,2),(2,4),(2,3),(2,2),(3,3),(3,2),0, (1,4),(2,4),(3,1),(3,3),(2,1),(2,3),(1,1),(1,3),0\\\cline{2-3}
&Type:&$(A_5,(c_{A_5})^{3}: A_4,s_3 s_1 (c_{A_4})^{2}:A_3,(c_{A_3})^{2})$\\\cline{2-3}
&$P_{\Lambda}$&$<(\c_{1,1},\c_{0}),(\c_{1,2},\c_{1,4}),(\c_{2,1},\c_{2,4}),(\c_{2,2},\c_{2,3}),(\c_{3,1},\c_{3,3})>$\\\cline{2-3}
&&(3,1),(1,1),(2,2),(3,3),(1,3),(2,4),0,(2,1),(3,2),(1,2),(2,3), (1,4),(3,1),(1,1),(2,2),(3,3),(1,3),(2,4),(2,1),(3,2),(1,2),(2,3)\\\cline{2-3}
&Type:&$(A_3,(c_{A_3})^{2}:A_3,(c_{A_3})^{2}:A_3,(c_{A_3})^{2}:A_1,s)$\\\cline{2-3}
&$P_{\Lambda}$&$<\c_{1,a},\c_{3,a}> a=1,2,3, (\c_{2,4},\c_{1,4})$\\
\hline
\hline
$Z_{13}$&&(3,2),(2,1),(2,3),(2,5),(1,2),(1,4),(2,2),(2,4),(1,1),(1,3),(1,5),(3,1), (2,1),(2,3),(2,5),(1,2),(1,4),(3,3),(2,2),(2,4),(1,1),(1,3),(1,5), (3,2),(2,1),(2,3),(2,5),(1,2),(1,4),(1,1),(3,3),(2,2),(2,4),(1,1),(1,3),(1,5)\\\cline{2-3}
&Type:&$(A_5,(c_{A_5})^{3}: A_5,(c_{A_5})^{3}:A_3,(c_{A_3})^{2})$\\\cline{2-3}
&$P_{\Lambda}$&$<(\gamma_{3,1},\gamma_{3,3}),(\gamma_{a,5},\gamma_{a,1}),(\gamma_{a,4},\gamma_{a,2}), a =1,2>$\\\cline{2-3}
&&(3,1),(3,3),(1,1),(1,3),(2,2),(1,5),(2,4),(3,2),(1,2),(2,1),(2,3),(2,5),(1,4), (3,1),(3,3),(1,1),(1,3),(2,2),(3,2),(1,2),(2,1),(2,3),(2,4),(1,5)\\\cline{2-3}
&Type:&$(A_3,(c_{A_3})^{2}: A_3,(c_{A_3})^{2}:A_3,(c_{A_3})^{2}: A_2, u:A_2,v)$\\\cline{2-3}
&$P_{\Lambda}$&$<(\gamma_{1,b},\gamma_{2,b}),b=4,5 ,(\gamma_{1,a},\gamma_{3,a}), a = 1,2,3> $\\
\hline
\end{tabular}
\end{center}
\caption{Weyl--factorized sequences for the quivers of figure \ref{Zquivers}.}
\end{table}

\begin{figure}
\begin{center}
\begin{equation*}
Q_{Q_{10}}\colon\quad\begin{gathered}\begin{xy} 0;<0.8pt,0pt>:<0pt,-0.8pt>:: 
(80,0) *+{3,1} ="0",
(143,0) *+{3,2} ="1",
(51,73) *+{2,1} ="2",
(123,73) *+{2,2} ="3",
(0,32) *+{4,1} ="4",
(95,32) *+{4,2} ="5",
(50,136) *+{1,1} ="6",
(121,136) *+{1,2} ="7",
(191,73) *+{2,3} ="8",
(191,136) *+{1,3} ="9",
"1", {\ar"0"},
"0", {\ar"2"},
"3", {\ar"1"},
"2", {\ar"3"},
"4", {\ar"2"},
"6", {\ar"2"},
"3", {\ar"5"},
"3", {\ar"7"},
"8", {\ar"3"},
"5", {\ar"4"},
"7", {\ar"6"},
"7", {\ar"9"},
"9", {\ar"8"},
\end{xy}\end{gathered}
\end{equation*}

\begin{equation*}
Q_{Q_{11}}\colon\quad\begin{gathered}\begin{xy} 0;<0.9pt,0pt>:<0pt,-0.9pt>:: 
(69,0) *+{3,1} ="0",
(127,0) *+{3,2} ="1",
(49,69) *+{2,1} ="2",
(106,69) *+{2,2} ="3",
(0,25) *+{4,1} ="4",
(84,25) *+{4,2} ="5",
(51,129) *+{1,1} ="6",
(108,129) *+{1,2} ="7",
(159,69) *+{2,3} ="8",
(161,129) *+{1,3} ="9",
(217,129) *+{0} ="11",
"1", {\ar"0"},
"0", {\ar"2"},
"3", {\ar"1"},
"2", {\ar"3"},
"4", {\ar"2"},
"6", {\ar"2"},
"3", {\ar"5"},
"3", {\ar"7"},
"8", {\ar"3"},
"5", {\ar"4"},
"7", {\ar"6"},
"7", {\ar"9"},
"9", {\ar"8"},
"11", {\ar"9"},
\end{xy}\end{gathered}
\end{equation*}

\begin{equation*}
Q_{Q_{12}}\colon\qquad
\begin{gathered}\begin{xy} 0;<0.8pt,0pt>:<0pt,-0.8pt>:: 
(85,0) *+{3,1} ="0",
(155,0) *+{3,2} ="1",
(54,80) *+{2,1} ="2",
(132,80) *+{2,2} ="3",
(0,37) *+{4,1} ="4",
(100,37) *+{4,2} ="5",
(55,144) *+{1,1} ="6",
(133,144) *+{1,2} ="7",
(202,80) *+{2,3} ="8",
(202,144) *+{1,3} ="9",
(270,80) *+{2,4} ="10",
(272,144) *+{1,4} ="11",
"1", {\ar"0"},
"0", {\ar"2"},
"3", {\ar"1"},
"2", {\ar"3"},
"4", {\ar"2"},
"6", {\ar"2"},
"3", {\ar"5"},
"3", {\ar"7"},
"8", {\ar"3"},
"5", {\ar"4"},
"7", {\ar"6"},
"7", {\ar"9"},
"9", {\ar"8"},
"8", {\ar"10"},
"11", {\ar"9"},
"10", {\ar"11"},
\end{xy}\end{gathered}
\end{equation*}
\caption{Quivers for the Q familiy.}\label{Qquivers}
\end{center}
\end{figure}

\begin{table}
\begin{center}
\begin{tabular}{|c|c|p{12cm}|}\hline
$Q_{10}$&&(4,1),(3,1),(2,2),(1,1),(2,3),(2,1),(1,3),(1,2),(4,2),(3,2),(2,2),(1,1), (2,3),(4,1),(3,1),(2,1),(1,3),(1,2)\\\cline{2-3}
&Type:&$(A_3,(c_{A_3})^{2}: A_3,(c_{A_3})^{2}:A_2,v: A_2, v)$\\\cline{2-3}
&$P_{\Lambda}$&$<(\gamma_{b,1},\gamma_{b,2}),b=3,4,(\gamma_{a,1},\gamma_{a,3}),a=1,2>$\\\cline{2-3}
&&(2,1),(3,2),(4,2),(1,2),(1,3),(4,1),(3,1),(1,1),(2,2),(2,3),(2,1),(4,2), (3,2),(1,2),(4,1),(3,1),(1,1),(2,2),(2,1),(4,2),(3,2),(1,2),(4,1),(3,1), (1,1),(2,2),(1,3)\\\cline{2-3}
&Type:&$(D_4, (c_{D_4})^{3}:D_4, (c_{D_4})^{3}:A_2, v)$\\\cline{2-3}
&$P_{\Lambda}$&$ <(\gamma_{1,3},\gamma_{2,3})>$\\
\hline
\hline
$Q_{11}$&&(1,1),(1,3),(2,2),(3,1),(4,1),(1,2),0,(2,1),(2,3),(1,3),(2,2),(3,2),(4,2),0, (1,1),(1,2),(2,1),(1,3),(2,3),(3,1),(4,1),(1,1)\\\cline{2-3}
&Type:&$(A_4,s_3 s_1 (c_{A_4})^{2}:A_3,(c_{A_3})^{2}:A_2,u:A_2,v)$\\\cline{2-3}
&$P_{\Lambda}$&$<\c_{a,1},\c_{a,2}> a=3,4, (\c_{2,1},\c_{2,3}), (\c_0,\c_{1,1}), (\c_{1,2},\c_{1,3})$\\\cline{2-3}
&&0,(2,3),(3,2),(1,2),(4,2),(2,1),(1,3),(2,2),(4,1), (1,1),(3,1),(3,2),(4,2),(1,2),(2,1),(2,2),(3,1),(4,1), (1,1),(3,2),(1,2),(4,2),(2,1),(2,2),(3,1),(4,1), (1,1),(2,3)\\\cline{2-3}
&Type:&$(D_4,(c_{D_4})^3:D_4,(c_{D_4})^3:A_2,u:A_1,s)$\\\cline{2-3}
&$P_{\Lambda}$&$<(\c_{1,3},\c_{2,3})>$\\
\hline
\hline
$Q_{12}$&&(1,1),(1,3),(2,4),(2,2),(3,1),(4,1),(2,3),(2,1),(1,2),(1,4),(2,4),(1,3), (2,2),(2,3),(3,2),(4,2),(1,1),(1,2),(2,1),(1,4),(1,1),(2,2),(1,3),(2,4), (4,1),(3,1)\\\cline{2-3}
&Type:&$(A_4,s_3 s_1(c_{A_4})^{2}: A_4,s_4 s_2(c_{A_3})^{2}:A_2,v: A_2, v)$\\\cline{2-3}
&$P_{\Lambda}$&$<(\gamma_{b,1},\gamma_{b,2}),b=3,4,(\gamma_{a,1},\gamma_{a,4}),(\gamma_{a,2},\gamma_{a,3}),a=1,2>$\\\cline{2-3}
&&(2,1),(4,2),(3,2),(1,2),(2,3),(1,4),(4,1),(3,1),(1,1),(2,2),(1,3),(2,4), (2,1),(4,2),(3,2),(1,2),(4,1),(3,1),(1,1),(2,2),(2,1),(4,2),(3,2),(1,2), (2,3),(1,4),(4,1),(3,1),(1,1),(2,2)\\\cline{2-3}
&Type:&$(D_4, (c_{D_4})^{3}:D_4, (c_{D_4})^{3}:A_2, u:A_2,v)$\\\cline{2-3}
&$P_{\Lambda}$&$ <(\gamma_{1,a},\gamma_{2,a}), a=3,4>$\\
\hline
\end{tabular}
\end{center}
\caption{Weyl--factorized sequences for the quivers of figure \ref{Qquivers}.}
\end{table}

\begin{figure}
\begin{center}
\begin{equation*}
\mu(Q_{Q_{10}})\colon\qquad\begin{gathered}\begin{xy} 0;<1pt,0pt>:<0pt,-1pt>:: 
(59,0) *+{4,1} ="0",
(102,0) *+{4,2} ="1",
(29,100) *+{1,1} ="2",
(72,100) *+{1,2} ="3",
(59,73) *+{2,1} ="4",
(102,73) *+{2,2} ="5",
(86,43) *+{3,1} ="6",
(131,43) *+{3,2} ="7",
(0,129) *+{2,3} ="8",
(45,129) *+{1,3} ="9",
"1", {\ar"0"},
"0", {\ar"4"},
"5", {\ar"1"},
"3", {\ar"2"},
"2", {\ar"4"},
"2", {\ar"8"},
"5", {\ar"3"},
"9", {\ar"3"},
"4", {\ar"5"},
"6", {\ar"4"},
"5", {\ar"7"},
"7", {\ar"6"},
\end{xy}\end{gathered}
\end{equation*}
\caption{\label{otherQquiv}The quiver in the mutation class of $Q_{10}$ of example
\eqref{exother}.}
\end{center}
\end{figure}
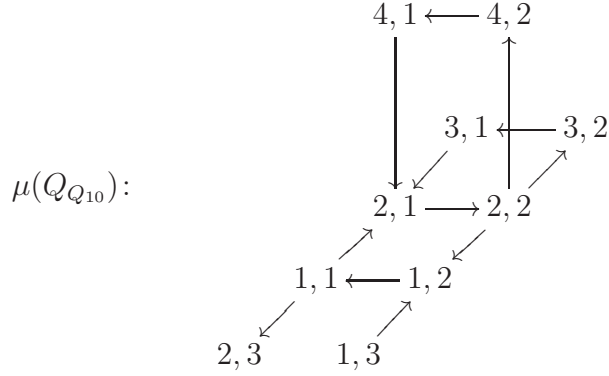
\begin{table}
\caption{Weyl--factorized sequences associated with the quiver of figure \ref{otherQquiv}. Both of them corresponds to mutations that are the identity on the quiver. The first one, being of order 1 corresponds to the full quantum monodromy, the second to the half--monodromy.}\label{otherQ}
\begin{center}
\begin{tabular}{|c|c|p{12cm}|}\hline
$Q_{10}$&&(4,1),(3,1),(1,1),(2,2),(4,2),(3,2),(1,3),(1,2),(2,1), (4,1),(3,1),(1,1),(2,2),(4,2),(3,2),(2,3),(1,2),(2,1), (4,1),(3,1),(1,3),(1,1),(2,2),(4,2),(3,2),(1,2),(2,1),(2,3)\\
\cline{2-3}
&Type:&$(A_2,c^3:A_2,c^3:A_2,c^3:A_2,c^3:A_1,c^2:A_1,c^2)$\\
\cline{2-3}
\cline{2-3}
&&(2,1),(4,2),(3,2),(1,3),(1,2),(4,1),(3,1),(1,1),(2,2),(2,1), (4,2),(3,2),(1,2),(4,1),(3,1),(1,1),(2,2),(2,1),(4,2),(3,2), (1,2),(4,1),(3,1),(1,1),(2,2),(2,3)\\
\cline{2-3}
&Type:&$(D_4,(c_{D_4})^3: D_4,(c_{D_4})^3: A_1,s :A_1,s ) $\\
\hline
\end{tabular}
\end{center}
\end{table}

\begin{figure}
\begin{center}
\begin{equation*}
Q_{S_{11}}\colon\qquad\begin{gathered}
\xymatrix{\begin{xy} 0;<0.8pt,0pt>:<0pt,-0.8pt>:: 
(64,0) *+{1,1} ="0",
(131,0) *+{1,2} ="1",
(48,70) *+{2,1} ="2",
(117,70) *+{2,2} ="3",
(0,29) *+{4,1} ="4",
(84,29) *+{4,2} ="5",
(47,125) *+{3,1} ="6",
(118,125) *+{3,2} ="7",
(186,70) *+{2,3} ="8",
(187,125) *+{3,3} ="9",
(202,0) *+{1,3} ="10",
"1", {\ar"0"},
"0", {\ar"2"},
"3", {\ar"1"},
"1", {\ar"10"},
"2", {\ar"3"},
"4", {\ar"2"},
"6", {\ar"2"},
"3", {\ar"5"},
"3", {\ar"7"},
"8", {\ar"3"},
"5", {\ar"4"},
"7", {\ar"6"},
"7", {\ar"9"},
"9", {\ar"8"},
"10", {\ar"8"},
\end{xy}}
\end{gathered}
\end{equation*}

\begin{equation*}
Q_{S_{12}}\colon\quad\begin{gathered}\begin{xy} 0;<0.8pt,0pt>:<0pt,-0.8pt>:: 
(69,0) *+{3,1} ="0",
(127,0) *+{3,2} ="1",
(49,69) *+{2,1} ="2",
(106,69) *+{2,2} ="3",
(0,25) *+{4,1} ="4",
(84,25) *+{4,2} ="5",
(51,129) *+{1,1} ="6",
(108,129) *+{1,2} ="7",
(159,69) *+{2,3} ="8",
(161,129) *+{1,3} ="9",
(184,0) *+{3,3} ="10",
(217,129) *+{0} ="11",
"1", {\ar"0"},
"0", {\ar"2"},
"3", {\ar"1"},
"1", {\ar"10"},
"2", {\ar"3"},
"4", {\ar"2"},
"6", {\ar"2"},
"3", {\ar"5"},
"3", {\ar"7"},
"8", {\ar"3"},
"5", {\ar"4"},
"7", {\ar"6"},
"7", {\ar"9"},
"9", {\ar"8"},
"10", {\ar"8"},
"11", {\ar"9"},
\end{xy}\end{gathered}
\end{equation*}

\begin{equation*}
Q_{W_{13}}\colon\quad\begin{gathered}
\xymatrix{3,1 \ar[r]& 3,2 \ar[d] & 3,3\ar[r]\ar[l] & 3,4\ar[d] &\\
2,1\ar[u]\ar[d] & 2,2\ar[l]\ar[r] & 2,3\ar[u]\ar[d] & 2,4\ar[l] &\\
1,1,\ar[r] & 1,2\ar[u] & 1,3\ar[l]\ar[r] & 1,4\ar[u] & 0\ar[l]}
\end{gathered}
\end{equation*}

\caption{Quivers for the S familiy and for the W13 theory.}\label{SWquivers}
\end{center}
\end{figure}

\begin{table}
\begin{center}
\begin{tabular}{|c|c|p{12cm}|}\hline
$W_{13}$&&(3,2),(3,4),(2,1),(2,3),(1,2),(1,4),(3,1),(3,3),(2,2),(1,1),(1,3),0, (3,2),(2,1),(2,4),(1,2),(1,4),(3,1),(3,4),(2,3),(1,1),(1,3),0, (3,3),(2,2),(2,4),(1,2),(1,4),(3,2),(3,4),(2,1),(2,3),(1,1),(1,3),0\\\cline{2-3}
&Type:&$(A_5,(c_{A_5})^{3}:A_4,s_3 s_1 (c_{A_4})^{3}:A_4,s_4 s_2(c_{A_4})^{3})$\\\cline{2-3}
&$P_{\Lambda}$&$<(\c_0,\c_{1,1}),(\c_{1,2},\c_{1,4}),(\c_{a,1},\c_{a,4}),(\c_{a,2},\c_{a,3}), a= 2,3>$\\\cline{2-3}
&&(3,1),(1,1),(2,2),(3,3),(1,3),(2,4),0,(2,1),(3,2),(1,2),(2,3),(3,4),(1,4), (3,1),(1,1),(2,2),(3,3),(1,3),(2,4),(2,1),(3,2),(1,2),(2,3),(3,4),(1,4)\\\cline{2-3}
&Type:&$(A_3, (c_{A_3})^2:A_3, (c_{A_3})^2:A_3, (c_{A_3})^2:A_3, (c_{A_3})^2:A_1:s)$\\\cline{2-3}
&$P_{\Lambda}$&$<(\c_{1,a},\c_{3,a}),a=1,3>$\\
\hline
\hline
$S_{11}$&&(1,1),(1,3),(3,1),(3,3),(2,2),(4,1),(1,2),(3,2),(2,1),(2,3),(4,2), (1,1),(1,3),(3,1),(3,3),(2,2),(1,2),(3,2),(2,1),(2,3),(4,1)\\\cline{2-3}
&Type:&$(A_3,(c_{A_3})^{2}: A_3,(c_{A_3})^{2}:A_3,(c_{A_3})^{2}: A_2, v)$\\\cline{2-3}
&$P_{\Lambda}$&$<(\gamma_{a,1},\gamma_{a,3}),a=1,2,3,(\gamma_{4,1},\gamma_{4,2})>$\\\cline{2-3}
&&(2,1),(4,2),(1,2),(3,2),(2,3),(4,1),(1,1),(3,1),(2,2),(1,3),(3,3), (2,1),(4,2),(1,2),(3,2),(2,3),(4,1),(1,1),(3,1),(2,2),(2,1),(4,2), (1,2),(3,2),(4,1),(1,1),(3,1),(2,2),(3,3),(1,3)\\\cline{2-3}
&Type:&$(D_4, (c_{D_4})^{3}:D_4, (c_{D_4})^{3}:A_3, (c_{A_3})^{2})$\\\cline{2-3}
&$P_{\Lambda}$&$ <(\gamma_{1,3},\gamma_{3,3})>$\\
\hline
\hline
$S_{12}$&&(3,1),(3,3),(1,1),(1,3),(2,2),(4,1),(3,2),(1,2),0,(2,3),(2,1),(4,2), (1,3),(1,1),(2,2),(3,1),(3,3),0,(1,2),(1,1),(1,3),(2,3),(2,1),(3,2),(4,1)\\\cline{2-3}
&Type:&$(A_4,s_3 s_1 (c_{A_4})^{2} :A_3,(c_{A_3})^{2}:A_3,(c_{A_3})^{2}:A_2,v)$\\\cline{2-3}
&$P_{\Lambda}$&$<(\c_{1,1},\c_{0}),(\c_{1,2},\c_{1,3}), (\c_{a,1},\c_{a,3}), a=2,3, (\c_{4,1},\c_{4,2})>$\\\cline{2-3}
&&(2,1),(4,2),(3,2),(1,2),(2,3),0,(4,1),(3,1),(1,1),(2,2),(3,3),(1,3),(2,1), (4,2),(3,2),(1,2),(2,3),(4,1),(3,1),(1,1),(2,2),(2,1),(4,2),(3,2),(1,2), (4,1),(3,1),(1,1),(2,2),(1,3),(3,3)\\\cline{2-3}
&Type:&$(D_4, (c_{D_4})^{3}:D_4, (c_{D_4})^{3}:A_3,(c_{A_3})^{2}:A_1,s)$\\\cline{2-3}
&$P_{\Lambda}$&$<\c_{1,3},\c_{3,3}>$\\
\hline
\end{tabular}
\end{center}
\caption{Weyl--factorized sequences for the quivers of figure \ref{SWquivers}.}
\end{table}

%%%%%%%%%%%%%%%%%%%%%%%%%%%%%%%%%%%%%%%%%


\newpage

\begin{thebibliography}{50}

\bibitem{Seiberg:1994rs}
N.~Seiberg and E.~Witten, ``Electric-magnetic duality, monopole condensation,
  and confinement in {$\N=2$} supersymmetric {Y}ang-{M}ills theory,'' {\em
  Nucl. Phys.} {\bf B426} (1994) 19--52,
\href{http://www.arXiv.org/abs/hep-th/9407087}{{\tt hep-th/9407087}}.
%%CITATION = NUPHA,B426,19;%%.

\bibitem{Seiberg:1994rs2}
N.~Seiberg and E.~Witten, ``Monopoles, dulaity and chiral symmetric breaking in $N=2$ supersymmetric QCD,'' {
  Nucl. Phys.} {\bf B431} (1994) 485--550,
\href{http://www.arXiv.org/abs/hep-th/9408099}{{\tt hep-th/9408099}}.

\bibitem{Gaiotto:2008cd}
D.~Gaiotto, G.~W. Moore, and A.~Neitzke, ``{Four-dimensional wall-crossing via
  three-dimensional field theory},''
\href{http://www.arXiv.org/abs/0807.4723}{{\tt 0807.4723}}.


\bibitem{Gaiotto:2009hg}
D.~Gaiotto, G.~W. Moore, and A.~Neitzke, ``{Wall-crossing, Hitchin Systems, and
  the WKB Approximation},''
\href{http://www.arXiv.org/abs/0907.3987}{{\tt 0907.3987}}.

\bibitem{gaiotto}
D.~Gaiotto, ``N=2 dualities,''
\href{http://www.arXiv.org/abs/0904.2715}{{\tt 0904.2715}}.


%%%%%%%%%%%%%%%%%


\bibitem{douglas1}
M.R.~Douglas, B.~Fiol and C.~R\"omelsberger,
``Stability and BPS branes,''
{\tt hep-th/0002037}.

\bibitem{douglas2}
M.R.~Douglas, B.~Fiol and C.~R\"omelsberger,
``The spectrum of BPS branes on a noncompact Calabi--Yau,''
{\tt hep-th/0003263}.


\bibitem{fiol1}
B.~Fiol and M.~Marino,
``BPS states and algebras from quivers,''
{\tt hep-th/0006189}.


\bibitem{lerche}
W. Lerche, ``On a boundary CFT description of nonperturbative $\cn=2$ Yang--Mills theory''
{\tt hep-th/0006100.}

\bibitem{fiol2}
B.~Fiol,
``The BPS spectrum of N=2 SU(N) SYM
and parton branes,''
{\tt hep-th/0012079.}



\bibitem{denef}
F.~Denef,
``Quantum quivers and Hall/Holes Halos,''
JHEP 0210 (2002) 023,
\href{http://www.arXiv.org/abs/hep-th/0206072}{{\tt hep-th/0206072}}.

\bibitem{cnv}
S.~Cecotti, A.~Neitzke and C.~Vafa, ``$R$--Twisting and $4d/2d$ correspondences,'' 
\href{http://www.arXiv.org/abs/1006.3435}{{\tt 1006.3435}}.



\bibitem{ks1}
M.~Kontsevich and Y.~Soibelman, ``{S}tability structures, motivic
  {D}onaldson-{T}homas invariants and cluster transformations,''
  \href{http://www.arXiv.org/abs/0811.2435}{{\tt 0811.2435}}.



\bibitem{Dimofte:2009bv}
T.~Dimofte and S.~Gukov, ``{Refined, Motivic, and Quantum},'' {\em Lett. Math.
  Phys.} {\bf 91} (2010) 1,
\href{http://www.arXiv.org/abs/0904.1420}{{\tt 0904.1420}}.
%%CITATION = 0904.1420;%%.

\bibitem{Dimofte:2009tm}
T.~Dimofte, S.~Gukov, and Y.~Soibelman, ``{Quantum Wall Crossing in N=2 Gauge
  Theories},''
\href{http://www.arXiv.org/abs/0912.1346}{{\tt 0912.1346}}.
%%CITATION = 0912.1346;%%.

\bibitem{Cecotti:2009uf}
S.~Cecotti and C.~Vafa, ``{BPS Wall Crossing and Topological Strings},''
\href{http://www.arXiv.org/abs/0910.2615}{{\tt 0910.2615}}.


%%%%%%%%%%%%%%%%%%%%%%%%%%



\bibitem{MR1887642}
S.~Fomin and A.~Zelevinsky, ``Cluster algebras. {I}. {F}oundations,'' {\em J.
  Amer. Math. Soc.} {\bf 15} (2002), no.~2, 497--529 (electronic).

\bibitem{MR2004457}
S.~Fomin and A.~Zelevinsky, ``Cluster algebras. {II}. {F}inite type
  classification,'' {\em Invent. Math.} {\bf 154} (2003), no.~1, 63--121.


\bibitem{fominIV}
S.~Fomin and A.~Zelevinsky, ``Cluster algebras IV: Coefficients,''
Compos, Mathh. \textbf{143} (2007) 112--164 {\tt math.RA/0602259.}


\bibitem{MR2383126}
S.~Fomin and N.~Reading, ``Root systems and generalized associahedra,'' in {\em
  Geometric combinatorics}, vol.~13 of {\em IAS/Park City Math. Ser.},
  pp.~63--131.
\newblock Amer. Math. Soc., Providence, RI, 2007.
\newblock \href{http://www.arXiv.org/abs/math/050551}{{\tt math/050551}}.

\bibitem{MR2132323}
S.~Fomin and A.~Zelevinsky, ``Cluster algebras: notes for the {CDM}-03
  conference,'' in {\em Current developments in mathematics, 2003}, pp.~1--34.
\newblock Int. Press, Somerville, MA, 2003.

\bibitem{cluster-intro}
B.~Keller, ``{C}luster algebras, quiver representations and triangulated
  categories,'' \href{http://www.arXiv.org/abs/0807.1960}{{\tt 0807.1960}}.

\bibitem{qd-cluster}Fock, V. V. and Goncharov, A. B. {``{T}he quantum dilogarithm and representations of quantum cluster varieties,''} {\tt math/0702397}.

\bibitem{clqd2}Fock, V. V. and Goncharov, A. {``{C}luster ensembles, quantization and the dilogarithm {II}: the intertwiner,''} {\tt math/0702398}. 
 
 \bibitem{qd-pentagon}Goncharov, A. B. {``{P}entagon relation for the quantum dilogarithm and quantized {${M}^{cyc}_{0,5}$},''} {\tt 0706.4054}. 
% 
% 

\bibitem{shapere}
A.D. Shaper and C. Vafa, ``BPS structure of Argyres--Douglas superconformal theories'',
{\tt arXiv:hep-th/9910182.}


\bibitem{Gukov:1999ya} S. Gukov, C. Vafa and E. Witten, {``{CFT}'s from {C}alabi-{Y}au four-folds,''} {\it Nucl. Phys.}  {\bf B584} (2000) 69-108,  {\tt hep-th/9906070}.  



\bibitem{min1} E.~Martinec, ``Algebraic geometry and effective Lagrangians'', Phys. Lett. \textbf{217B} (1989) 431.
\bibitem{min2} C.~Vafa and N.P.~Warner, ``Catastrophes and the classification of con formal theories'', Phys. Lett. \textbf{43} (1989) 730.
 \bibitem{rings} W. Lerche, C. Vafa, abd N.P. Warner, ``Chiral rings in $\cn=2$ superconformal theories'', Nucl. Phys. {\bf 324 B} (1989) 427--474.

\bibitem{Cecotti:1993rm}
S.~Cecotti and C.~Vafa, ``On classification of {$\N=2$} supersymmetric theories,'' { Commun. Math. Phys.} {\bf 158} (1993) 569--644,
\href{http://www.arXiv.org/abs/hep-th/9211097}{{\tt hep-th/9211097}}.

\bibitem{ar}
V.I. Arnold, S. Gusejn--Zade, and A. Varchenko, \textit{Singularities of differentiable maps,} Monographs in Mathematics 82. Birkh\"auser (1985).

\bibitem{eb} 
W. Ebeling, \textit{Functions of several complex variables and their singularities,} Graduate Studies in Mathematics 83, American Mathematical Society, Providence RI, (2007).

\bibitem{CV11}
S. Cecotti and C. Vafa, ``Classification of complete $\cn=2$ supersymmetric theories in $4$ dimensions'', 
{\tt arXiv:1103.5832[hep-th].}


\bibitem{le1}
H. Lenzing and J.A. De La Pe\~na, ``Extended canonical algebras and Fuchsian singularities'', 
{\tt arXiv:math/0611532 [math.RT]}.


\bibitem{le2}
H. Lenzing and J.A. De La Pe\~na, ``Spectral analysis of finite dimensional algebras and singularities'', 
{\tt arXiv:0805.1018 [math.RT]}.

\bibitem{le3} H. Lenzing, ``Rings of singularities'', Advanced school and conference on homological and geometrical methods in representation theory, (ICTP, January 18--February 5, 2010). Notes available from the ICTP website.


\bibitem{perrev}
A. Skowronski, ``Periodicity in representations theory of algebras'', 
Lecture Notes of the \textit{ICTP Advanced School and Conference on Representation Theory and Related Topics,} (9--27 January 2006)  Available from the ICTP website.

\bibitem{keller-periodicity}
B.~Keller, ``{T}he periodicity conjecture for pairs of {D}ynkin diagrams,''
  \href{http://www.arXiv.org/abs/1001.1531}{{\tt 1001.1531}}.

 \bibitem{zamolodchikovTBA} Al. B. Zamolodchikov, ``On the thermodynamical Bethe ansatz for reflectionless ADE scattering theories'', Phys. Lett. \textbf{B 253} (1991) 391--394.


\bibitem{milnor}
J. Milnor, ``Singular points of complex hypersurfaces,''
Annals of Math. Studies \textbf{61}, Princeton, 1968.


\bibitem{kac}
V.G.~Kac, \textit{Infinite dimensional Lie algebras,}
Third edition, Cambridge University press, 1990.


\bibitem{cachazo}
F. Cachazo, B. Fiol, K. Intriligator, S. Katz, and C. Vafa,
``A geometric unification of dualities'', {\tt arXiv:hep-th/0110028.}

\bibitem{HIV}K.~Hori, A.~Iqbal and C.~Vafa,
 {``D-branes and mirror symmetry,''}
{\tt  arXiv:hep-th/0005247.}


\bibitem{Ring}
C.M. Ringel, \textit{Tame algebras and integral quadratic forms,} Springer Lectures Notes in Mathematics 1099, Springer (1984).


\bibitem{AL1} P. Gabriel and A.V. Roiter, \textit{Representations of finite--dimensional algebras},
Encyclopaedia of Mathematical Sciences, \textsc{Algebra VIII}, vol. 73, A.I. Kostrikin and I.R. Shafarevich Eds., Springer--Verlag (1991). 

\bibitem{AL2} M. Auslander, I. Reiten, and S.O. Smalo, \textit{Representation theory for Artin algebras,}
Cambridge studies in advanced mathematics, vol 36. Cambridge University Press (1995).

\bibitem{AL3} I. Assem, D. Simson, and A. Skowronski, \textit{Elements of the representation theory of associative algebras. Volume 1. Techniques of representation theory,} London Mathematical Society Student Texts
\textbf{65}, Cambridge University Press (2006). 


\bibitem{kelP} 
B. Keller, ``Deformed Calabi--Yau completions'', 
{\tt arXiv:0908.3499 [math.RT]}.


\bibitem{kelapp}
B.~Keller, ``Quiver mutation in Java'', available from the author's homepage, {\tt http://www.institut.math.jussieu.fr/$\widetilde{\phantom{-}}$keller/quivermutation}.

\bibitem{eb2}
W. Ebeling, ``On Coxeter--Dynkin diagrams of hypersurface singularities'', 
J. Math. Sciences \textbf{82} (1996) 3657--3664.

\bibitem{kacthm}
V. Kac, ``Infinite root systems, representations of graphs, and invariant theory'',
Invent. Math. \textbf{56} (1980) 57--92.\\
V. Kac, ``Infinite root systems, representations of graphs, and invariant theory. II'',
J. Algebra \textbf{78} (1982) 141--162.\\
V. Kac, \textit{Root systems, representations of quivers and invariant theory}, 
(Montecatini, 1982). Lecture Notes in Mathemathics \textbf{996} Springer (1983) pages 74--108.

\bibitem{sko}
D. Simson, A. Skowronski, \textit{Elements of the representation theory of associative algebras. Volume 2: Tubes and concealed algebras of Euclidean type}, London Mathematical Society Student Texts \textbf{71} Cambridge University Press (2007).

\bibitem{cox}
R. Stekolshchik,
\textit{Notes on the Coxter transformations and the McKay correspondence},
Springer Monographs in Mathematics (2008).

\bibitem{craw}
W. Crawley--Boevey,
``Lectures on representations of quivers''
available from author's web--page
{\tt www.amsta.lead.ac.uk/~pmtwc/quivlecs.pdf.}

\bibitem{zel1}
H. Derksen, J. Weyman, and A. Zelevinsky,
``Quivers with potentials and their representations I: mutations'',
{\tt arXiv:0704.0649 [math.RA]}.


\bibitem{zel2}
A. Zelevinsky,
``Mutations for quivers with potentials: Oberwolfach talk, April 2007'',
{\tt arXiv:0706.0822 [math.RA]}.


\bibitem{zel3}
H. Derksen, J. Weyman, and A. Zelevinsky,
``Quivers with potentials and their representations II: applications to cluster algebras'',
J. Amer. Math. Soc. \textbf{23} (2010) 749--790.
{\tt arXiv:0904.0676 [math.RA]}.

\bibitem{kel}
B. Keller,
``On cluster theory and quantum dilogarithm identities'',
{\tt arXiv:1102.4148 [math.RT].}

\bibitem{BGP}
I.N. Bernstein, I.M. Gelfand, and V.A. Ponomariev,
``Coxeter functors, and Gabriel's theorem'', Uspehi Mat. Nauk \textbf{28} (1973) 19--33.
English translation: Russian Math. Surveys \textbf{28} (1973) 17--32.


\bibitem{Ravanini:1992fi}
F.~Ravanini, R.~Tateo, and A.~Valleriani, ``{Dynkin TBAs},'' {\em Int. J. Mod.
  Phys.} {\bf A8} (1993) 1707--1728,
\href{http://www.arXiv.org/abs/hep-th/9207040}{{\tt hep-th/9207040}}.
%%CITATION = HEP-TH/9207040;%%.

\bibitem{Kuniba:1993cn}
A.~Kuniba, T.~Nakanishi, and J.~Suzuki, ``{Functional relations in solvable
  lattice models. 1: Functional relations and representation theory},'' {\em
  Int. J. Mod. Phys.} {\bf A9} (1994) 5215--5266,
\href{http://www.arXiv.org/abs/hep-th/9309137}{{\tt hep-th/9309137}}.
%%CITATION = HEP-TH/9309137;%%.

\bibitem{dolgachev}
I.V. Dolgachev and V.A. Iskovskikh, ``Finite subgroups of the plane Cremona group'', 
in \textit{Algebra, arithmetic and geometry: in honor of Yu.I. Manin,} Vol. I, 443--548, Progress in Mathematics \textbf{269} Birkh\"auser, Boston (2009).


\bibitem{FomLecture}
S.~Fomin, Lectures at Aarhus University, (June 14th--18th 2010).
Video available from 
{\tt http://qgm.au.dk/video/clusalg/.}
\end{thebibliography}
 \end{document}